\pdfoutput=1
\documentclass[12pt]{article}
\usepackage[bookmarks=true]{hyperref}
\usepackage{graphicx,epsfig,color}
\usepackage{cite}
\usepackage[english]{babel}
\usepackage{amssymb,amsfonts,amsmath}
\usepackage{verbatim}
\usepackage{physics}

\usepackage[top=2.5cm, bottom=2.5cm, left=2cm, right=2cm]{geometry}	

\newcommand{\be}{\begin{equation}}
\newcommand{\ee}{\end{equation}}
\newcommand{\beq}{\begin{equation}}
\newcommand{\eeq}{\end{equation}}
\newcommand{\bea}{\begin{eqnarray}}
\newcommand{\eea}{\end{eqnarray}}

\newcommand{\ba}{\begin{eqnarray}}
\newcommand{\ea}{\end{eqnarray}}

\def\ie{{\em i.e.},}

\newcommand{\rc}{\nonumber\\}
\newcommand{\bear}{\begin{eqnarray}}
\newcommand{\eear}{\end{eqnarray}}

\newcommand{\mt}[1]{\textrm{\scriptsize #1}}
\def\Nf{N_\mt{f}}
\def\nf{n_\mt{f}}
\def\Nc{N_\mt{c}}
\def\Qf{Q_\mt{f}}
\def\Qc{Q_\mt{c}}
\def\nq{n_\mt{q}}
\def\nb{n_\mt{b}}

\def\a{\alpha}

\def\d{\delta}
\def\dd{{\rm{d}}}
\def\eps{\epsilon}           
\def\f{\phi}               

\def\g{\gamma}

\def\k{\kappa}                    
\def\l{\lambda}
\def\m{\mu}
\def\n{\nu}
  \def\w{\omega}
\def\p{\pi}
\def\r{\rho}                                    
\def\rh{\rho_\mt{h}}                                    
\def\s{\sigma}                                   
\def\t{\tau}

\def\tt{\hat{\theta}}

\def\atanh{{\rm arctanh}}

\def\rt{\tilde{\rho}}

\graphicspath{{figures/}}
\numberwithin{equation}{section}
\renewcommand{\theequation}{{\rm\thesection.\arabic{equation}}}

\begin{document}

\begin{flushright}
HIP-2022-20/TH
\end{flushright}

\begin{center}

\centerline{\Large {\bf  Flavored anisotropic black holes}}

\vspace{8mm}

\renewcommand\thefootnote{\mbox{$\fnsymbol{footnote}$}}
Ana Garbayo,${}^{1,2}$\footnote{ana.garbayo.peon@usc.es} Carlos Hoyos,${}^{3,4}$\footnote{hoyoscarlos@uniovi.es} 
Niko Jokela,${}^{5,6}$\footnote{niko.jokela@helsinki.fi} \\
Jos\'e Manuel Pen\'\i n,${}^{5,6}$\footnote{jmanpen@gmail.com} and Alfonso V. Ramallo${}^{1,2}$\footnote{alfonso@fpaxp1.usc.es}

\vspace{4mm}

${}^1${\small \sl Departamento de  F\'\i sica de Part\'\i  culas and} \\
${}^2${\small \sl Instituto Galego de F\'\i sica de Altas Enerx\'\i as (IGFAE)} \\
{\small \sl Universidade de Santiago de Compostela} \\
{\small \sl E-15782 Santiago de Compostela, Spain} 
\vskip 0.2cm

${}^3${\small \sl Department of Physics and } \\
${}^4${\small \sl Instituto de Ciencias y Tecnolog\'\i as espaciales de Asturias (ICTEA) }\\
{\small \sl Universidad de Oviedo} \\
{\small \sl E-33007/33004, Oviedo, Spain} 
\vskip 0.2cm

${}^5${\small \sl Department of Physics} and ${}^6${\small \sl Helsinki Institute of Physics} \\
{\small \sl P.O.Box 64} \\
{\small \sl FIN-00014 University of Helsinki, Finland} 
\vskip 0.2cm

\end{center}

\vspace{6mm}
\numberwithin{equation}{section}
\setcounter{footnote}{0}
\renewcommand\thefootnote{\mbox{\arabic{footnote}}}

\begin{abstract}
We construct a black hole geometry dual to a (2+1)-dimensional defect in an ambient (3+1)-dimensional gauge theory at non-zero temperature and quark density. The geometry is a solution to the equations of motion of type IIB supergravity with brane sources, a low energy limit of an intersection of stacks of color D3-branes and flavor D5-branes. We consider the case in which the number of D5-branes is large and they can be homogeneously distributed along the directions orthogonal to the defect, creating in this way a multilayer structure. The quark density is generated by exciting a gauge field in the worldvolume of the dynamic brane sources.  We study the thermodynamics of the anisotropic black hole and compute the energy density of the dual theory, as well as the pressures and speeds of sound along the directions parallel and orthogonal to the defect. We also calculate transport coefficients in the shear channel, quark-antiquark potentials, and the entanglement entropies for slab subregions. These analyses give us a good overview on how the degrees of freedom are spread, entangled, and behave in this unquenched system in the deconfining phase at strong coupling.

\end{abstract}

\newpage
\tableofcontents

\newpage

\section{Introduction}

The holographic AdS/CFT correspondence\cite{Maldacena:1997re} is one of the most powerful tools to study strongly coupled systems in a large variety of situations in high-energy  and condensed matter  physics\cite{Casalderrey-Solana:2011dxg,McGreevy:2009xe,Ramallo:2013bua}. One interesting and physically relevant situation arises in anisotropic states in which one or more of the spatial directions are distinguished from the rest while remaining homogeneous. In the condensed matter realm, nematic phases appear in a variety of systems with strongly correlated electrons (see \cite{2010ARCMP...1..153F} for a review on the topic). If an anisotropic phase was developed in QCD at large baryon density this could lead to more compact neutron stars, leading to interesting connections with black hole thermodynamics \cite{Yagi:2015upa,Yagi:2016ejg}. In such condensed matter and of QCD phases there are two basic ingredients: a finite density and a source of anisotropy. Our goal is to present a top-down holographic dual that realizes both and could be applied to model systems with similar characteristics. We will also study thermodynamics and other properties of the anisotropic phases.

Many gravity backgrounds dual to anisotropic field theories have been studied in the literature. These include initial anisotropic states of the quark-gluon plasma formed in heavy ion collisions \cite{Janik:2008tc,Chesler:2008hg}, states with axionic/dilatonic sources \cite{Azeyanagi:2009pr,Mateos:2011ix,Mateos:2011tv,Koga:2014hwa,Jain:2014vka,Cheng:2014qia,Banks:2015aca,Roychowdhury:2015cva,Roychowdhury:2015fxf,Jahnke:2015obr,Banks:2016fab,Avila:2016mno,Donos:2016zpf,Giataganas:2017koz,Itsios:2018hff,Arefeva:2018hyo}, electromagnetic fields \cite{Karch:2007pd,Albash:2007bk,Albash:2007bq,Erdmenger:2007bn,DHoker:2009mmn,DHoker:2009ixq,Jensen:2010vd,Evans:2010hi,DHoker:2010zpp,Kim:2010pu,Hoyos:2011us,Ammon:2012qs,Gursoy:2018ydr,Evans:2011tk,Kharzeev:2011rw,Donos:2011pn,Jokela:2015aha,Itsios:2016ffv}, p-wave superfluids \cite{Gubser:2008wv,Ammon:2008fc,Basu:2008bh,Iizuka:2012iv,Donos:2012gg,Iizuka:2012wt}, and multilayered systems \cite{Conde:2016hbg,Penin:2017lqt,Jokela:2019tsb,Gran:2019djz,Hoyos:2020zeg,Hoyos:2021vhl}.

In this paper we will construct a new anisotropic top-down supergravity background based on a low energy limit of D-brane intersections. These types of models contain several stacks of branes and anisotropy is naturally induced due to different orientations of the intersecting stacks. Actually, we will be dealing with two classes of branes. The first class consists of color D3-branes placed at the tip of a cone over a five-dimensional Sasaki-Einstein space. These branes contain the color degrees of freedom of the gauge theory, \ie\  fields that transform in the adjoint representation of the gauge group, and are dual to supergravity fluxes.

The second class are flavor branes \cite{Karch:2002sh}, which are dual to degrees of freedom transforming in the fundamental representation of the gauge group. These branes are governed by the Dirac-Born-Infeld (DBI) plus Wess-Zumino (WZ) actions, which act as sources for the supergravity fields. When the number $\Nc$ of color branes is much greater than the number $\Nf$ of flavor branes, it is justified to work in the so-called probe approximation, in which one neglects the contribution on the supergravity background that would otherwise be induced by the flavor branes. However, when $\Nf\sim \Nc$ the probe approximation breaks down and one has to take into account the backreaction of the flavor branes on the geometry. 

In this paper we consider the case of the intersection of color D3-branes and flavor D5-branes. In this setup the flavor branes share two spatial directions with the color branes, giving rise to an anisotropy between the two common spatial directions and the third one. The field theory dual to this configuration is well-known and consists of a theory with hypermultiplets living in a $(2+1)$-dimensional defect, coupled to the ambient four-dimensional super Yang-Mills theory \cite{DeWolfe:2001pq,Erdmenger:2002ex,Domokos:2022lvn}.  We construct the (backreacted) geometry for this D3-D5-brane intersection by using the smearing technique, in which the stack of flavor branes is considered as a continuous distribution. This technique has been used in a large variety of setups (see \cite{Nunez:2010sf} for a review and references). The smearing of the branes makes the equations of motion easier to solve since Einstein equations are ODEs instead of PDEs that appear when the flavor branes are localized.  This approach is accurate in the Veneziano limit, where $\Nf,\Nc\to\infty$ while the ratio $\Nf/\Nc$ is kept fixed.  The D5-branes are smeared both in the cartesian direction orthogonal to the defect and on the internal directions and, therefore, this configuration is dual to a system of multiple $(2+1)$-dimensional layers in a $(3+1)$-dimensional ambient gauge theory. The supersymmetric ten-dimensional supergravity equations corresponding to this multilayer setup was first introduced in \cite{Conde:2016hbg} and further studied in \cite{Penin:2017lqt,Jokela:2019tsb,Gran:2019djz,Hoyos:2020zeg,Hoyos:2021vhl}. However, in this paper we present a different branch of solutions, those that can be continuously connected with flavorless backgrounds.\footnote{In \cite{Penin:2017lqt,Jokela:2019tsb,Gran:2019djz,Hoyos:2020zeg,Hoyos:2021vhl} the limit $\Nf\to 0$ is singular, a blight which however does not blemish the applicability of finite flavor solutions.}

The background we find in this paper solves the equations of motion of ten-dimensional type IIB supergravity with (smeared) D5-brane sources. The corresponding metric has a horizon, which means that our solution is a black hole that is dual to a field theory at non-zero temperature. Moreover, since our sources are dynamical objects, rather than fluxes, they have a worldvolume action which can contain a gauge field. This feature can be used to construct gravity duals of gauge theories at non-zero baryon density.\footnote{We identify the baryon symmetry with the $U(1)$ that acts equally at each defect D5-brane on the fields localized there. Due to the smearing, this effectively becomes a global $U(1)$ symmetry in 3+1 dimensions.} Indeed, as was shown in \cite{Kobayashi:2006sb, Mateos:2007vc} for the D3-D7 system in the probe approximation, in order to introduce a  non-vanishing baryon density,  one should switch on a radially dependent time component of a gauge potential on the worldvolume of the flavor brane. To obtain the backreacted geometry we have to determine how the ten-dimensional metric and fluxes of type IIB supergravity are modified by the presence of this gauge field in the DBI plus WZ action. 

The rest of this paper is organized as follows. In section~\ref{setup} we present our ansatz  for the different fields of type IIB supergravity. The equations of motion are derived in detail in appendix~\ref{appendix_A}. These equations are complicated and we have not been able to integrate them analytically. Instead, we followed the same procedure as in refs. \cite{Bigazzi:2009bk,Bigazzi:2011it} and found a perturbative solution at first order in certain expansion parameters. In section~\ref{BH_without_mu} we present our solution in the case in which the chemical potential vanishes. Details of the calculations leading to this results are given in appendix~\ref{appendix_B}. The solution found is first-order in an expansion in a temperature dependent parameter, proportional to $\lambda^{{1\over 2}}\,\nf /(\Nc\,T)$, where $\lambda$ is the 't Hooft coupling, $\nf$ is the density of flavors per unit length in the direction orthogonal to the defects, and $T$ is the temperature.

In section~\ref{BH_witth_mu} we continue to the case of nontrivial gauge potential and thereby present the solution with nonzero chemical potential at finite temperature. The calculations are detailed in appendix~\ref{appendix_C}. In this case the solution is (doubly) perturbative. The expansion parameters are the one we introduced above for zero chemical potential plus another one proportional to $\lambda^{{1\over 2}}\,\nb^2/(\Nc\,\nf\,T^5)$, where $\nb$ is the baryon density per unit volume. 

In section~\ref{Thermo} we analyze the thermodynamic properties of the background. We first compute the temperature and entropy and then we obtain the internal energy density by calculating the ADM energy of the background. In order to completely characterize the anisotropic thermodynamics of our model we employ the formalism of \cite{Caldarelli:2010xz}, which requires introducing a brane potential that measures the energy cost of adding flavor branes in the system. This analysis allows us to compute the pressures and the speeds of sound along the  different field theory directions.  Moreover, we check these results by computing directly the vacuum expectation value of the stress-energy tensor and the Euclidean on-shell action.

Section~\ref{applications} is devoted to the calculation of several observables. First of all we calculate the hydrodynamic coefficients in the shear channel at zero  quark density. The calculation requires dealing with the dimensionally reduced theory. The dimensionally reduced action was derived in \cite{Penin:2017lqt} and is reviewed in appendix~\ref{appendix_D}, where the detailed calculation of the hydrodynamic transport coefficients is performed. In section~\ref{applications}  we compute the quark-antiquark potentials in directions parallel and orthogonal to the D5-brane layers. Similarly, we calculate the entanglement entropy of parallel and orthogonal slabs. 

Finally, in section~\ref{conclusions} we summarize our results and discuss possible extensions and further applications of our work.

\section{Brane setup and the ansatz}\label{setup}

In this section we will briefly describe our general ten-dimensional background geometry. The specific characteristics of our ansatz  are presented in the subsequent sections, while the details of the calculations are relegated to the appendices. 

Our top-down model is based on the following array representing the intersection of $\Nc$ D3-branes and $\Nf$ D5-branes:
\beq
\begin{array}{cccccccccccl}
 &0&1&2&3& 4& 5&6 &7&8&9 &  \\
(\Nc)\,\,\text{D}3: & \times &\times &\times &\times & \cdot &\cdot & \cdot &\cdot &\cdot &\cdot    \\
(\Nf)\,\,\text{D}5: & \times &\times&\times&\cdot&\times&\times&\times&\cdot&\cdot&\cdot 
\end{array}
\label{D3D5intersection}
\eeq
The branes are extended in directions marked by $\times$ and localized otherwise.
In (\ref{D3D5intersection}) the D3-branes are color branes which, in the absence of the D5-branes, generate the 
$AdS_5\times {\cal M}_5$ geometry dual to a supersymmetric Yang-Mills theory in four dimensions. In the following we will assume that ${\cal M}_5$ is a compact Sasaki-Einstein manifold and, therefore, the D3-branes are placed at the tip of a six-dimensional cone over ${\cal M}_5$.

The D5-branes are flavor branes dual to matter hypermultiplets living in (2+1)-dimensional defects of the ambient (3+1)-dimensional theory. 
We want to obtain a black hole geometry incorporating the backreaction of the flavor branes. We will assume that $\Nf$ is large and, following the approach reviewed in \cite{Nunez:2010sf}, we will consider a continuous distribution of flavor branes. 
The total action of the system is the sum of the one corresponding to ten-dimensional type IIB supergravity action $S_{{\rm{IIB}}}$ and the action of the branes:
\beq
S\,=\,S_{{\rm{IIB}}}+S_{\text{branes}}\,\,.
\label{total_action}
\eeq
The ten-dimensinal metric $g$, the dilaton $\phi$, the Ramond-Ramond (RR) forms $F_1$, $F_3$, and $F_5$ and the Neveu-Schwarz (NS) three-form $H_3$ enter in the supergravity action:
\bear
&&S_{{\rm{IIB}}}\,=\,\frac{1}{2 \kappa_{10}^2} \int \dd^{10}x\sqrt{-g} \Big( R-\frac{1}{2} (\partial \phi)^2-\frac{e^{-\phi}}{2 \cdot 3!} H_3^2 -\frac{e^{2\phi}}{2}F_1^2-\frac{e^{\phi}}{2 \cdot 3!}F_3^2-\frac{1}{4 \cdot 5!}F_5^2 \Big) \rc\rc
&&\qquad\qquad\qquad\qquad\qquad
-\frac{1}{2 \kappa_{10}^2}\int \frac{1}{2} C_4 \wedge H_3 \wedge F_3\,\,,
\label{IIB_action}
\eear
and  $S_{\text{branes}}$ is the sum of the Dirac-Born-Infeld (DBI) and Wess-Zumino (WZ) action of the flavor D5-branes:
\beq
S_{\text{branes}}\,=\,-T_5 \sum_{\Nf} \int_{{\cal M}_6}\dd^6\xi e^{\frac{\phi}{2}} \sqrt{-\det(\hat{g}-e^{-\frac{\phi}{2}} \mathcal{F})}\,+\,
S_{\rm{WZ}}\,\,.
\label{Brane_action}
\eeq
In (\ref{Brane_action})  $\hat g$ is the pullback of the ten-dimensional metric to the D5-brane worldvolume, 
$\mathcal{F}=\hat B_2+F$, where $\hat B_2$ is the pullback of the two-form potential for $H_3$ ($H_3=\dd F_2$) and the two-form $F=\dd A$ is the worldvolume gauge field strength. Moreover, the WZ term $S_{\rm{WZ}}$ can be written as:
\beq
S_{\rm{WZ}}\,=\,T_5 \int \Xi \wedge \Big(C_6-C_4 \wedge \mathcal{F}+\frac{1}{2!} \,C_2 \wedge \mathcal{F} \wedge \mathcal{F}-\frac{1}{3!} \,C_0\, \mathcal{F}\wedge \mathcal{F}\wedge \mathcal{F}\Big)\,\,,
\label{WZ_action}
\eeq
where $C_i$ ($i=0,2,4,6)$ are RR potentials and the four-form $\Xi$ is the so-called smearing form, which encodes the distribution of the D5-brane charge of the stack of flavor branes. The ten-dimensional gravitational constant in (\ref{IIB_action}) and the D5-brane tension in (\ref{Brane_action}) and (\ref{WZ_action}) are given by:
\beq
2\,\kappa_{10}^2\,=\,(2\pi)^7\,g_\mt{s}^2\,\alpha'^4 \ , \ T_5\,=\,{1\over (2\pi)^5\,g_\mt{s}\,\alpha'^3}\,\,,
\eeq
with $g_\mt{s}$ and $\alpha'$ being the string constant and the Regge slope, respectively.

Let us now specify our ansatz  to solve the equations of motion stemming from the action (\ref{total_action}). The ten-dimensional metric in Einstein frame is written as
\bear
&&\dd s_{10}^2\,=\,h^{-{1\over 2}}\,\Big[
-b\,\dd t^2\,+\,\dd x_1^2\,+\,\dd x_2^2\,+\,\alpha^2\,\dd x_3^2\,\Big]\nonumber \\
&&\qquad\qquad\qquad\qquad
+\,h^{{1\over 2}}\,\Big[{G^2\over b}\,\dd \rho^2\,+\,S^2\,\dd s_{\rm{KE}}^2\,+\,F^2\,(\dd \tau+{\cal A})^2\Big]\,\,,
\label{metric_ansatz }
\eear
where all the functions $h$, $b$, $\alpha$, $G$, $S$, and $F$ depend on the non-compact holographic radial coordinate $\rho$; the boundary, corresponding to the UV, is at $\r\to\infty$. Notice that the third cartesian coordinate $x_3$ is  distinguished from $x_1$ and $x_2$. This anisotropy along the gauge theory directions is to be expected from the brane setup (\ref{D3D5intersection}).

The function $G(\rho)$ in the metric (\ref{metric_ansatz }) can be chosen at will by selecting the radial coordinate $\rho$. It is quite convenient to choose the gauge in which $G(\rho)$ is related to the other functions of the metric as:
\beq
G\,=\,{b\,\alpha\,S^4\,F\,\over \rho\,(\rho^4-\rh^4)}\,\,,
\label{G_gauge}
\eeq
where $\rh$ is a constant corresponding to the radius the event horizon of the black hole. Indeed, the blackening function $b(\rho)$ vanishes at $\rho=\rh$ and the Hawking temperature of the black hole can be obtained from the surface gravity, and it is given by:
\beq
T\,=\,\left.{1\over 2\p}{1\over \sqrt{g_{\r\r}}}{\dd \over \dd \r}\left(\sqrt{-g_{tt}}\right)\right|_{\r\to\rh}=
{1\over 4\pi\,G(\rh)}\,\partial_{\rho}\Big(b h^{-{1\over 2}}\Big)\Big|_{\rho=\rh}\,\,.
\label{T_general}
\eeq

Let us now describe the internal part of the metric (\ref{metric_ansatz }). Before adding the deformation induced by the flavor branes, the internal space is a five-dimensional Sasaki-Einstein (SE) manifold ${\cal M}_5$, whose line element can be written as a $U(1)$-bundle over a K\"ahler-Einstein (KE) space ${\cal M}_4$: 
$\dd s_{\rm{SE}}^2\,=\,\dd s_{\rm{KE}}^2+(\dd \tau +{\cal A})^2$. Let $\{e^i\}$, $i=1, \ldots, 4$, be a canonical basis of vielbein one-forms for the KE space ($\dd s_{\rm{KE}}^2\,=\,\sum_i\,(e^i)^2$). In this basis the K\"ahler two-form $J_{\rm{KE}}$ of ${\cal M}_4$ is:
\beq
 J_{\rm{KE}}\,=\,e^1\wedge e^2\,+\,e^3\wedge e^4\,\,,
\eeq
and is related to the one-form ${\cal A}$ of the metric (\ref{metric_ansatz }) as $J_{\rm{KE}}=\dd {\cal A}/2$. Let us next define the two-form $\hat\Omega_2$ as:
\beq
\hat\Omega_2\,=\,e^{3i\tau}\big(e^1+i e^2)\wedge (e^3+ie^4)\,\,.
\eeq
We will use the real and imaginary parts of $\hat\Omega_2$ in our ansatz  for the RR forms $F_5$ and $F_3$. They satisfy:
\beq
\dd \,{\rm Re}\,(\hat\Omega_2)\,=\,-3\,{\rm Im} (\hat\Omega_2)\wedge (\dd \tau+{\cal A})\,\,,
\qquad\qquad
\dd \,{\rm Im}\,(\hat\Omega_2)\,=\,3\,{\rm Re} (\hat\Omega_2)\wedge (\dd \tau+{\cal A})\,\,.
\label{d_Re_Im_Omega_2}
\eeq
We are now ready to write our ansatz for the different forms of the type IIB background. First of all, we will take $F_1=H_3=0$. 
Moreover, the RR five-form $F_5$ can be decomposed as:
\beq
F_5=F^{(0)}_5+F^{\text{cp}}_5+\star F^{\text{cp}}_5\,\,,
\label{F5_ansatz }
\eeq
where $\star$ is the ten-dimensional Hodge star operator and $F^{(0)}_5$ is the standard five-form sourced by the D3-branes:
\bea
F^{(0)}_5=K(\rho) (1+\star) \,\dd^4x \wedge \dd \rho\,\,.
\eea
We will require that  $F^{(0)}_5$  is a closed form, which in turn implies that the function $K(\rho)$ must be related to the other functions of the metric as:
\beq
K\,=\,\Qc\,{b\,\alpha^2\over h^2\,\rho(\rho^4-\rh^4)}\,\,,
\eeq
with $\Qc$ being a constant. Using standard  arguments to quantize the flux of $F_5$, we can relate the constant $\Qc$ to the number of color D3-branes $\Nc$ as:
\beq
\Qc\,=\,{(2\pi)^4\,g_\mt{s}\,\alpha'^2\,\Nc\over {\rm Vol} \big({\cal M}_5\big)}\,\,.
\eeq

In order to construct a gravity dual of a flavored theory with  non-zero chemical potential we need to have a non-vanishing worldvolume gauge field $A$ of the form:
\beq
A\,=\,A_t(\rho)\,\dd t\,\,.
\label{A_ansatz }
\eeq
This worldvolume gauge field acts as a source of the RR forms due to the coupling of $A$ to the RR potentials in the WZ action (\ref{WZ_action}). In particular, the presence of the worldvolume gauge field (\ref{A_ansatz }) induces extra components of $F_5$ that should be added to $F^{(0)}_5$ as in (\ref{F5_ansatz }). As our ansatz, we will take (below cp stands for chemical potential):
\beq
F^{\text{cp}}_5=\dd  C^{\text{cp}}_4\,\,,
\label{F_5_cp_C_4}
\eeq
where the four-form $C^{\text{cp}}_4$ depends on a new function $J(\rho)$ and is given by:
\beq
C^{\text{cp}}_4=J(\rho) \dd x^1\wedge \dd x^2 \wedge \mathrm{Re}  (\hat{\Omega}_2)\,\,.
\label{C_4_cp}
\eeq
We will also adopt the following ansatz  for the RR three-form $F_3$:
\beq
F_3=\Qf \,\dd x^3 \wedge \mathrm{Im}(\hat{\Omega}_2)\,+\,F_{123} \,\dd x^1 \wedge \dd x^2 \wedge \dd x^3\,\,,
\label{F_3_ansatz }
\eeq
where $F_{123}$ is constant. The first term in (\ref{F_3_ansatz }) enforces the violation of the Bianchi identity which, according to the WZ action (\ref{WZ_action}), takes the form:
\beq
\dd F_3\,=\,2 \kappa_{10}^2\,T_5\,\Xi\,\,.
\label{Bianchi_F3}
\eeq
By using (\ref{d_Re_Im_Omega_2}) we can find the explicit expression of the smearing form $\Xi$, namely:
\bea
2 \kappa^2_{10} T_5\,\Xi=-3\Qf \,\dd x^3 \wedge \mathrm{Re} (\hat{\Omega}_2) \wedge (\dd \tau+{\cal A})\,\,.
\label{smearnig_form}
\eea
Notice that $\Xi$ has the same form as the one introduced in \cite{Conde:2016hbg} to smear massless flavors. The component of $F_3$ proportional to $F_{123}$ in (\ref{F_3_ansatz })  is needed in order to satisfy the equations of motion when the worldvolume gauge field (\ref{A_ansatz }) is present. A similar term was  introduced in \cite{Bigazzi:2011it} in the D3-D7 system.

Notice that $\Xi$ determines the D5-brane charge distribution of the smeared stack of flavor branes. As the four-form $\Xi$ does not depend on the coordinate $x^3$, although it contains $\dd x^3$ in its expression, our stack of flavor branes is a homogeneous continuous distribution of D5-branes along the $x^3$-direction, which is dual to a system of multiple $(2+1)$-dimensional layers. As shown in \cite{Jokela:2019tsb}, the parameter $\Qf$ is proportional to the density of smeared branes along the $x^3$-direction. The precise relation between $\Qf$ and this density depends on the internal Sasaki-Einstein manifold ${\cal M}_5$. If we distribute $\Nf$ D5-branes along a distance $L_3$ in the third cartesian direction, we can define the density of flavor branes $\nf$ as:
\beq
\nf\,=\,{\Nf\over L_3}\,\,.
\eeq
The D5-branes wrap a two-dimensional submanifold ${\cal M}_2$ within the five-dimensional compact manifold ${\cal M}_5$. 
Let us define the transverse volume $v_{\perp}$ of ${\cal M}_2$ as the ratio:
\beq
v_{\perp}\,=\,{{\rm Vol} ({\cal M}_5)\over 
{\rm Vol} ({\cal M}_2)}\,\,.
\eeq
Then, $\Qf$ is just given by:
\beq
\Qf\,=\,{2\pi^2\,g_\mt{s}\,\alpha'\,\over 3}\,{\nf\over v_{\perp}}\,\,.
\eeq
Interestingly, $\Qf$ can be rewritten as:
\beq
\Qf\,=\,{2\kappa_{10}^2\,T_5\over 3!}\,{\nf\over v_{\perp}}\,\,,
\eeq
and it is just proportional to the ratio between the tension of the D5-brane and the gravitational constant $(2\kappa_{10}^2)^{-1}$. 
When ${\cal M}_5={\mathbb S}^5$, the embedding of the D5-branes has been analyzed in detail in \cite{Jokela:2019tsb}, where it was shown that:
\beq
\Qf({\mathbb S}^5)\,=\,{4\pi\,g_\mt{s}\,\alpha'\,\nf\over 9\sqrt{3}}\,\,.
\label{Q_f_n_ff}
\eeq
Similarly, one can study  the embeddings in the case in which ${\cal M}_5=T^{1,1}$, leading to the result:
\beq
\Qf(T^{1,1})\,=\,{3\pi\,g_\mt{s}\,\alpha'\,\nf\over 8}\,\,.
\eeq

The equations of motion for our ansatz are worked out in appendix \ref{appendix_A}. In appendix \ref{appendix_B} we solve these equations in the case in which $J=A_t=F_{123}=0$. We explore this particular solution in the next section.

\section{Black holes without charges}\label{BH_without_mu}

In this section we consider the background in which the worldvolume gauge field vanishes, the gravity dual of the flavored theory at zero chemical potential but at non-zero temperature. The details of the integration of the equations of motion leading to this solution are relegated to appendix \ref{appendix_B}. Our solution is a black hole which has a horizon at $\rho=\rh$. The blackening factor in the metric (\ref{metric_ansatz }) is given by:
\beq
b\,=\,1\,-\,{\rh^4\over \rho^4}\,\,.
\label{blackening}
\eeq
In these  zero chemical potential solutions the function $\alpha(\rho)$ encoding the anisotropy is related to the dilaton $\phi(\rho)$ as:
\beq
\alpha\,=\,e^{-\phi}\ . \label{alpha_phi}
\eeq
Moreover, the warp factor $h$ can be written in terms of the dilaton $\phi$ as:
\beq
h\,=\,{\Qc\over 4\rho^4}\,e^{-\phi}\,\,.\label{h_rho_phi}
\eeq
The equations for the remaining functions of the ansatz are quite involved and are in general only amenable for numerical study. To prosecute toward analytical solutions, we introduce a small expansion parameter $\epsilon$,
\beq
\epsilon\,=\,{\Qf\over 5\,\rh}\ ,
\label{epsilon_def}
\eeq
where the factor 5 has been included for convenience. This allows us to find explicit solutions, perturbatively in $\epsilon$.  To this end, let $\tilde \rho$ be the following rescaled radial coordinate:
\beq
\tilde \rho\,=\,{\rho\over \rh}\,\,,\label{tilde_rho_def}
\eeq
and $\Omega(\tilde \rho)$ the following function:
\beq
\Omega(\tilde \rho)\,\equiv \,{5\over 4}\,\Bigg[
2\,\arctan \tilde \rho\,+\,\log\Big({\tilde \rho^4\over (\tilde \rho+1)^2 (\tilde\rho^2+1)}\Big)\,-\,\pi\Big]\,\,.
\label{Omega_def}
\eeq
Then, at first order in $\epsilon$, we have:
\bea\label{eq:OrderEpsMetricFuncSubsConst}
&&b(\tilde \rho)=1-{1\over \tilde\rho^4}\,,\qquad\qquad\qquad\qquad
\f(\tilde \rho)=\epsilon  \,\Omega\left(\tilde\rho\right)\,\,,\rc\rc
&&\a(\tilde \rho)=1-\epsilon  \,\Omega\left(\tilde\rho\right)\,,
\qquad\qquad
h(\tilde \rho)=\frac{\Qc}{4 \rh^4 \tilde\rho^4}\left(1-\epsilon  \,\Omega\left(\tilde\rho\right)\right)\,\,,\rc\rc
&&G(\tilde \rho)\,=\,1+\epsilon \Big(\,{1\over 2}\,\tilde\rho^4\Omega(\tilde\rho)\,+\,
{5\over 8}\big(4\tilde\rho^3-1\big)\Big)\,\,,
\eea
while the functions $F(\tilde\rho)$ and $S(\tilde\rho)$, that characterize the deformation of the internal manifold, can be written as:
\beq
{F(\tilde\rho)\over \rh}\,=\,\tilde \rho\,\Big(1+\epsilon \,F_1(\tilde \rho)\Big)\,,
\qquad\qquad
{S(\tilde\rho)\over \rh}\,=\,\tilde \rho\,\Big(1+\epsilon \,S_1(\tilde \rho)\Big)\,\,.
\label{F_S_BHsolution}
\eeq
Due to their lengthy nature, the explicit expressions of $F_1(\tilde \rho)$ and $S_1(\tilde \rho)$ are collected inside (\ref{F_1_S_1_explicit}) of appendix \ref{appendix_B}, however.

\section{Charged black holes}\label{BH_witth_mu}

When the worldvolume gauge field $A_t(\rho)$ is non-trivial our background is dual to a gauge theory at non-zero chemical potential. The presence of this non-trivial worldvolume gauge field induces in extra components of the RR fields $F_3$ and $F_5$ written in eqs. (\ref{F_5_cp_C_4})-(\ref{F_3_ansatz }), which involve the constant $F_{123}$ and the function
$J(\tilde \rho)$. As shown in appendix \ref{appendix_A}, the constant $F_{123}$ is proportional to the quark density $\nq$ (see eq. (\ref{n_q_F123})).  We proceed to seek for a perturbative solution of the equations of motion in which these new components are small. 
With this purpose, let us redefine $F_{123}$  and $J(\tilde \rho)$ as:
\bea\label{eq:ReDefDoubleExp}
F_{123}=\eps\,\delta \ \ , \ \  J(\tilde \rho)= \eps\,\delta \, j (\tilde\rho)\,\,,
\eea
where $\delta$ is a new expansion parameter. By using (\ref{n_q_F123}) we can relate $\nq$ to $\epsilon$ and $\delta$ as:
\beq
\nq\,=\,-{\Nc\over 4\pi^2\,g_\mt{s}\,\alpha'}\,\,\epsilon\,\delta\,\,.
\eeq

Actually, in our expansion of the solution for non-vanishing chemical potential, the different functions of the background get an additional term proportional to $\epsilon\,\tilde\delta^2$, where $\tilde\delta$ is related to $\delta$ as:
\beq
\tilde{\d}=\frac{\Qc}{10\sqrt{6}}\,{\delta\over \rh^3} \,\,.
\eeq
To make the physics in the expansion parameters more transparent, let us relate $\tilde\delta$ to the parameters of the dual theory. A simple calculation using  eq. (\ref{n_q_F123}) shows that  $\tilde\delta$ is related to the temperature $T$ and quark density $\nq$ as:
\beq
\tilde{\d}\,=\,-{4\sqrt{2}\,\alpha'\,g_\mt{s}\over \sqrt{3}}\,
{\nq\over \Nc}\,{1\over \nf\,T^2}\Big(1+{\cal O}\Big({\nf\over T}\Big)\Big)\,=\,-{4\sqrt{2}\,\alpha'\,g_\mt{s}\over \sqrt{3}}\,
{\nb\over  \nf}\,{1\over T^2}\Big(1+{\cal O}\Big({\nf\over T}\Big)\Big)\,\,,
\label{tilde_delta_n_q}
\eeq
where, in the last step, we have introduced the baryon density $\nb=\nq/\Nc$.

To method of obtaining the perturbative solution to the equations of motion is explained in detail in appendix \ref{appendix_C}. As compared with the one found in section \ref{BH_without_mu}, the functions of the ansatz  get an extra correction of order $\epsilon\,\tilde\delta^2$.  For example, the blackening factor is now:
\beq
b(\tilde\rho)=\left(1-{1\over \rt^4}\right)\Bigg(1-4\,\epsilon\,\tilde{\d}^2  \left(\Omega\left(\rt\right)+{5\over \rt}\right)\Bigg)\,\,,
\label{b_chemical}
\eeq
where $\Omega\left(\rt\right)$ is the function defined in (\ref{Omega_def}). Notice that, at the order we are working, the apparent horizon read of from the metric (\ie\ the zero of $b(\rt)$) still occurs at $\tilde\rho=1$. The expressions for the functions $\f(\tilde\rho)$, 
$\a(\tilde\rho)$, $h(\tilde\rho)$, and $G(\tilde\rho)$ are given by:
\bear
&&\f(\tilde\rho)=\epsilon  \,\Omega\left(\rt\right)-\epsilon\,\tilde{\d}^2   \left(\Omega\left(\rt\right)+{5\over \rt}\right)
\,\,\rc\rc
&&\a(\tilde\rho)=1-\epsilon  \,\Omega\left(\rt\right)-\epsilon\,\tilde{\d}^2   \left(\Omega\left(\rt\right)+{5\over \rt}\right)
\,\,\rc\rc
&&h(\tilde\rho)=\frac{\Qc}{4 \rh^4 \rt^4}\Bigg(1-\epsilon  \,\Omega\left(\rt\right)-3\,\epsilon\,\tilde{\d}^2   \left(\Omega\left(\rt\right)+{5\over \rt}\right)\Bigg)
\,\,\rc\rc
&&G(\tilde\rho)=1-\frac{1}{2} \epsilon  \left(\frac{5}{4} \left(1-4 \rt^3\right)-\rt^4\, \Omega (\rt)\right)
\rc\rc
&&\qquad\qquad\qquad\qquad
-\epsilon\,\tilde{\d}^2  \left(\,{5\over 8} \left(20\rt^3-5+{4\over \rt}
\right)+{5\over 2}\,\rt^4\,\Omega(\tilde\rho)
\right)\,\,.
\label{phi_alpha_h_G_chemical}
\eear
Moreover, the functions $F(\tilde\rho)$ and $S(\tilde\rho)$ can now be written as:
\bear
&&{F(\tilde\rho)\over \rh}\,=\,\tilde \rho\,\Big(1+\epsilon \,F_1(\tilde \rho)\,+\,
\epsilon\,\tilde{\d}^2 \,F_2(\tilde \rho)
\Big)\rc\rc
&&{S(\tilde\rho)\over \rh}\,=\,\tilde \rho\,\Big(1+\epsilon \,S_1(\tilde \rho)\,+\,
\epsilon\,\tilde{\d}^2 \,S_2(\tilde \rho)
\Big)\,\,,
\label{F_S_chemical}
\eear
where $F_1(\tilde \rho)$ and $S_1(\tilde \rho)$ are the same as in (\ref{F_S_BHsolution}) and the new functions 
$F_2(\tilde \rho)$ and $S_2(\tilde \rho)$ are written in (\ref{F_2_S_2_explicit}). 

To complete the solution we must give the flux function $j(\tilde\rho)$ and the worldvolume gauge field 
$A_t'(\tilde\rho)$. They are displayed in eqs. (\ref{eq:jSol}) and (\ref{At_prime_sol}) respectively. 

It is interesting to write down the asymptotic forms of the different functions of the background in the UV region $\tilde \rho\to\infty$. In this region, the metric becomes $AdS_{5}\times {\cal M}_5$, with subleading flavor corrections. These corrections can be straightforwardly obtained from our equations. The expansions of the functions $\phi$, $\alpha$, $h$, and $G$ can be easily obtained by expanding the function $\Omega(\rt)$ for large $\rt$, finally leading to the expansions
\bear
&&\phi(\rt)\,=\,-{5\,\epsilon\over \rt}\,+\,{5\over 4}\,{\epsilon\,-\,\epsilon\,\tilde \delta^2\over \rt^4}\,+\ldots\,\,,
\qquad\qquad
\alpha(\rt)\,=\,1\,+\,{5\,\epsilon\over \rt}\,-\,{5\over 4}{\epsilon+\epsilon\,\tilde \delta^2\over \rt^4}
+\ldots\rc\rc
&&h(\rt)\,=\,{\Qc\over 4\rh^4\,\rt^4}\,\Big(1\,+\,{5\,\epsilon\over \rt}\,-\,
{5\over 4}\,{\epsilon\,+\,3\epsilon\,\tilde \delta^2\over \rt^4}\,+\ldots\Big)\rc\rc
&&G(\rt)\,=\,1\,-\,{\epsilon\over 2\rt}\,+\,{5\over 16}\,{\epsilon\,-\,5\,\epsilon\,\tilde \delta^2\over \rt^4}\,+\ldots\,\,.
\label{UV_expansion}
\eear
Moreover, the UV expansion of the functions $F$ and $S$ can be found from (\ref{F_1_S_1_explicit}) and (\ref{F_2_S_2_explicit}) and is given by:
\bear
&&{F(\rt)\over \rh}\,=\,\rt\Big(1\,-\,{7\over 10}\,{\epsilon\over \tilde\rho}\,+\,
{5\over 16}\,{\epsilon\,+\,3\epsilon\,\tilde \delta^2\over \rt^4}\,+\,\ldots\Big)\rc\rc
&&{S(\rt)\over \rh}\,=\,\rt\Big(1\,-\,{6\over 5}\,{\epsilon\over \tilde\rho}\,+\,
{5\over 48}\,{8\epsilon\,+\,9\epsilon\,\tilde \delta^2\over \rt^4}\,+\,\ldots\Big)\,\,.
\eear

\section{Thermodynamics}\label{Thermo}

Now that we have found the solution for our geometry in the presence of flavor and in particular the non-trivial gauge potential, let us work out the corresponding thermodynamics of the field theory. First of all, we obtain the temperature of the system by equating it with the Hawking temperature, using (\ref{T_general}). We get:
\beq
T=\frac{2 \,\rh}{\pi\,  \Qc^{1/2}} \left(1\,-\, {15\over8}\,\eps\,-\,{5\over 8}\,\eps\,\tilde{\delta^2}\right)\,\,.
\label{t_rho_h_chemical}
\eeq
The entropy density $s$ is identified with the entropy of the black hole, given by the Bekenstein-Hawking formula:
\beq
s\,=\,{2\pi\over \kappa_{10}^2}\,{A_8\over V_3}\,\,,
\eeq
where $A_8$ is the volume at the horizon of the eight-dimensional space orthogonal to the $(t, \tilde \rho)$-plane and $V_3$ is the infinite constant volume of the three-dimensional Minkowski directions $(x^1,x^2,x^3)$. For our geometry, we obtain:
\beq
s\,=\,{{\rm Vol} \big({\cal M}_5\big)\over (2\pi)^6\,\alpha'^4\,g_\mt{s}^2}\,
\Qc^{1/2} \,\rh^3\, \left(1\,+\, {15\over8}\,\eps\,+\,{5\over 8}\eps\,\tilde{\delta^2}\right)\,\,.
\label{s_rho_h}
\eeq
In order to obtain the internal energy density ${\cal E}$ of our system, we calculate the ADM energy of the background, which is given by the standard formula \cite{Bigazzi:2009bk}:
\bea
E_{ADM}=-{1\over \k_{10}^2}\sqrt{-g_{tt}}\int_{{\cal M}_{t, \rho_{\infty}}}
 \dd^8x\sqrt{g_8}\left(K_{T,\mu}-K_0\right)\,\,.
\label{E_ADM}
\eea
In (\ref{E_ADM}) $K_{T,\mu}$ is the extrinsic curvature of the eight-dimensional subspace within the nine-dimensional (constant time) space.  For our metric ansatz  (\ref{metric_ansatz }), $K_{T,\mu}$ is given by:
\bea
K_{T,\mu}={1\over\sqrt{\det g_9}}\partial_\m\left(\sqrt{\det g_9\,}{1\over \sqrt{g_{\r\r}}}\delta_\r^\m\right)\,=\,
{\sqrt{b}\over h^{{3\over 4}}\,G\,\alpha\,F\,S^4}\,\partial_{\rho}\Big(
h^{{1\over 2}}\,\alpha\,F\,S^4\Big)\,\,.
\eea
Moreover, $K_0$ is the extrinsic curvature at zero temperature and zero quark density. At first order  in $\Qf$, $K_0$ is given by:
\beq
K_0\,=\,{3\,\sqrt{2}\over \Qc^{{1\over 4}}}\,\Big(1\,-\,
{17\over 60}\,{\Qf\over \rho}\Big)\,\,.
\eeq
The integral in (\ref{E_ADM})  is taken at constant time and at a fixed radial position (which is then sent to infinity).  We get:
\beq
\mathcal{E} ={E_{ADM}\over V_3}\,=\,{3\,{\rm Vol} \big({\cal M}_5\big)\over (2\pi)^7\,\alpha'^4\,g_\mt{s}^2}\,
\rh^4
\left(1\,-\,{5\over 6}\,\eps\,+\,{25\over6}\eps\,\tilde{\delta}^2\right)\,.
\eeq
We can now compute the free energy density as:
\beq
f=\mathcal{E} -T\,s\,=\,-{{\rm Vol} \big({\cal M}_5\big)\over (2\pi)^7\,\alpha'^4\,g_\mt{s}^2}\,
\rh^4 \left(1\,+\,\,\frac{5}{2}\,\eps\,-\,\frac{25}{2}\eps\,\tilde{\delta}^2   \right)\,\,.
\eeq
All these expressions contain the combination $\eps\,\tilde{\delta}^2$, which can be written in terms of the quark density $\nq$ as:
\beq
\eps\,\tilde{\delta}^2 \,=\,\gamma\,{\nq^2\over \Qf\,\rh^5}\,\,,
\label{epsilon_delta2_nq}
\eeq
where the coefficient $\gamma$ depends on the internal  manifold ${\cal M}_5$ and  is given by:
\beq
\gamma\,=\,{2^{9}\,\pi^{12}\,g_\mt{s}^4\,\alpha'^6\over
15\,\Big({\rm Vol} \big({\cal M}_5\big)\Big)^2}\,\,.
\label{gamma_coeff}
\eeq
Let us next obtain $s$, $\mathcal{E}$, and $f$ in terms of physical quantities. With this purpose, let us define $\lambda$ as:
\beq
\lambda\,=\,4\pi\,g_\mt{s}\,\Nc\,\,,
\eeq
which, in the case in which the internal manifold ${\cal M}_5$ is the five-sphere ${\mathbb S}^5$ and the unflavored geometry it is the dual of ${\cal N}=4$ super Yang-Mills theory, is just the 't Hooft coupling written in terms of string variables. Furthermore, the `$a$' coefficient of the conformal anomaly is inversely proportional to the internal volume \cite{Gubser:1998vd}, so we can define the ratio between the $a$ coefficients of the field theory dual to the D3-branes on the cone and  ${\cal N}=4$ super Yang-Mills as the ratio between the volume of $\mathcal{M}_5$ with the five-sphere
\beq
\bar{a}=\frac{a_{\mathcal{M}_5}}{a_{{\cal N}=4}}=\frac{\pi^3}{\text{Vol}\left(\mathcal{M}_5\right)} \ .
\eeq

It will be convenient recall the definition of the baryon density
\beq
\nb={\nq\over \Nc}
\eeq
so that we can then invert (\ref{t_rho_h_chemical}) and write $\rh$ in terms of $\Nc$, $\lambda$, $\nf$, and $T$ as:
\beq
\rh=\pi\, \l^{{1\over 2}} \bar{a}^{{1\over 2}}\, \a'\, T\left[1+\frac{\l^{{1\over 2}}}{\Nc\bar{a}^{{1\over 2}}}\left(\frac{1}{16\,v_{\perp} }{\nf \over T}+\frac{v_{\perp}}{2\, \pi ^4}{\nb^2\over \nf\, T^5}\right)\right]\,\,.
\label{rho_h_T_chemical}
\eeq
Thereby the expansion parameters $\epsilon$ and $\epsilon\tilde{\delta}^2$ in terms of physical quantities read
\be
\eps  =  {\l^{1\over 2}\over 30\,v_{\perp}\,\Nc\, \bar{a}^{1\over 2}}{\nf\over T}+\ldots\\ \ \ , \ \ \
\eps\,\tilde{\delta}^2  =  {4\,v_{\perp}\over 5\,\p^{4}}{\l^{1\over2}\,\over \Nc\,\bar{a}^{1\over2}}{\nb^2\over \nf\,T^5}+\ldots
 \ .
\label{epsilon_delta_physical}
\ee

Using (\ref{rho_h_T_chemical}),  (\ref{epsilon_delta_physical}),  and (\ref{epsilon_delta2_nq}) we finally arrive at the following expressions for the entropy and energy densities:
\bear
s & = & \frac{\pi ^2 \Nc^2 }{2\, }\,\bar{a}\,T^3 \left[1+\frac{\l^{{1\over 2}}}{\Nc\bar{a}^{{1\over 2}}}\left(\frac{1}{4\,v_{\perp} }{\nf \over T}+\frac{2v_{\perp}}{ \pi ^4}{\nb^2\over \nf\, T^5}\right)\right]\nonumber \\
\mathcal{E} & = & \frac{3 \pi ^2 \Nc^2 }{8 }\, \bar{a}\, T^4\left[1+\frac{\l^{{1\over 2}}}{\Nc\bar{a}^{{1\over 2}}}\left(\frac{2}{9\,v_{\perp} }{\nf \over T}+\frac{16v_{\perp}}{3\, \pi ^4}{\nb^2\over \nf\, T^5}\right)\right] \ ,
\qquad\qquad
\label{s_E_physical_chemical}
\eear
while the free energy density is given by:
\beq
f=-\frac{\pi ^2\, \Nc^2 }{8 }\, \bar{a}\, T^4\left[1+\frac{\l^{{1\over 2}}}{\Nc\bar{a}^{{1\over 2}}}\left(\frac{1}{3\,v_{\perp} }{\nf \over T}-\frac{8v_{\perp}}{\pi ^4}{\nb^2\over \nf\, T^5}\right)\right]\,\,.
\qquad
\label{f_physical_chemical}
\eeq

In the case in which ${\cal M}_5\,=\,{\mathbb S}^5$,  we have $\bar a=1$ and the leading terms in (\ref{s_E_physical_chemical}) and (\ref{f_physical_chemical}) reproduce the standard values of $s$, ${\cal E}$, and $f$ for ${\cal N}=4$ super Yang-Mills theory. Notice that the flavor corrections in these equations depend on the ratios 
$\nf/T$ and $\nb^2/(\nf T^5)$, which are dimensionless, since $[T]=[\nf]=[{\rm length}]^{-1}$ and $\nb\,=\,[\rm{length}]^{-3}$. The flavor contribution increases the value of the entropy and the energy densities, indicating an increase in the number of degrees of freedom. This is expected since each D5-brane introduces additional fields localized at the intersection with the color D3-branes.

Following the approach of \cite{Caldarelli:2010xz,Mateos:2011tv}, to unravel a better understanding of the thermodynamics of the anisotropic system it is convenient to consider the situation in which the number of flavor D5-branes change. Therefore, we allow $\nf$ to vary and write the first law of thermodynamics as:
\beq
\dd \mathcal{E}\,=\,T\,\dd s+\Phi\,\dd \nf+\m\,\dd \nq\,\,,
\eeq
where $\mu$ is the ordinary (baryon) chemical potential and $\Phi$ is a brane potential which measures the energy cost of introducing additional flavor branes in the system. Taking into account that $f=\mathcal{E} -T\,s$, 
we  can write the variation of the free energy as:
\bea
\dd f=-s\,\dd T+\Phi\,\dd \nf+\m\,\dd \nq\,\,.
\label{df_chemical}
\eea
It follows from (\ref{df_chemical})  that:
\beq
s=-\left({\partial f\over \partial T}\right)_{\nf,\nq}\,\,,
\label{s_derivative_f}
\eeq
which can be easily checked directly using (\ref{s_E_physical_chemical}) and (\ref{f_physical_chemical}) and can be regarded as a consistency check of our expression (\ref{f_physical_chemical}) for $f$. Similarly, the brane potential  $\Phi$ can be obtained from $f$ as:
\beq
\Phi=\left({\partial f\over \partial \nf}\right)_{T,\nq}\,\,.
\label{Phi_derivative_f}
\eeq
Computing the derivative in (\ref{Phi_derivative_f}) using (\ref{f_physical_chemical}), we arrive at the following expression for the brane potential $\Phi$:
\beq
\Phi=-\frac{\pi ^2 \,\l^{1\over 2} \,\Nc \,\bar{a}^{1\over 2} }{24\,v_{\perp}}T^3\left(1+\frac{24\,v_{\perp}^2}{\pi ^4 }\,\,{\nb^2\over \nf^2\,T^4}\right)\,\,.
\eeq
Moreover, the baryonic chemical potential $\mu$ can be obtained as:
\beq
\m=\left({\partial f\over \partial \nq}\right)_{T,\nf}\,\,,
\eeq
and is given by:
\beq
\m=\frac{2 \,\l^{{1\over 2}}\,\bar{a}^{1\over 2}\,v_{\perp} }{\p^2}\,{\nb\over \nf\,T}\
\,\,.
\eeq
For Gibbs free energy $g = f-\Phi \nf - \mu \nq$ we get
\bea
g & = & -\frac{\pi ^2 \,\Nc^2}{8 }\bar{a} \,T^4\,. 
\eea
From $f$ and $g$ we get the pressures along the $xy$- and $z$-directions as:
\beq
p_{xy}\,=\,-f\,+\,\mu\,\nq\,=\,-g\,-\,\Phi\,\nf\,\,,
\qquad\qquad
p_z\,=\,-g\,\,.
\eeq
More explicitly, these two pressures can be written as:
\bear
 p_{xy} & = & \frac{\pi ^2 \,\Nc^2}{8}\,\bar{a}\, T^4\left[1+\frac{\l^{{1\over 2}}}{\Nc\bar{a}^{{1\over 2}}}\left(\frac{1}{3\,v_{\perp} }{\nf \over T}+\frac{8v_{\perp}}{\pi ^4}{\nb^2\over \nf\, T^5}\right)\right]\nonumber \\
 p_z & = & \frac{\pi ^2 \,\Nc^2}{8}\,\bar{a}\, T^4\ .
\label{pressures}
\eear

Notice that $p_{xy}$ becomes larger, while $p_z$ in (\ref{pressures}) is not affected by the addition of D5-branes, which is consistent with having flavors living in layers extended along the $xy$-directions. Actually, we can easily check that:
\beq
{\partial p_{xy}\over \partial\mu}\,=\,\nq\,\,,
\qquad\qquad\qquad
{\partial p_{z}\over \partial\mu}\,=\,0\,\,.
\eeq

We have checked these thermodynamic results in several ways. First of all, we verified that the regulated euclidean on-shell action divided by the inverse temperature equals the grand canonical potential $\Omega=f-\mu\,\nq$. 
To perform this calculation we have to regulate the on-shell action by subtracting its zero temperature value, both for the bulk and Gibbons-Hawking terms (for details, see appendix B in \cite{Bigazzi:2009bk}). 

We have also checked the expressions of the energy density and pressures that we have just found using the following thermodynamic arguments. Indeed, one can compute the vacuum expectation value (VEV) of the stress-energy tensor from the Brown-York tensor at the boundary, by using the proposal of \cite{Balasubramanian:1999re}.  The Brown-York tensor of the ten-dimensional gravity theory is:
\beq
\tau_{ij}\,=\,{1\over \kappa_{10}^2}\,\big(K_{ij}\,-\,K\,\gamma_{ij}\big)\,\,,
\label{BY_tensor_def}
\eeq
where $\gamma_{ij}$ is the induced metric at a $r={\rm constant}$ surface, $K_{ij}$ is the extrinsic curvature of the surface, and $K=\gamma^{ij}\,K_{ij}$.  The VEV of the stress-energy tensor of  the dual theory is related to the Minkowski components of the  Brown-York tensor evaluated at the boundary \cite{Balasubramanian:1999re}:
\beq
\langle T^{\mu}_{\,\,\,\,\nu}\rangle\,=\,V_{\rm{SE}}\,\sqrt{-\gamma_{{\rm{min}}}}\,\,\tau^{\mu}_{\,\,\,\,\nu\,}\Big|_{\,\rm{reg}\,,\,\rho\to\infty}\,\,,
\label{em_Tensor-BY}
\eeq
where $V_{\rm{SE}}=h^{{5\over 4}}\,F\,S^{4}\,{\rm Vol}({\cal M}_5)$ is the volume for the compact 5d part of the metric. The right-hand side of (\ref{em_Tensor-BY}) is singular at the UV boundary. To regulate it we subtract its zero temperature value (given by supersymmetry), similarly as was done in \cite{Bigazzi:2011it}. Therefore, we obtain the VEV of the stress-energy tensor as:
\bea
\langle T^\m{}_\n \rangle=V_{\rm{SE}}\lim_{\r_\Lambda\to\infty}\Bigg[
\sqrt{-\g_{{\rm{min}}}}\t^\m{}_\n-\left[b^{1/2}\lim_{\rh\to0}\left(\sqrt{-\g_{{\rm{min}}}}\t^\m{}_\n\right)\right]_{\tilde{\delta}\to 0}\Bigg]_{\r=\r_\Lambda},
\eea
where $\g_{{\rm{min}}}$ is the determinant of the Minkowski part of the induced metric and the $b^{{1\over 2}}$ factor, needed to match the geometries at the cutoff, is evaluated at zero chemical potential. After some calculation, we get:
\bear
\langle T^t{}_t\rangle & = & -{3\,{\rm Vol} \big({\cal M}_5\big)\over (2\pi)^7\,\alpha'^4\,g_\mt{s}^2}\,
\rh^4\left(1\,-\,{5\over 6}\,\eps\,+\,{25\over6}\eps\,\tilde{\delta}^2\right)\nonumber\\
\langle T^{x^1}{}_{x^1}\rangle & = & \langle T^{x^2}{}_{x^2}\rangle\,=\,
{{\rm Vol} \big({\cal M}_5\big)\over (2\pi)^7\,\alpha'^4\,g_\mt{s}^2}\,\rh^4
\left(1\,+\,{5\over 2}\,\eps\,+\,{15\over 2}\eps\,\tilde{\delta}^2\right)\nonumber \\
\langle T^{x^3}{}_{x^3}\rangle & = & 
{{\rm Vol} \big({\cal M}_5\big)\over (2\pi)^7\,\alpha'^4\,g_\mt{s}^2}\,\rh^4
\left(1\,-\,{15\over 2}\,\eps\,-\,{15\over 6}\eps\,\tilde{\delta}^2\right)\ .
\label{VEV_T}
\eear
We can check that ${\cal E}=-\langle T^t{}_t\rangle$ and that the pressures along the parallel and transverse directions are indeed given by the expectation values of the corresponding components of the stress-energy tensor, namely:
\beq
p_{xy}=\langle T^{x^1}{}_{x^1}\rangle\,=\,\langle T^{x^2}{}_{x^2}\rangle
\,\,,
\qquad\qquad\qquad
p_{zz}=\langle T^{x^3}{}_{x^3}\rangle\,\,.
\eeq

In addition, in appendix \ref{appendix_D} we present a further check of the VEV of $T^\mu{}_\nu$ in the case of zero chemical potential. In this case we can make use of the reduction to five dimensions found in \cite{Penin:2017lqt}. The reduced theory contains gravity and three scalars, whose contribution to the VEV of the stress-energy tensor must be regulated by using their superpotential. The final results coincide with those in (\ref{VEV_T}) for $\tilde\d=0$.

As in \cite{Penin:2017lqt} the anisotropy of the system is manifested through a pressure difference, measured by the potential $\Phi$. Indeed, one can readily check that:
\bea
p_z-p_{xy}=\Phi\,\nf\,.
\eea
Actually, using (\ref{VEV_T}) we can write, at ${\cal O}(\eps^2)$,  the two pressures in terms of the energy density as:
\beq
p_{xy}=\Bigg({1\over 3}\,+\,{10\over 9}\left(1+\tilde{\delta}^2\right)\eps\Bigg)\mathcal{E}\,\,,
\qquad
p_{xy}=\Bigg({1\over 3}\,-\,{20\over 9}\left(1+\tilde{\delta}^2\right)\eps\Bigg)\mathcal{E}\,\,.
\eeq
The speeds of sound along the two different directions  are now easily computed, with the result:
\beq
v^2_{xy}=\left({\partial p_{xy}\over\partial\mathcal{E}}\right)={1\over 3}\,+\,{10\over 9}\left(1+\tilde{\delta}^2\right)\eps\,\,,
\qquad v^2_{z}=\left({\partial p_{z}\over\partial\mathcal{E}}\right)={1\over 3}\,-\,{20\over 9}\left(1+\tilde{\delta}^2\right)\eps\,\,,
\eeq
where we have calculated the derivatives by keeping the expansion parameters $\epsilon,\tilde\delta$ fixed. Recall that this means to keep the density of flavor branes and/or the baryon density, appropriately scaled with the powers of the temperature, fixed, see \eqref{epsilon_delta_physical}. Rewriting these speeds in this vein leads to
\bear
v_{xy}^2 & = & \frac{1}{3}\left[1+\frac{\l^{{1\over 2}}}{\Nc\bar{a}^{{1\over 2}}}\left(\frac{1}{9\,v_{\perp} }{\nf \over T}+\frac{8v_{\perp}}{3\pi ^4}{\nb^2\over \nf\, T^5}\right)\right]
\nonumber\\
v_z^2 & = & \frac{1}{3} \left[1-2\frac{\l^{{1\over 2}}}{\Nc\bar{a}^{{1\over 2}}}\left(\frac{1}{9\,v_{\perp} }{\nf \over T}+\frac{8v_{\perp}}{3\pi ^4}{\nb^2\over \nf\, T^5}\right)\right]
\ .
\label{speeds_physical}
\eear

To gain some intuition on the above expressions, let us first recall how the speed of sound works out for massless probe D5-branes immersed in the geometry induced by D3-branes. There the speed of sound for collective modes localized on a (single) defect D5-brane coincides with that of a conformal $(2+1)$-dimensional theory $v_\text{s}^2=1/2$ \cite{Itsios:2016ffv}. One should bear in mind though that there are actually two types of sound waves: these faster ones propagating along the directions of the defects, but also those slower that are blind to defects, propagating in the whole 3+1 dimensions with innate conformal speed $v_\text{s}^2=1/3$. 

In the current setup, where the backreaction is taken into account and the defects are smeared in the transverse directions, there is an interesting interplay of collective phenomena leading to speeding and braking from the conformal value $1/3$. The speed along the defect directions is larger than in the orthogonal directions: The interaction between the adjoint and fundamental degrees of freedom leads to slowing down the defect degrees of freedom, while at the same time speeding up the fields mixing mode that moves at an intermediate speed. This suggests that the interaction inborn from defect degrees of freedom with the degrees of freedom outside the defects is such that it ``pushes''  the motion of the collective modes to higher speeds along the defect directions. On the contrary, the motion in the direction transverse to the defects slows down, which can be seen in part as a consequence of the equation of state below \eqref{eq:eos}. In a sense, the total kinetic energy of the fluid is conserved but more of it is put in the direction of the defects, so that there is less available for the transverse motion.

The deviation from the value $1/3$ in (\ref{speeds_physical}) naively reflects the breaking of conformality along the ($x^1,x^2$)-plane and the $x^3$-direction. Values for the speed of sound above the conformal value in holography \cite{Hoyos:2016cob,Ecker:2017fyh} are now abound in various settings \cite{Hoyos:2021uff}. However, here the full equation of state
\bea\label{eq:eos}
2p_{xy}+p_z=\mathcal{E}
\eea
should be regarded as the ``conformal''  analog in homogeneous, but anisotropic system. Indeed, we have not introduced any further energy scale in the system as the D3- and D5-branes are not separated and the corresponding fundamental quark masses vanish. One can also easily check the following relations:
\bea
&& p_{xy}={1\over3}\left(\mathcal{E}-\Phi\,\nf\right)\,\,,\qquad\qquad\qquad p_{z}={1\over3}\left(\mathcal{E}+2\Phi\,\nf\right)\,\,,\rc\rc
&&\mathcal{E}={1\over 4}\left(3\,T\,s+3\m\,\nq+\Phi\,\nf\,\right)\,,\rc\rc
&&p_{xy}=T\,s+\m\,\nq-\mathcal{E}\,\,,\qquad\qquad\qquad p_z=T\,s+\m\,\nq+\Phi\,\nf-\mathcal{E}\,\,.
\eea

We now compare  the chemical potential $\mu$ with the UV value of the worldvolume gauge field $A_t$. For this purpose, it is quite convenient to rewrite $\mu$ as:
\beq
\mu\,=\,-{\rh\over \sqrt{6}\,\pi\,\alpha'}\,\tilde\delta\,
\Big(1+{15\over 8}\,\epsilon\Big)\,\,.
\label{mu_eps_delta}
\eeq
On the other hand, in appendix \ref{appendix_C} we found, by imposing the condition that $A_t$ vanishes at the horizon, that:
\beq
{A_{t, \text{UV}}\over 2\pi\alpha'}\,=\,-{\rh\over \sqrt{6}\,\pi\,\alpha'}\,\,\tilde\delta\,
\Big(1-\Delta_1\,\epsilon\Big)\,\,,
\eeq
where $\Delta_1\approx 1.06$. These two results coincide at leading order in $\tilde \delta$, but they differ at ${\cal O}(\epsilon\tilde \delta)$. Indeed, one can easily demonstrate that:
\beq
{A_{t, \text{UV}}\over 2\pi\alpha'}\,=\,\mu\,
\Big(1\,-\,\Big({15\over 8}+\Delta_1\Big)\epsilon\Big)\,\,.
\label{A_t_mu}
\eeq
This result is similar to the one found in \cite{Bigazzi:2011it} for the D3-D7 system at finite baryon density and may be regarded as an effect due to the backreaction of the flavors.

\section{Applications}\label{applications}

In this section we make use of the background geometry to compute several observable quantities in the dual field theory. We are specifically interested in determining the dependence on the flavor deformation parameters of these observables. We start by analyzing the transport properties at zero chemical potential.

\subsection{Hydrodynamics at zero chemical potential}\label{hydro_results}

Let us start by studying the hydrodynamic properties of our system. We will compute the transport coefficients of perturbations propagating along the ($x^1,x^2$)-plane. In order to do that we consider the dimensionally reduced theory to four dimensions and perform a suitably chosen perturbation of the 4d metric. This calculation is done in detail in appendix \ref{hydro_appendix}, following the lines of ref. \cite{Penin:2017lqt}, whereas in this section we will summarize the results. We will restrict our analysis to the so-called shear channel. Our main purpose is to obtain the dispersion relation $\omega=\omega(q)$ of these shear modes at low momentum $q$. Let us assume that, as in the 
$AdS_5\times {\mathbb S}^5$ geometry, this dispersion relation has the form:
\beq
\omega\,=\,-i\,D_{\eta}\,q^2\,\big(1+\tau_s\,D_{\eta}\,q^2\big) +\ldots \ ,
\label{disp_rel_shear_q4}
\eeq
where  we are keeping terms up to quartic power of $q$. The dispersion relation (\ref{disp_rel_shear_q4}) depends on two transport coefficients $D_{\eta}$  and $\tau_s$, which we  calculate next.  We will work in dimensionless variables  $\hat{q}$ and $\hat{\w}$, defined as:
\beq
\hat{q}={q\over 2\p\,T}\,\,,\qquad\qquad \hat{\w}={\w\over 2\p\,T}\,\,.
\label{hat_q_hat_omega_def}
\eeq
Moreover, we define rescaled coefficients $\hat D_{\eta}$ and $\hat\tau_s$ as:
\beq
\hat D_{\eta}\,=\,2\pi\,T\,D_{\eta}\,\,,
\qquad\qquad
\hat\tau_s\,=\,2\pi\,T\,\tau_s\,\,.
\eeq
In terms of the rescaled quantities, the dispersion relation (\ref{disp_rel_shear_q4}) takes the form:
\beq
\hat\omega\,=\,-i\,\hat D_{\eta}\,\hat q^2\,\big(1+\hat\tau_s\,\hat D_{\eta}\,\hat q^2\big)+\ldots \ .\label{hatted_disp_rel_shear}
\eeq
The coefficient $\hat D_{\eta}$ determines the ratio of the  shear viscosity $\eta$ to the entropy density $s$,  namely: 
\beq
{\eta\over s}\,=\,{\hat D_{\eta}\over 2\pi}\,\,.
\eeq
From the results of appendix \ref{appendix_D} we get:
\beq
\hat{D}_\eta={1\over 2}\,\,,\qquad\qquad\hat{\t}_s=1-\log\, 2\,+\, \frac{5}{2} (\pi -3)\,\eps\,\,.
\label{transport_coeff_results}
\eeq
Notice that the coefficient $\hat D_{\eta}$ is not corrected by the flavor. Its value  corresponds to having 
$\eta/s=1/(4\pi)$. On the contrary, the coefficient $\hat\tau_s$ is modified by the flavor. The $\epsilon=0$ value 
$\hat\tau_s=1-\log\, 2\simeq 0.307$ is the result found for the $AdS_{5}\times {\mathbb S}^5$ geometry. The ${\mathcal O}(\epsilon)$ correction written in (\ref{transport_coeff_results}) encodes the effect of flavors, and of the corresponding anisotropy, on this transport coefficient. At the intersection between the D3 and D5 branes there is $(2+1)$-dimensional CFT, which it would be expected to have a holographic dual $AdS_4$ geometry. In this case the coefficient $\hat{D}_\eta$ would take the same value as in $AdS_5$, but $\hat{\tau}_s$ takes a larger value \cite{Natsuume:2008gy}
\beq
\hat{\tau}_s^{CFT_{2+1}}=\frac{3}{4} \left(1 - \log\, 3\right)+\frac{\pi}{4\sqrt{3}}\simeq 0.379.
\eeq
The flavor correction in \eqref{transport_coeff_results} increases the value of $\tau_s$ relative to the the $AdS_5$ result thus maybe indicating that it is tending towards the $AdS_4$ value.

It would be interesting to calculate the transport coefficients for perturbations propagating along the $x^3$-direction. This calculation must be done in the five-dimensional supergravity theory constructed in \cite{Penin:2017lqt}. Besides 5d gravity and scalars, this theory contains a codimension one defect whose fluctuations are very difficult to analyze. For this reason we will not attempt to carry out this calculation here.

\subsection{Quark-antiquark potentials}\label{quark_potentials}

In this section we will be using the (dimensionful) radial coordinate $\r$. We will follow the standard holographic prescription \cite{Maldacena:1998im,Rey:1998ik} to find 
the potential between a quark and an antiquark and will solve the equations of motion of a fundamental string with its two endpoints lying at the UV boundary. These equations are obtained by extremizing the Nambu-Goto action:
\bea\label{ec:NGaxction}
S={1\over 2\pi\alpha'}\int_{\Sigma} \dd \t\, \dd \s\, e^{\f/2}\sqrt{-\det(g_2)}\,\,,
\eea
where $\tau$ and $\sigma$ are coordinates of the string worldsheet $\Sigma$ and $g_2$ is the  induced metric on $\Sigma$ in the Einstein frame.  Since our geometry is anisotropic, we have to consider two different cases, depending on whether the string is extended along the directions  $x^1$ and $x^2$ of the layers of D5-branes or along the orthogonal direction $x^3$. These two cases are treated in different subsections in what follows.

\subsubsection{Intra-layer potential}

\begin{figure}[!ht]
\center
 \includegraphics[width=0.47\textwidth]{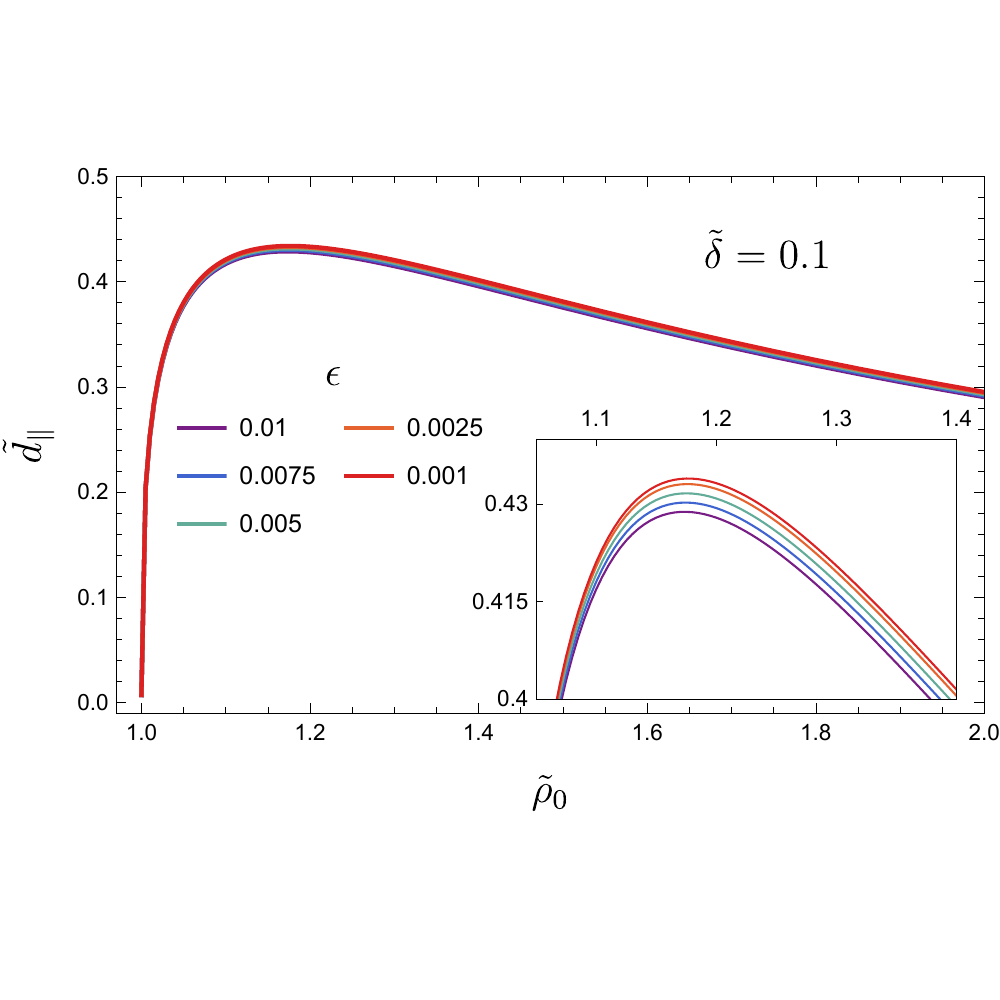} 
 \includegraphics[width=0.47\textwidth]{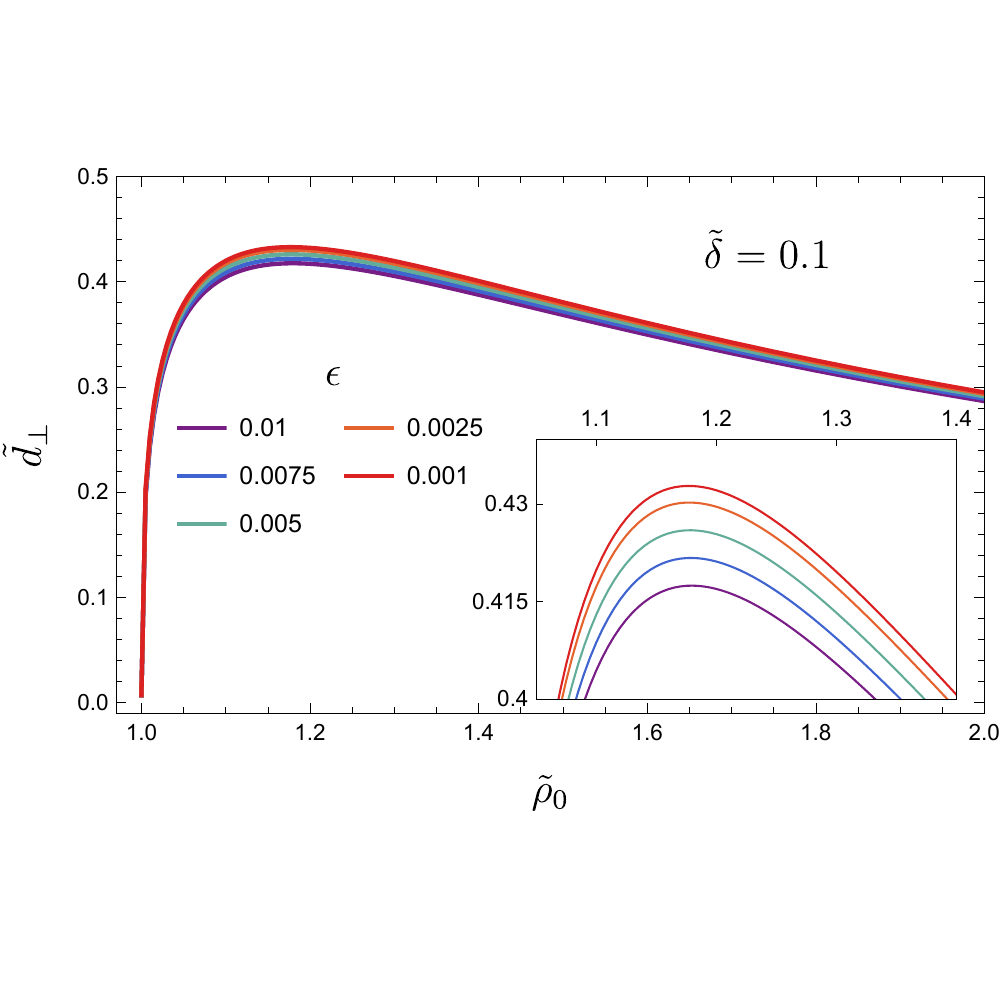} 
  \caption{Quark-antiquark distances $\tilde{d}_\parallel=\pi\,T\,d_\parallel$ and 
   $\tilde{d}_\perp=\pi\,T\,d_\perp$ versus  the turning point $\tilde{\r}_0={\r_0\over \rh}$ for different $\eps$ at fixed $\tilde{\delta}$.}
  \label{fig:d}
\end{figure}

We first consider a fundamental string hanging from the UV boundary and extended along $x^1$ with no dependence on the other two cartesian coordinates  $x^2,x^3$; otherwise one ends up with partial differential equations in general. Let us take $\tau=x^0$,  $\sigma=x^1$, and $\rho=\rho(x^1)$. Then, \eqref{ec:NGaxction} becomes:
\bea\label{ec:NGaxctionx}
S_\parallel={1\over 2\pi \alpha'}\int \dd x^0\, \dd x^1\, e^{\f/2}\,\sqrt{\frac{b^2 F^2 S^8 \alpha^2 }{\rho ^2 \left(\rho^4-\rh^4\right)^2}\,\rho'^2+\frac{b}{h}}\,\,,
\eea
where $\rho'$ denotes the derivative of $\rho(x^1)$ with respect to $x^1$. 
From the action  (\ref{ec:NGaxctionx})  we obtain the following equation of motion for $\r'$:
\bea
\r'=\pm\sqrt{b\over b_0}\frac{\sqrt{h_0\, b \,e^{\phi-\phi(\r_0)}-b_0 \,h}}{\sqrt{B_1}\, h}\,\,,
\label{rho_prime_parallel}
\eea
with $\r_0$  being  the turning point of $\rho$ (\ie\ the minimal value of the coordinate $\r$) and $h_0$ and $b_0$ are the metric functions $b$ and $h$ evaluated at $\r=\r_0$. Moreover,  $B_1=B_1(\rho)$ is defined as 
\beq
B_1=\frac{b^2 F^2 S^8 \alpha^2}{\rho^2 \left(\rho^4-\rh^4\right)^2}\,\,.
\eeq
We can invert (\ref{rho_prime_parallel})  to obtain $x^1$ as a function of $\rho$, namely:
\bea
x_1(\r)=\pm \int_{\r_0}^\r \dd \r \,\sqrt{b_0\over b}\,\frac{\sqrt{B_1} \,h}{\sqrt{h_0\, b \,e^{\phi-\phi(\r_0)}-b_0 \,h}}\,\,.
\eea
It follows that the quark-antiquark distance $d_\parallel$ at the boundary is:
\bea\label{eq:dparGen}
d_\parallel=2 \int_{\r_0}^\infty \dd \r \,\sqrt{b_0\over b}\,\frac{\sqrt{B_1} \,h}{\sqrt{h_0\, b \,e^{\phi-\phi(\r_0)}-b_0 \,h}}\,\,.
\eea
Numerical integration of $d_\parallel$ is shown in Fig.~\ref{fig:d}, where we plot the dimensionless quark-antiquark distance $\tilde{d}_\parallel=\pi\,T\,d_\parallel$ versus the turning point coordinate $\tilde{\r}_0={\r_0\over \rh}$. We fast forward with representing the numerical results and leave the physical interpretation later in Sec.~\ref{sec:physintqqbar}.

\begin{figure}[!ht]
\center
 \includegraphics[width=0.47\textwidth]{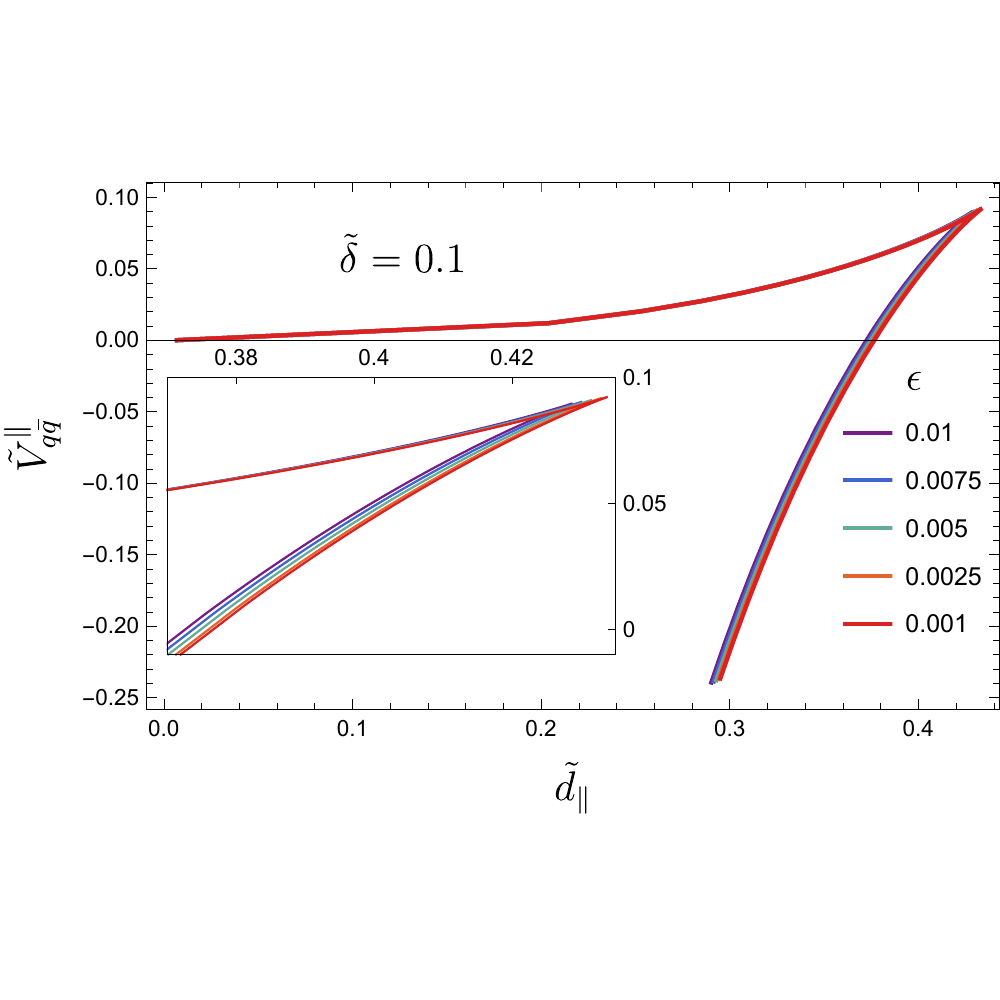} 
 \includegraphics[width=0.47\textwidth]{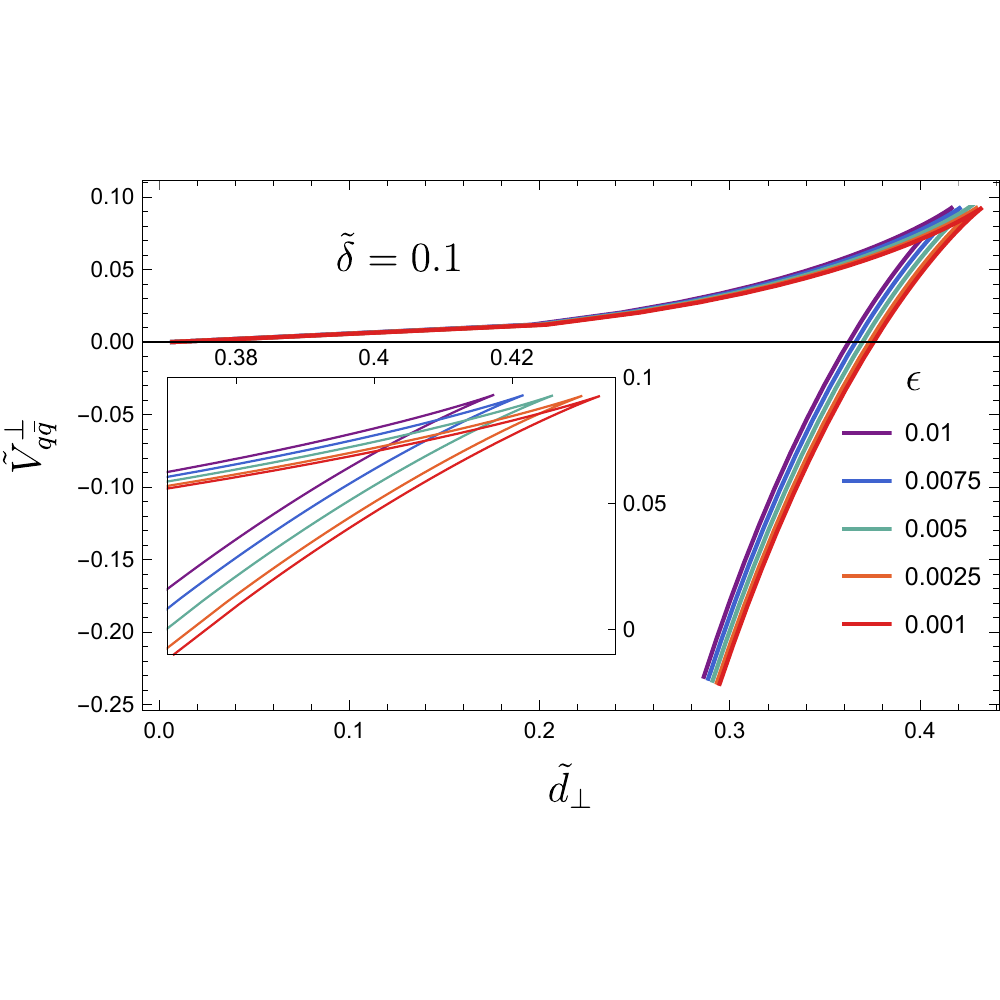} 
  \includegraphics[width=0.47\textwidth]{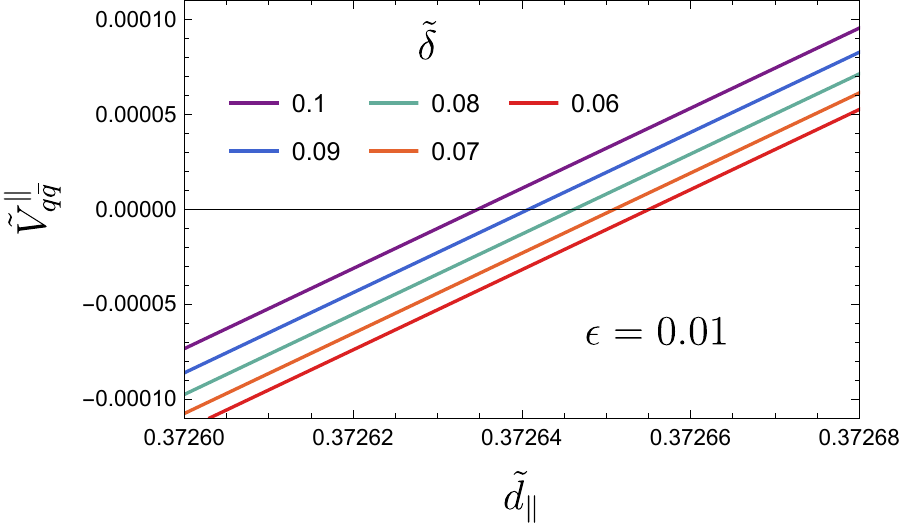} 
 \includegraphics[width=0.47\textwidth]{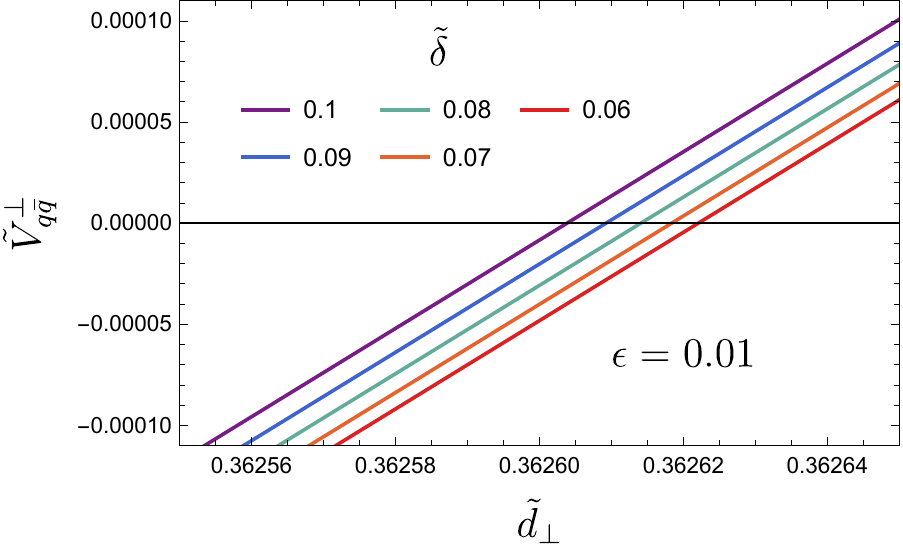} 
  \caption{Quark-antiquark potential $\tilde{V}_{q\bar{q}}^a={{\rm Vol}(\mathcal{M}_5)^{{1\over2}}\over \p^{3\over2}\,\lambda^{1\over2}T}\,V_{q\bar{q}}^a$ versus $\tilde{d}_a=\p\,T\,d_a$ ($a=\ \parallel,\perp$). The upper plots are for  for different $\eps$ and fixed $\tilde{\delta}$, while the lower ones correspond to different values of $\tilde{\delta}$ and fixed $\eps$. The left plot shows the intra-layer potential, while the right one is the inter-layer one.}
\label{fig:V}
\end{figure}

In order to find the quark-antiquark potential we have to evaluate the on-shell action. Plugging (\ref{rho_prime_parallel}) into (\ref{ec:NGaxctionx}), we obtain:
\beq
2\pi\,\alpha'S_\parallel^{\rm{on-shell}}=2\int_{\r_0}^{\r_{\rm{max}}} \dd \r \,\frac{\sqrt{h_0}\,b^{3/2}\, F \,S^4\, \alpha \,e^{\phi -\frac{\phi (\r_0)}{2}}}{\r\,\left(\r^4-\rh^4 \right) \,\sqrt{h_0\, b \,e^{\phi-\phi (\r_0)}-b_0\, h}}\equiv\int_{\r_0}^{\r_{\rm{max}}} \dd \r \,\mathcal{L}_\parallel^{\rm{on-shell}}\,\,.
\eeq
It is straightforward to see that the on-shell Lagrangian $\mathcal{L}_\parallel^{\rm{on-shell}}$ diverges at the UV boundary  $\rho\to\infty$. We regularize this divergence by subtracting the action of two fundamental strings going straight from $\r =\rh$ to the boundary at $\r\to\infty$. The regularized on-shell action is identified with the quark-antiquark potential. We get:
\bea
2\pi\,\alpha'\,V_{q\bar{q}}^\parallel & = & 2\pi\,\left(S_\parallel^{\rm{on-shell}}-\int_{\rh}^{\infty} \dd \r \,\mathcal{L}_{\parallel}^{\rm{straight}}\right)\rc
& =& 2\int_{\r_0}^{\infty} \dd \r \,\frac{\sqrt{h_0}\,b^{3/2}\, F \,S^4\, \alpha \,e^{\phi -\frac{\phi (\r_0)}{2}}}{\r\,\left(\r^4-\rh^4 \right) \,\sqrt{h_0\, b \,e^{\phi-\phi (\r_0)}-b_0\, h}}-2\int_{\rh}^{\infty} \dd \r \,\,e^{\frac{\f}{2}}\, \frac{\a\, F \,S^4}{\rho ^5}\rc
& = & 2\int_{\r_0}^{\infty} \dd \r \left(\,\frac{\sqrt{h_0}\,b^{3/2}\, F \,S^4\, \alpha \,e^{\phi -\frac{\phi (\r_0)}{2}}}{\r\,\left(\r^4-\rh^4 \right) \,\sqrt{h_0\, b \,e^{\phi-\phi (\r_0)}-b_0\, h}}-e^{\frac{\f}{2}}\, \frac{\a\, F \,S^4}{\rho ^5}\right)-2\int_{\rh}^{\r_0} \dd \r \,\,e^{\frac{\f}{2}}\, \frac{\a\, F \,S^4}{\rho ^5}\,\,.\rc
\eea
As for the quark-antiquark distance, these integrals cannot be performed analytically. The numerical results are shown in Fig.~\ref{fig:V}. 

\begin{figure}[!ht]
\center
 \includegraphics[width=0.47\textwidth]{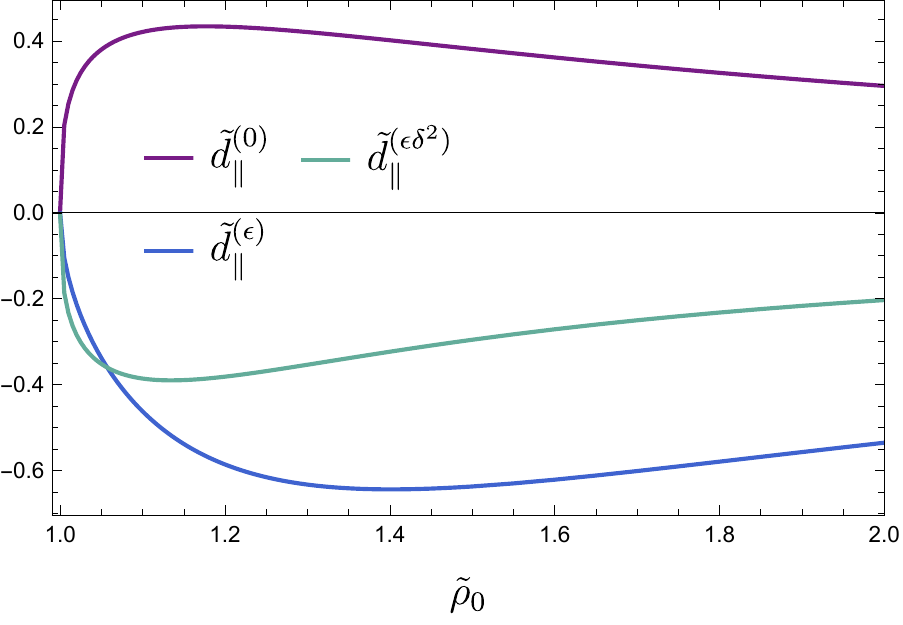} 
 \includegraphics[width=0.47\textwidth]{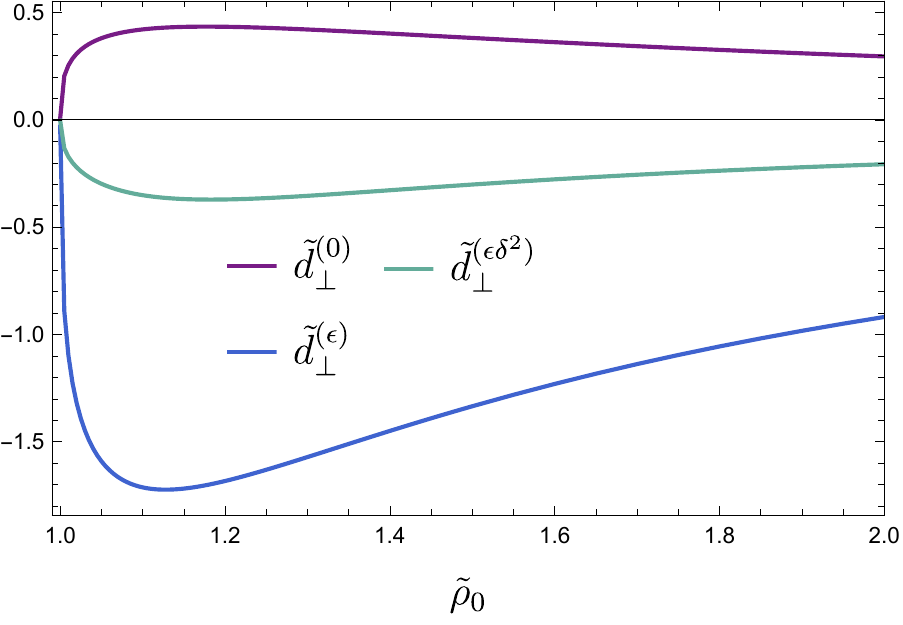} 
 \includegraphics[width=0.47\textwidth]{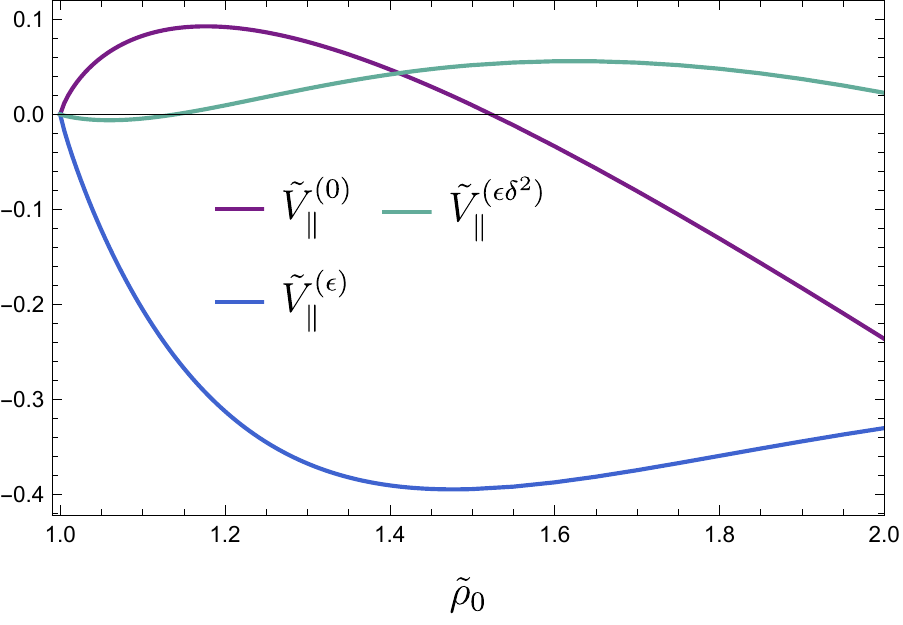} 
 \includegraphics[width=0.47\textwidth]{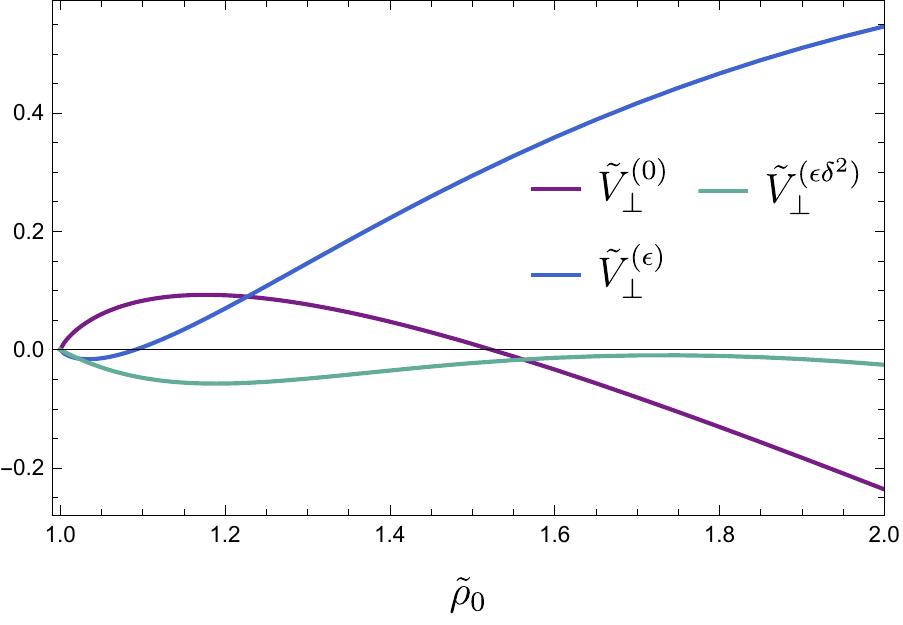} 
  \caption{Comparison between the corrections at different orders. In these plots $\tilde V_a={{\rm Vol}(\mathcal{M}_5)^{{1\over2}}\over \p^{3\over2}\,\lambda^{1\over2}T}\,V_a$  and $\tilde d_a\,=\,\pi\,T\,d_a$ for $a=\parallel, \perp$ and $\tilde\rho_0=\rho_0/\rh$.} 
 \label{fig:corrections}
\end{figure}

Examining the anisotropic and finite density contribution separately, the order $\eps$ correction increases the maximum parallel distance $\tilde{d}_\parallel^{\rm{max}}$, while the order $\eps\tilde{\delta}^2$ correction does not change significantly the numerical results. We represent these contributions as:
\beq
d_{\parallel}\,=\,d_{\parallel}^{(0)}\,+\,\epsilon\,
d_{\parallel}^{(\epsilon)}\,+\,\epsilon\,\tilde\delta^2\,
d_{\parallel}^{(\epsilon\,\tilde\delta^2)}\,\,,
\qquad\qquad
V_{q\bar{q}}^{\parallel}\,=\,V_{\parallel}^{(0)}\,+\,\epsilon\,
V_{\parallel}^{(\epsilon)}\,+\,\epsilon\,\tilde\delta^2\,
V_{\parallel}^{(\epsilon\,\tilde\delta^2)}\,\,.
\label{d_V_contributions}
\eeq
In Fig.~\ref{fig:corrections} we plot separately these contributions to the distance and to the potential.

\subsubsection{Inter-layer potential}

We now consider a fundamental string hanging from the UV boundary and extended along $x^3$ with the other two cartesian coordinates being constant. We now take $\rho=\rho(x^3)$. Eq.  \eqref{ec:NGaxction} becomes:
\bea\label{ec:NGaxctionz}
S_\perp={1\over 2\pi\alpha'}\int \dd \t \,\dd x^3 \,e^{\f/2}\,\sqrt{\frac{b^2 F^2 S^8 \alpha^2 }{\rho ^2 \left(\rho^4-\rh^4\right)^2}\,\rho'^2+\frac{b\,\a^2}{h}}\,\,,
\eea
from which we obtain the following equation for $\r'$:
\bea
\r'=\pm \sqrt{{b\over b_0}}\frac{\a\sqrt{h_0\, b \,\a^2\,e^{\phi-\phi(\r_0)}-b_0\,\a_0^2 \,h}}{\sqrt{B_2} \, h\,\a_0}\,\,,
\eea
where $B_2$ is the following function:
\beq
B_2=\frac{b^2 F^2 S^8 \alpha \,^2}{\rho^2 \left(\rho^4-\rh^4\right)^2}\,\,.
\eeq
As before, we can invert it to obtain $x^3$ as a function of $\rho$:
\bea
x_3(\r)=\pm \int_{\r_0}^\r \dd \r\,\sqrt{{b_0\over b}}\, \frac{\sqrt{B_2} \, h\,\a_0}{\a\sqrt{h_0\, b \,\a^2\,e^{\phi-\phi(\r_0)}-b_0\,\a_0^2 \,h}}\,\,.
\eea
It follows that the quark-antiquark distance $d_\perp$ at the boundary is given by the following integral:
\bea
d_\perp=2 \int_{\r_0}^\infty \dd \r \,\sqrt{{b_0\over b}}\, \frac{\sqrt{B_2} \, h\,\a_0}{\a\sqrt{h_0\, b \,\a^2\,e^{\phi-\phi(\r_0)}-b_0\,\a_0^2 \,h}}\,\,,
\eea
which, as in the intra-layer case must be performed numerically. The result is shown in Fig.~\ref{fig:d}. We next calculate the on-shell action  for these configuration, which reads:
\bea
2\pi\,\alpha'\,S_\perp^{\rm{on-shell}}=2\int_{\r_0}^{\r_{\rm{max}}} \dd \r \frac{\sqrt{h_0} \,b^{3/2} F S^4 \alpha^2 e^{\phi -\frac{\phi (\r_0)}{2}}}{\r\left(\r^4-\rh^4 \right) \sqrt{h_0 \,\a^2\,b e^{\phi-\phi (\r_0)}-b_0\,\a_0^2\, h}}=\int_{\r_0}^{\r_{\rm{max}}} \dd \r \,\mathcal{L}_\perp^{\rm{on-shell}}\,\,,\rc
\eea
which is again divergent at the UV. As above, we regularize this divergence by subtracting the action of two fundamental strings going straight from $\r =\rh$ to the boundary at $\r\to\infty$. The regularized on-shell action is identified with the quark-antiquark potential along the $x^3$-direction, and is given by
\bea
 2\pi\,\alpha'\,V_{q\bar{q}}^\perp & = & 2\pi\,\left(S_\perp^{\rm{on-shell}}-\int_{\rh}^{\infty} \dd \r \,\mathcal{L}_\perp^{\rm{straight}}\right)\rc\rc
&= & 2\int_{\r_0}^{\infty} \dd \r \,\frac{\sqrt{h_0} \,b^{3/2} F S^4 \alpha^2 e^{\phi -\frac{\phi (\r_0)}{2}}}{\r\left(\r^4-\rh^4 \right) \sqrt{h_0 \,\a\,b e^{\phi-\phi (\r_0)}-b_0\,\a_0\, h}}-2\int_{\rh}^{\infty} \dd \r \,\,e^{\frac{\f}{2}}\, \frac{\a\, F \,S^4}{\rho ^5}\rc\rc
& = & 2\int_{\r_0}^{\infty} \dd \r \left(\,\frac{\sqrt{h_0} \,b^{3/2} F S^4 \alpha^2 e^{\phi -\frac{\phi (\r_0)}{2}}}{\r\left(\r^4-\rh^4 \right) \sqrt{h_0 \,\a\,b e^{\phi-\phi (\r_0)}-b_0\,\a_0\, h}}-e^{\frac{\f}{2}}\, \frac{\a\, F \,S^4}{\rho ^5}\right)\rc\rc
&&\qquad\qquad-2\int_{\rh}^{\r_0} \dd \r \,\,e^{\frac{\f}{2}}\, \frac{\a\, F \,S^4}{\rho ^5}\,\,.
\eea
These integrals can be performed numerically. The results are shown in Fig.~\ref{fig:V}. In addition, we consider the anisotropic and finite density corrections similarly, introducing an expansion in $\epsilon$ and $\tilde\delta$ similar to \eqref{d_V_contributions}, but for the inter-layer separation and potential. The results are represented in Fig.~\ref{fig:corrections}.

\subsubsection{Physical interpretation}\label{sec:physintqqbar}

Let us now comment on the results shown in Fig.~\ref{fig:V}. At small separations between the quark and the antiquark, the potential follows the colored curves, until it reaches the horizontal line. At this point the disconnected configuration becomes dominant and the potential is flat, thus the color charges are screened by the plasma.  We observe that as the density of defects, parameterized by $\epsilon$, increases, the separation where the charges are screened becomes smaller, and that the effect is more pronounced in the direction transverse to the defects. The enhanced screening could be taken as a natural consequence of having more color nonsinglet degrees of freedom in the plasma, from the fields at the defects. 
Note also that strings can end on D5-branes so a connected configuration can break in two separate strings with the new endpoints localized at the D5-branes. In principle the breaking might be expected to be more favorable if the string is extended in the direction transverse to the D5-brane, since the string should be orthogonal to the D5-brane worldvolume at the endpoint. This would serve as a possible explanation for the difference in screening between the transverse and parallel directions. 

Dialing the baryon charge density, through changes in $\tilde\delta$, also affects the screening. Although the effect is relatively mild (recall that the baryon density is small compared to the defect density), increasing the baryon density seems to also increase the screening. Note that the purple and blue curves in the lower plots of Fig.~\ref{fig:V} are slightly displaced to the right compared to the upper plots, and that $\tilde \delta$ should be smaller for those. This is natural in view that increasing the baryon density would increase the number of degrees of freedom that can contribute to the screening. Notice also that in this case the effect seems to be stronger in the directions parallel to the defect, as might have been expected since in principle the charges would be localized along these directions.

\subsection{Entanglement entropy}

The holographic entanglement entropy \cite{Ryu:2006bv,Ryu:2006ef} between a spatial region $A$ in the gauge theory and its complement is obtained by finding the eight-dimensional spatial surface $\Sigma$ whose boundary coincides with the boundary of A and minimizes the functional:
\bea
S_A={1\over 4G_{10}}\int_\Sigma\,\dd^8\xi\,\sqrt{\text{det}g_8}\,\,,
\eea
where $G_{10}=8\,\pi^6\,g_\mt{s}^{2}\, \alpha'^4$ is the ten-dimensional Newton constant satisfying $16\pi\,G_{10}=2\kappa_{10}^2$ and $g_8$ is the induced metric on $\Sigma$.  We will minimize $S_A$ for the case in which the region $A$ is a slab of infinite extent in two  of the spatial gauge theory directions and has a finite width in the third one. There are clearly two different cases to be studied, depending on whether the direction with finite width is parallel or transverse to the defect. We study these two possibilities in the two subsections that follow.

\subsubsection{Parallel slab}

Let $A$ be the region:
\beq
A\,=\,\left\{-\frac{l_\parallel}{2}<x^1<\frac{l_\parallel}{2}, -\frac{L_2}{2}\leq x^2<\frac{L_2}{2},\, -\frac{L_3}{2}\leq x^3<\frac{L_3}{2} \right\}\,\,.
\eeq
We consider both $x^2$- and $x^3$-directions to be periodic with periods $L_2$ and $L_3$, otherwise one ends up with partial differential equations. Eventually, we are interested in the $L_2,L_3\to \infty$ limit. We will consider a surface $\Sigma$ ending on the boundary of $A$ at the UV $\r\to\infty$, which penetrates into the bulk  and reaches a minimal value $\rho_0$ of the holographic coordinate $\rho$. Actually, in this section we will use the dimensionless variable $y$ defined as:
\beq
y\,=\,{\rho\over \rho_0}\,\,.
\label{y_def}
\eeq
We will parameterize $\Sigma$ by a function
$y=y(x^1)$ and, to compute the entanglement functional $S_A$, 
we will  integrate  over all variables except $x^1$. We get:
\bea\label{eq:Spar}
S_\parallel={ L_2\,L_3\,{\rm Vol} \big({\cal M}_5\big)\over 4 G_{10}}\int \dd x^1\,\sqrt{\frac{b\, F^4 \,h^2 \,S^{16} \alpha^4 }{y^2\left(\r_0^4 y^4-\rh^4 \right)^2}\,y'^2+F^2\,h \,S^8 \,\alpha^2}\,\,,
\eea
The functional $S_\parallel$ has a first integral that allows to obtain $y'$ as:
\bea\label{eq:dypar}
\frac{\dd y\,}{\dd x^1}=\pm\,F\,h^{1/2}\,S^4\,\a\frac{\sqrt{ F^2\,h \,S^{8} \,\alpha^2-\alpha_0^2\, F_0^2\, h_0 \,S_0^8}}{B_3^{1/2}\alpha _0\, \,F_0 \,h_0^{1/2}\, S_0^4}\,\,,
\eea
where the subscript $0$ implies the functions are evaluated at $y=y_0=1$ and $B_3$ is given by:
\beq
B_3=\frac{b\, F^4 \,h^{2} \,S^{16}\, \alpha^4}{y^2 \left(\r_0^4 y^4-\rh^4\right)^2}\,\,.
\eeq
Thus, we have:
\bea
x^1=\pm\int_1^y\dd y\,{B_3^{1/2}\alpha _0\, \,F_0 \,h_0^{1/2}\, S_0^4\over F\,h^{1/2}\,S^4\,\a\sqrt{F^2\,h \,S^{8} \,\alpha^2-\alpha_0^2\, F_0^2\, h_0 \,S_0^8}}\,\,,
\eea
and therefore, the length $l_\parallel$ in this direction is:
\bea\label{eq:lpar}
l_\parallel=2\int_1^\infty \dd y\,{B_3^{1/2}\alpha _0\, \,F_0 \,h_0^{1/2}\, S_0^4\over F\,h^{1/2}\,S^4\,\a\sqrt{F^2\,h \,S^{8} \,\alpha^2-\alpha_0^2\, F_0^2\, h_0 \,S_0^8}}\,\,.
\eea
Plugging this result in \eqref{eq:Spar}, we obtain that $S_\parallel$ has the following on-shell value:
\bea\label{eq:SparNew}
{4G_{10}\over  L_2\,L_3\,{\rm Vol} \big({\cal M}_5\big)}S_\parallel=2\int_1^\infty \dd y\,\frac{b^{1/2} F^3 h^{3/2} S^{12} \alpha^3}{y\,\left(\r_0^4 y^4-\rh^4 \right) \sqrt{F^2\, h\, S^8 \alpha^2-\alpha_0^2 F_0^2 h_0\, S_0^8}}\equiv 2\int_1^\infty \dd y\,\mathcal{L_\parallel}\,\,,
\eea
where, in the last step, we have defined the function $\mathcal{L_\parallel}$. 
If we now want to evaluate the entropy for this configuration, we need to subtract the divergent contributions that come from the upper limit of the integral, $y\to\infty$. When $y\to\infty$, $\mathcal{L_\parallel}$ behaves as
\bea
\mathcal{L}_\parallel=\r_0\, \Qc\,\left(\frac{\r_0}{4} y +\eps\,\rh \right)+{\cal O}(y^{-3})
\eea
leading to asymptotic behavior for $S_\parallel$ as
\bea
{4G_{10}\over  L_2\,L_3\,{\rm Vol} \big({\cal M}_5\big)}{S_\parallel}=\,2\,\r_0\, \Qc\left(\frac{\r_0}{8} y_{\rm{max}}^2 +\eps\,\rh\,y_{\rm{max}} \right)+{\cal O}(y^{-2}) \ ,
\eea
where the last term vanishes when $y_{\rm{max}}\to\infty$. Defining the cutoff (in units of length) as $\varepsilon_{\text{\scriptsize{UV}}}=Q_c^{1/2}/(2\rho_{\rm{max}})\to 0$, the divergent part is
\be \label{eq:Sdivpar}
S_\parallel^{\rm{div}} = \frac{N_c^2}{2\pi}\bar{a\, }\frac{L_2 L_3}{\varepsilon_{\text{\scriptsize{UV}}}^2}+ \Nf\,\Nc\, \bar{a}^{1\over 2} \frac{2}{15} {\l^{1\over 2} \over  v_{\perp}\, }\frac{L_2}{\varepsilon_{\text{\scriptsize{UV}}}} \,.
\ee
The first term reproduces the area law for the theory in the absence of flavors. The second term is a subleading correction proportional to the number of flavors that has the form of an area law for a ($2+1$)-dimensional theory, so it naturally captures the contribution to the entanglement entropy from the fields localized at the defects.

Subtracting the divergent part, we can write:
\bea
{4G_{10}\over L_2\,L_3\,{\rm Vol} \big({\cal M}_5\big)}S_\parallel^{\rm{reg}} & = & {4G_{10}\over  L_2\,L_3\,{\rm Vol} \big({\cal M}_5\big)}\left(S_\parallel-S_\parallel^{\rm{div}}\right)\rc\rc
& = & 2\int_1^\infty \dd y\,\left(\mathcal{L_\parallel}-\mathcal{L}_\parallel^{\rm{div}}\right)-2\,\r_0\, \Qc\,\left(\frac{\r_0}{8}+\eps\,\rh \right)\,\,.
\eea

\begin{figure}[!ht]
\center
 \includegraphics[width=0.47\textwidth]{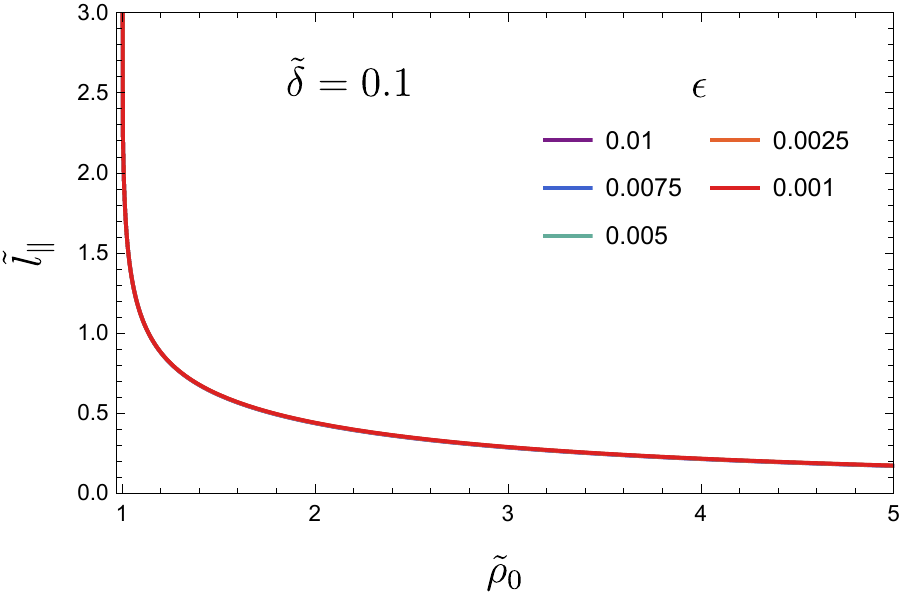} 
  \includegraphics[width=0.47\textwidth]{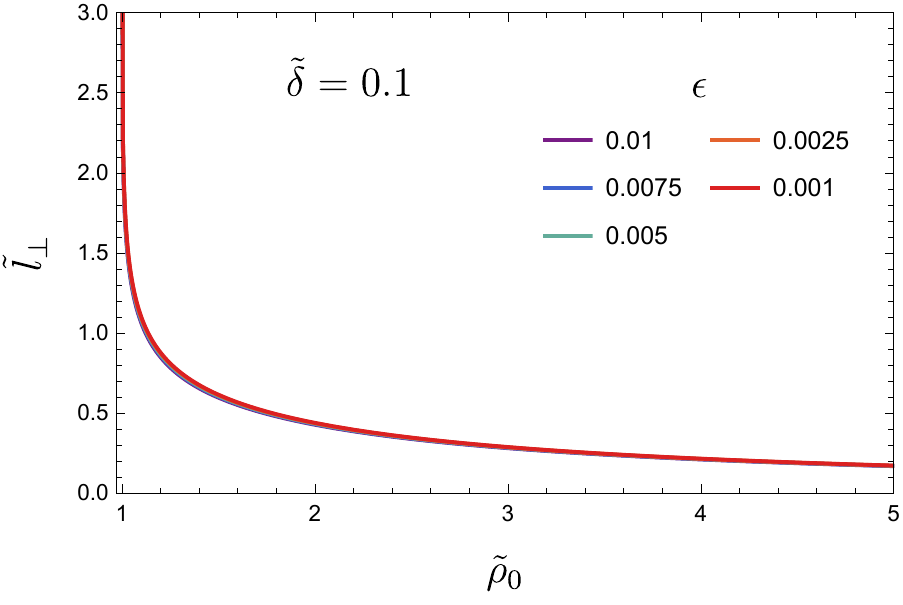} 
   \includegraphics[width=0.47\textwidth]{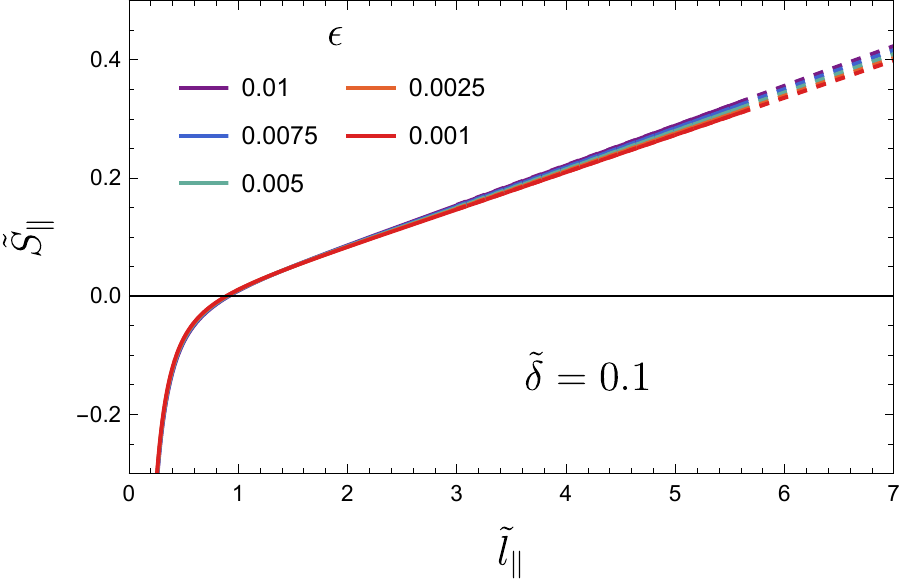} 
    \includegraphics[width=0.47\textwidth]{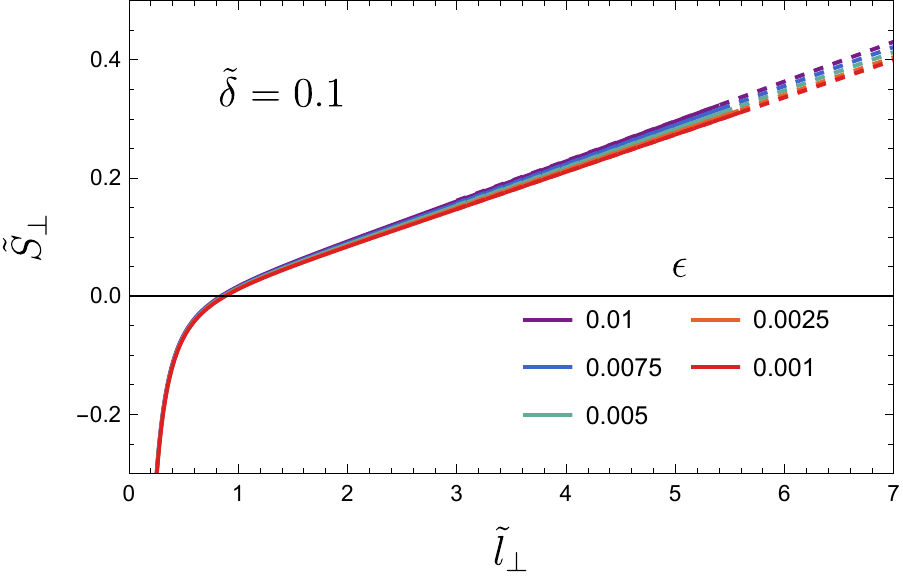} 
   \includegraphics[width=0.47\textwidth]{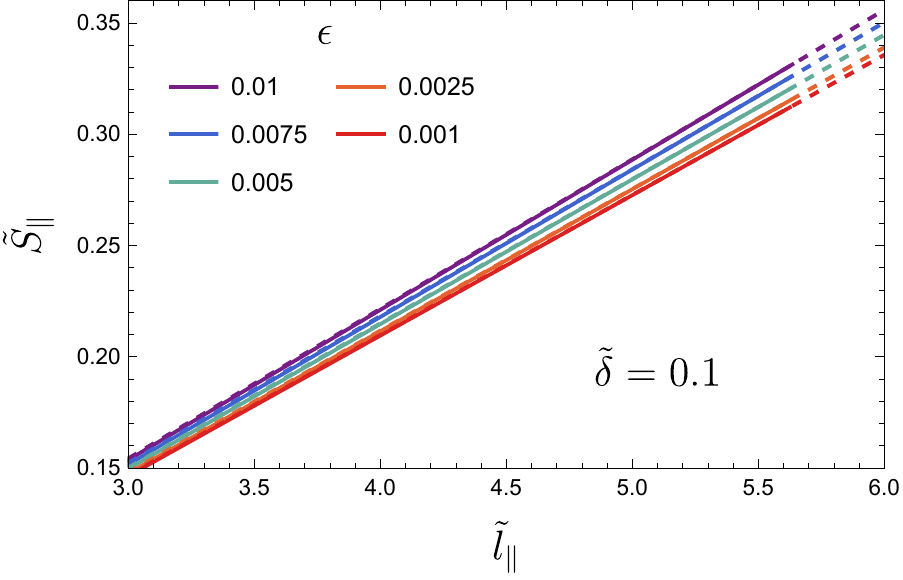}
    \includegraphics[width=0.47\textwidth]{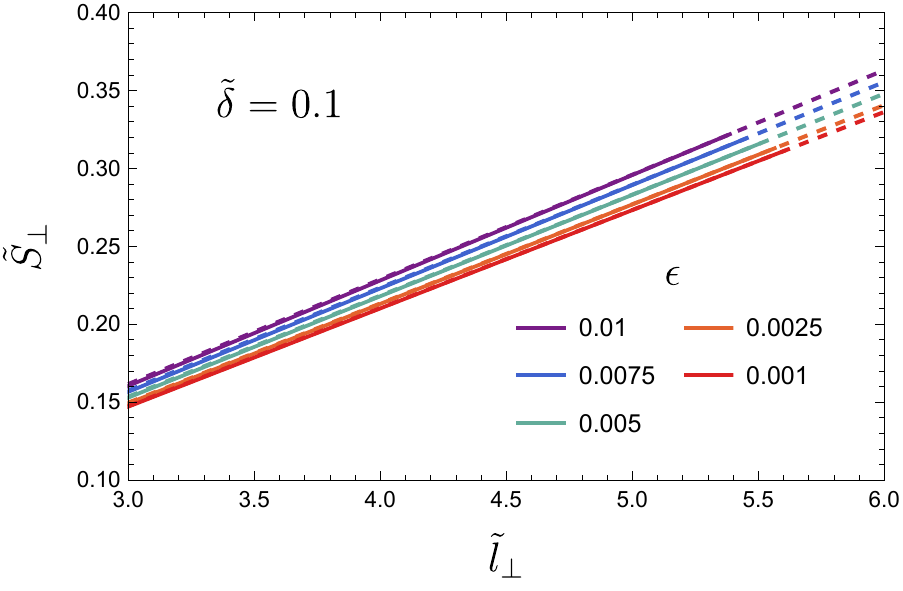} 
  \caption{First row: rescaled parallel (left) and transverse (right)  width $\tilde{l}=\p\,T\, l$ versus the rescaled turning point $\tilde{\r}_0={\r_0\over \rh}$. Second and third rows: and rescaled parallel (left) and transverse (right) entanglement entropy $\tilde{S}={{\rm Vol}(\mathcal{M}_5)\over 8\,\p^4\,L_2\,L_3\,\Nc^2\,T^2}S^{\rm{reg}}$ versus the rescaled width. The slope of the dashed lines are fixed by the thermal entropy density according to \eqref{entanglementS_themalS_parallel}and \eqref{entanglementS_themalS_perp}.}
\label{fig:Sl}
\end{figure}

\begin{figure}[!ht]
\center
 \includegraphics[width=0.47\textwidth]{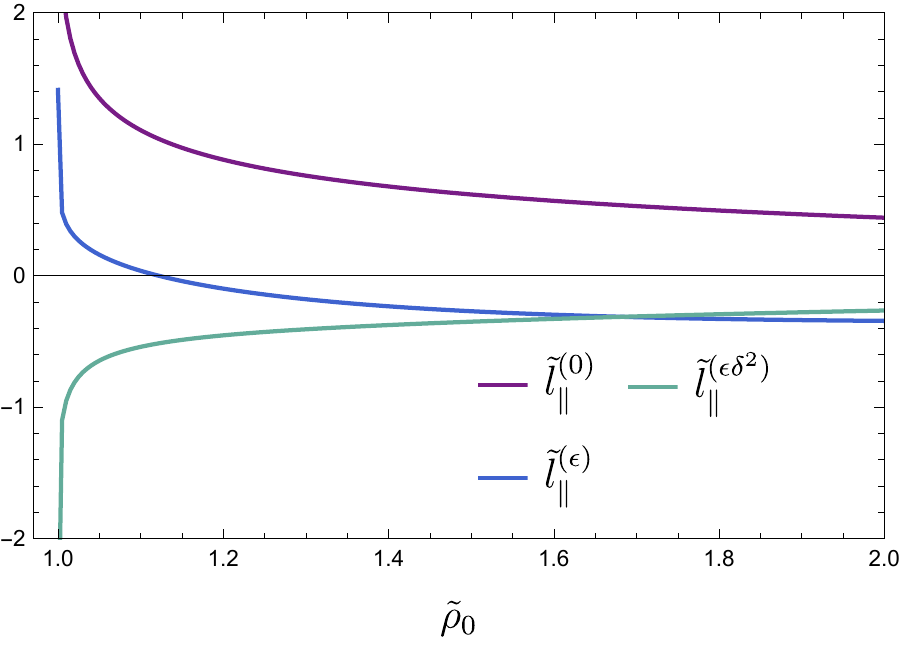} 
  \includegraphics[width=0.47\textwidth]{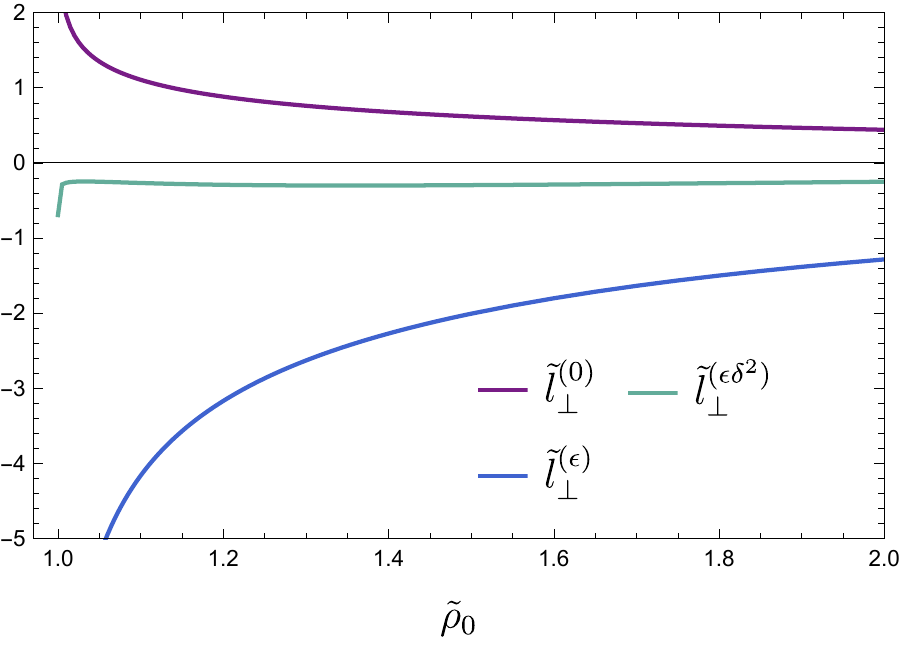} 
 \includegraphics[width=0.47\textwidth]{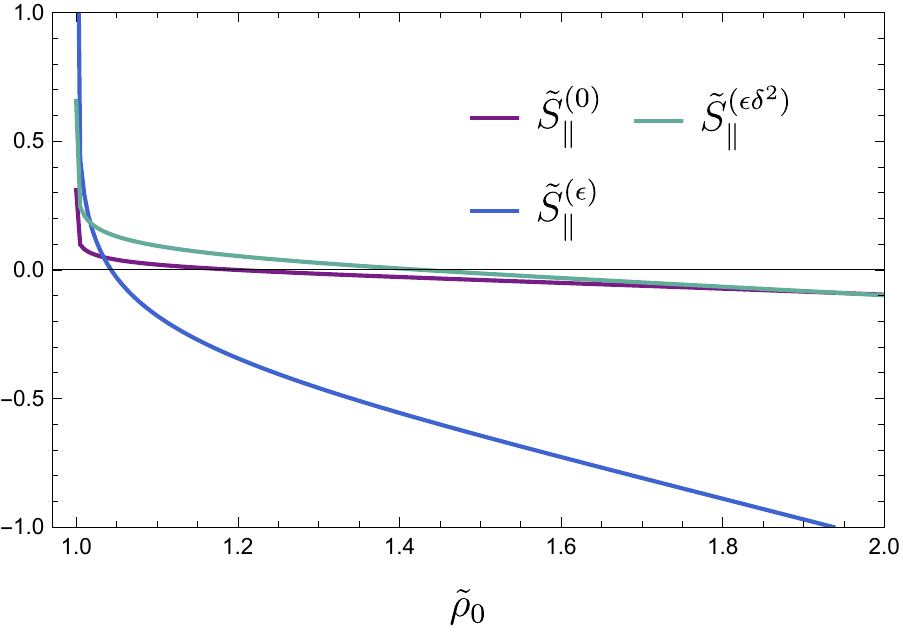} 
 \includegraphics[width=0.47\textwidth]{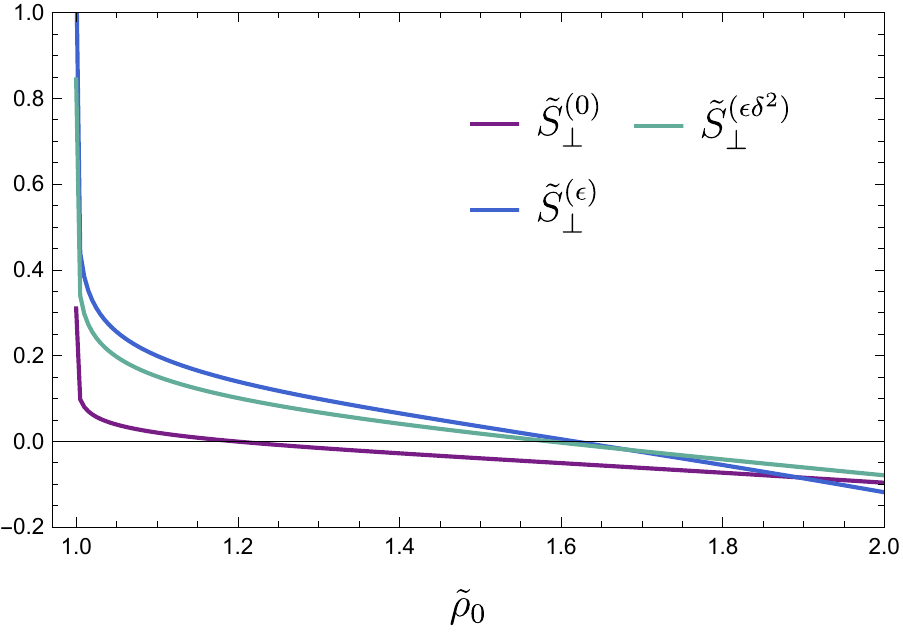} 
  \caption{Comparison between the corrections at different orders of the widths and entanglement entropies, defined in way similar to (\ref{d_V_contributions}). }
\label{fig:corrections2}
\end{figure}
\begin{figure}[!ht]
\center
 \includegraphics[width=0.47\textwidth]{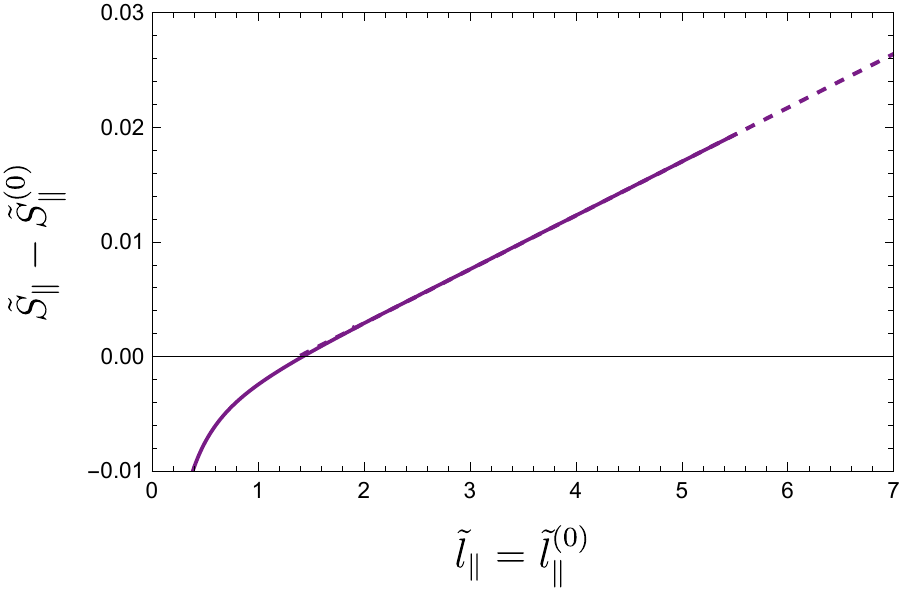} 
  \includegraphics[width=0.47\textwidth]{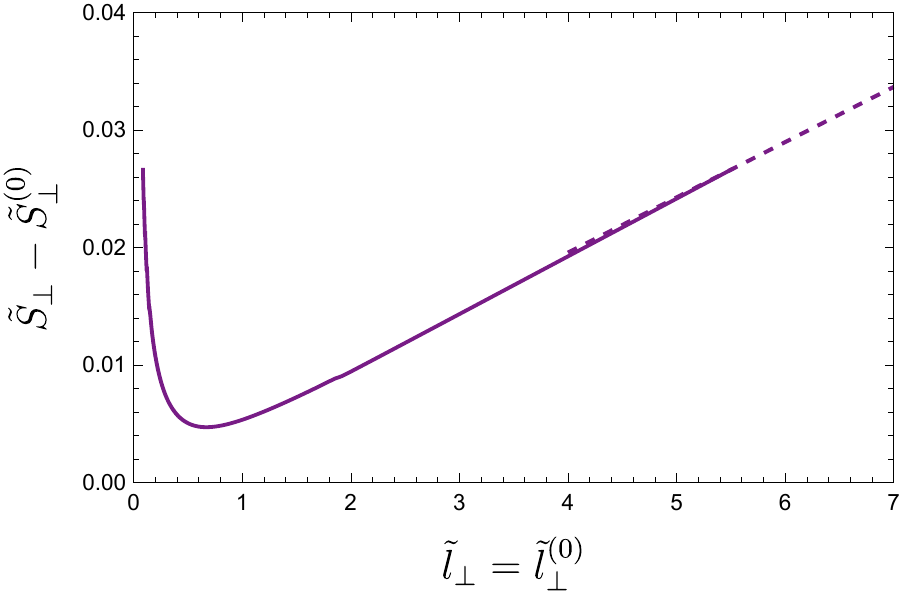} 
  \caption{Differences in the finite contributions to the entanglement entropy between the flavored and unflavored theories. We have set $\epsilon=0.01$ and $\tilde{\delta}=0.1$. The slope of the dashed lines are fixed by the flavor contribution to the thermal entropy density.}
\label{fig:differences2}
\end{figure}

The parallel distance $l_\parallel$ and the regulated entanglement entropy $S_\parallel^{\rm{reg}}$ are plotted in Fig.~\ref{fig:Sl}. The $\epsilon$ correction increases the value of the width, while it decreases slightly the value of the entanglement entropy for small $\tilde{l}_\parallel$, and increases it for big $\tilde{l}_\parallel$. The contributions of the different orders are represented in Fig.~\ref{fig:corrections2}, using a notation similar to the one employed in (\ref{d_V_contributions}). See also Fig.~\ref{fig:differences2}, where we compare the entanglement entropies of the flavored and unflavored theories in the two directions at the same slab widths.

The entanglement entropy found above matches the results of section \ref{Thermo} for the thermal entropy density of the black hole in the limit in which 
$T\,l_{\parallel}$ is large \cite{Fischler:2012uv,Erdmenger:2017pfh,Jokela:2019ebz}. Indeed, let us use the prescription given in \cite{Jokela:2020auu} to compute the derivative of the entanglement entropy with respect to $l_\parallel$ from the derivatives of both $l_\parallel$ and $S_\parallel$ with respect to the turning point $y_0$  when $y_0=1$. From \eqref{eq:lpar} we get:
\bea\label{eq:derlpar}
{\dd  l_\parallel\over \dd  y_0}\Bigg|_{y_0=1}\,
=\,\frac{2\,b_0^{1/2}\, F_0^2\,h_0 \,S_0^{8}\, \alpha_0^2}{\r_0^4 -\rh^4}\lim_{y\to 1}\left({1\over \sqrt{F^2\,h \,S^{8} \,\alpha^2-\alpha_0^2\, F_0^2\, h_0 \,S_0^8}}\right)\,\,.
\eea
Moreover,  from \eqref{eq:SparNew} we arrive at:
\bea
{4\,G_{10}\over  L_2\,L_3\,\,{\rm Vol} \big({\cal M}_5\big)}{\dd  S_\parallel^{\rm{reg}}\over \dd  y_0}\Bigg|_{y_0=1}
\,=\,\frac{2\,b_0^{1/2} F_0^3 h_0^{3/2} S_0^{12} \alpha_0^3}{\rho_0^4-\rh^4 }\lim_{y\to 1}\left({1\over \sqrt{F^2\,h \,S^{8} \,\alpha^2-\alpha_0^2\, F_0^2\, h_0 \,S_0^8}}\right)\,\,.
\eea
Thus, we can write:
\bea
 {4G_{10}\over {\rm Vol} \big({\cal M}_5\big)\, L_2\,L_3}{\dd  S_\parallel^{\rm{reg}}\over \dd  l_\parallel} & = & {4G_{10}\over {\rm Vol} \big({\cal M}_5\big)\, L_2\,L_3}{\dd  S_\parallel^{\rm{reg}}\over \dd  y_0}{ \dd  y_0\over\dd  l_\parallel} \nonumber \\
 & = & F_0\,h_0^{1/2} S_0^{4} \alpha_0\nonumber \\
& = & \frac{\r_0^3 }{2}  \,\Qc^{1/2}\Bigg(1+\frac{5 \epsilon}{2 \rh^4}\Bigg(\frac{1}{5} \left(\r_0^4-\rh^4\right) \Omega \left(\frac{\r_0}{\rh}\right)+\r_0^3 \rh-\frac{\rh^4}{4}\rc
&&+\tilde{\delta}^2 \left(-\left(\r_0^4-\rh^4\right) \Omega \left(\frac{\r_0}{\rh}\right)-5 \r_0^3 \rh+\frac{4 \rh^5}{\r_0}+\frac{5 \rh^4}{4}\right)\Bigg)\Bigg)\,\,.
\label{dS_parallel_dl}
\eea
When $\r_0\to\rh$ this last expression becomes:
\beq
{4G_{10}\over {\rm Vol} \big({\cal M}_5\big)\, L_2\,L_3}{\dd  S_\parallel^{\rm{reg}}\over \dd  l_\parallel}=\frac{\rh^3 }{2} \,\Qc^{1/2}\left(1+\eps\, {15\over 8}+\eps\,\tilde{\delta}^2\,{5\over8}\right)\,\,,
\label{entanglement_density}
\eeq
which agrees with the thermal entropy density of the  black hole obtained in section \ref{Thermo}.  Actually, taking into account the value of $G_{10}$, one can rewrite 
(\ref{s_rho_h}) as:
\beq
{4G_{10}\over {\rm Vol} \big({\cal M}_5\big)}s\,=\,{\rh^3\over 2}\,
\Qc^{1/2}\left(1+\eps\,\frac{15 }{8}+\eps\,\tilde{\delta}^2\,{5\over8} \right) \,\,.
\label{thermal_s}
\eeq
By comparing (\ref{entanglement_density}) and (\ref{thermal_s}, it follows that, when $\rho_0\to\rh$ and $T\,l_\parallel$ is large, we have the following approximate equality  between the entanglement entropy density and the thermal entropy density $s$:
\beq
  {1\over L_2\,L_3} {\dd  S_\parallel\over \dd  l_\parallel}\Big|_{\rho_0\to\rh}  = s \label{entanglementS_themalS_parallel}
\eeq
in agreement with the analysis of \cite{Fischler:2012uv,Erdmenger:2017pfh,Jokela:2019ebz}. 
By looking at  the slope of the entanglement entropy versus the parallel width, we have also verified that, for different values of the expansion parameters $\epsilon$ and $\epsilon\,\tilde\delta^2$,  our numerical results fulfill (\ref{entanglementS_themalS_parallel}) in the limit in which $T l_{\parallel}$ is large (see Fig.~\ref{fig:Sl}). Both \eqref{entanglementS_themalS_parallel} and the numerical results capture the increase in the number of degrees of freedom that enter through the fields localized at the defects, consistently with the thermodynamic quantities we computed previously.

\subsubsection{Transverse slab}

Let us now consider the region:
\beq
 A\,=\, \left\{-{l_\perp\over 2}<x^3<{l_\perp\over 2},-\frac{L_1}{2}\leq x^1< \frac{L_1}{2}, -\frac{L_2}{2}\leq x^2<\frac{L_2}{2}\right\}\ ,
\eeq
which is a slab with finite width in the transverse direction. As before, we take $x^1$ and $x^2$ to be periodic with periods $L_1$ and $L_2$ respectively, and we are ultimately interested in the $L_1,L_2\to \infty$ limit. Let us use again the variable $y$ defined in (\ref{y_def}) and take $y=y(x^3)$. By integrating over all variables except $x^3$, we get:
\bea\label{eq:Sper}
S_\perp={ L_1\,L_2\,{\rm Vol} \big({\cal M}_5\big)\over 4 G_{10}} \int \dd x_3\,F\,h^{1/2}\,S^4\, \alpha \, \sqrt{1+\frac{b\, F^2\, h\, S^8\, }{y^2\,\left(\r_0^4\, y^4-\rh^4\right)^2}y'^2}\,\,,
\eea
where now $y'$ denotes the derivative with respect to $x^3$. 
From the  first integral of $S_\perp$ we can get the value of $y'$, namely:
\bea\label{eq:dyper}
{\dd y\over \dd x_3}=\pm \frac{F \,h^{1/2}\, S^4\, \alpha\, \sqrt{F^2\, h\, S^8\, \alpha^2-\alpha_0^2\, F_0^2\,h_0\, S_0^8}}{\alpha_0\,B_4^{1/2} \,F_0\, h_0^{1/2} S_0^4}\,\,,
\eea
where the subscript $0$ implies the functions are evaluated at $y=y_0=1$ and 
$B_4$ is the function:
\beq
B_4=\frac{b\, F^4 \,h^2\, S^{16} \,\alpha^2}{y^2\,\left(\r_0^4\, y^4-\rh^4\right)^2}\,\,.
\eeq
Thus, we have:
\bea
\dd x^3=\dd y\,\frac{\alpha_0\, F_0\,h_0^{1/2} \,S_0^4 \,b^{1/2}\, F\, h^{1/2}\, S^4}{y\,\left(\r_0^4 \,y^4-\rh^4 \right) \sqrt{F^2\, h\, S^8 \,\alpha^2-\alpha_0^2\, F_0^2\, h_0\, S_0^8}}\,\,,
\eea
and the length $l_\perp$ in the transverse direction is:
\bea
l_\perp=2\int_1^\infty \dd y\,\frac{\alpha_0\, F_0\,h_0^{1/2} \,S_0^4 \,b^{1/2}\, F\, h^{1/2}\, S^4}{y\,\left(\r_0^4 \,y^4-\rh^4 \right) \sqrt{F^2\, h\, S^8 \,\alpha^2-\alpha_0^2\, F_0^2\, h_0\, S_0^8}}\,\,.
\label{l_perp}
\eea
Moreover, by using (\ref{eq:dyper}) in \eqref{eq:Sper}, the on-shell entanglement entropy functional $S_\perp$ becomes:
\bea\label{eq:SperNew}
{4G_{10}\over  L_1\,L_2\,{\rm Vol} \big({\cal M}_5\big)}S_\perp & = & 2\int_1^\infty \dd y\,\frac{b^{1/2}\, F^3\, h^{3/2}\, S^{12}\, \alpha^2}{y\,\left(\r_0^4 \,y^4-\rh^4 \right)\sqrt{F^2\, h\, S^8\, \alpha^2-\alpha_0^2\, F_0^2\, h_0 \,S_0^8}}\rc\rc
& \equiv & 2\int_1^\infty \dd y\,\mathcal{L_\perp}\,\,,
\eea
where, in the last step, we have defined $\mathcal{L_\perp}$. 
If we now want to evaluate the finite entropy for this configuration, we need to subtract the divergent contributions that come from the upper limit of the integral, $y\to\infty$. As $y\to\infty$, $\mathcal{L_\perp}$ goes like:
\bea
\mathcal{L}_\perp=\frac{1}{4} \,\Qc \,\r_0\,\rh\,\left( y\, \r_0-\eps\rh\right)+{\mathcal{O}}(y^{-2})\,\,.
\eea
Thus, the leading behavior of $S_\perp$ reads as
\bea
{4G_{10}\over  L_1\,L_2\,{\rm Vol} \big({\cal M}_5\big)}{S_\perp}=\frac{1}{2} \,\Qc \,\r_0\,\rh\,\left( {y_{\rm{max}}^2\over2}\, \r_0-\eps\,\rh\,y_{\rm{max}}\right)+{\mathcal{O}}(y^{-1})\,\,,
\eea
where the last term vanishes when $y_{\rm{max}}\to\infty$. In terms of field theory quantities, the divergent part is therefore
\be
S_\perp^{\rm{div}}= \frac{N_c^2}{2\pi}\bar{a}\frac{L_1 L_2}{\varepsilon_{\text{\scriptsize{UV}}}^2}-\Nc\, \bar{a}^{1\over 2} \frac{1}{30} {\l^{1\over 2}  \over  v_{\perp}\, }\frac{\nf\, L_1\, L_2}{\varepsilon_{\text{\scriptsize{UV}}}} \,.
\ee
As in the case of a parallel slab, the first term reproduces the area law in the absence of flavors. The cutoff dependence of the second term is similar to the one expected for the area law in a ($2+1$)-dimensional theory, but it gives a negative contribution and depends on the flavor density, rather than on the number of flavors. Comparing with \eqref{eq:Sdivpar}, the negative contribution suggests that the number of degrees of freedom that propagate in the directions transverse to the defects is effectively reduced, at least concerning their contribution to the entanglement entropy.

Removing the divergent pieces, we define the regulated entanglement entropy
$S_\perp^{\rm{reg}}$ as:
\bea
{4G_{10}\over  L_1\,L_2\,{\rm Vol} \big({\cal M}_5\big)}S_\perp^{\rm{reg}} & = & {4G_{10}\over \, L_1\,L_2{\rm Vol} \big({\cal M}_5\big)}\left(S_\perp-S_\perp^{\rm{div}}\right)\rc\rc
& = & 2\int_1^\infty \dd y\,\left(\mathcal{L_\perp}-\mathcal{L}_\perp^{\rm{div}}\right)-\frac{1}{2} \,\Qc \,\r_0\,\rh\,\left( {1\over2}\, \r_0-\eps\right) \ .
\eea
Both $l_\perp$ and $S_\perp^{\rm{reg}}$ are plotted in Fig.~\ref{fig:Sl}. In this transverse case, the width decreases with the epsilon corrections, and the entanglement entropy increases for big $l_\perp$ and decreases for small $l_\perp$. 

As we did for the parallel slab, one can compute the entanglement entropy density for large $T l_{\perp}$ and compare the result with the thermal entropy density of the flavored black hole. By following the same steps as before, we  calculate the derivative of the entanglement entropy with respect to $l_\perp$ from the derivatives of both $l_\perp$ and $S_\perp$ with respect to the turning point. From \eqref{l_perp}, we have:
\bea\label{eq:derlper}
{\dd  l_\perp\over \dd  y_0}\Bigg|_{y_0=1}=\frac{2\,b_0^{1/2}\, F_0^2\,h_0 \,S_0^{8}\, \alpha_0}{ \left(\r_0^4-\rh^4 \right)}\lim_{y\to 1}\left({1\over \sqrt{F^2\,h \,S^{8} \,\alpha^2-\alpha_0^2\, F_0^2\, h_0 \,S_0^8}}\right)\,\,,
\eea
and from \eqref{eq:SperNew} we get:
\bea
{4G_{10}\over  L_1\,L_2{\rm Vol} \big({\cal M}_5\big)
}{\dd  S_\perp^{\rm{reg}}\over \dd  y_0}\Bigg|_{y_0=1}=
\frac{2\,b_0^{1/2} F_0^3 h_0^{3/2} S_0^{12} \alpha_0^2}{\r_0^4-\rh^4 }\lim_{y\to 1}\left({1\over \sqrt{F^2\,h \,S^{8} \,\alpha^2-\alpha_0^2\, F_0^2\, h_0 \,S_0^8}}\right)\,\,.
\eea
Thus, for large $T l_{\perp}$ we can write:
\bea
&&{4G_{10}\over  L_1\,L_2{\rm Vol} \big({\cal M}_5\big)}{\dd  S_\perp^{\rm{reg}}\over \dd  l_\perp}={4G_{10}\over L_1\,L_2
{\rm Vol} \big({\cal M}_5\big)}{\dd  S_\perp^{\rm{reg}}\over \dd  y_0}\Bigg|_{y_0=1}{ \dd  y_0\over\dd  l_\perp}=F_0\,h_0^{1/2} S_0^{4} \alpha_0\,\,,
\eea
which is the same  result as the one found for the parallel slab (see \ref{dS_parallel_dl}). Thus, when $\r_0\to\rh$  we find a result which coincides with the thermal entropy of the black hole that we obtained in Sec.~\ref{Thermo},  
\beq
  {1\over L_1\,L_2} {\dd  S_\perp\over \dd  l_\perp}\Big|_{\rho_0\to\rh}  = s \ . \label{entanglementS_themalS_perp}
\eeq
The agreement between the entanglement entropy and thermal entropy densities for large $T\,l_\perp$ can also be verified from the results plotted in Fig.~\ref{fig:Sl}. 

Similarly to the parallel case in \eqref{entanglementS_themalS_parallel},  \eqref{entanglementS_themalS_perp} and the numerical results for the entanglement entropy in the transverse slab also capture the increase in the number of degrees of freedom. Although asymptotically the transverse and parallel entanglement entropies have the same slopes, in the plots of Fig.~\ref{fig:Sl} the transverse entanglement entropy increases faster for smaller widths.

\subsubsection{Mutual information and correlations among flavors}

To gain complementary information on the entanglement, it is useful to consider the mutual information of a bipartite system. This quantity is better suited to capture quantum entanglement than the entanglement entropy which is prone to temperature effects. In addition, there are no subtleties in regard to regularization.  

Given two regions $A$ and $B$, their mutual information $I(A,B)$ is defined through the entanglement entropies as
\beq
I(A,B)=S(A)+S(B)-S(A \cup B) \ .
\eeq
The value of the mutual information is finite, positive semidefinite, and it provides an upper bound on the correlations between the two regions \cite{Wolf:2007tdq}. If we consider the regions $A$ and $B$ to be two slabs of some fixed size, the value of the mutual information is usually determined by two competing configurations \cite{Headrick:2010zt,Fischler:2012uv}. If the separation between the two slabs is relatively small, then the minimal surface area that determines $S(A \cup B)$ consists of the union of a surface joining the closer edges of the two region with another surface connecting the edges that are further away. When the separation becomes larger, then the minimal area is the union of the minimal surface associated to region $A$ with the minimal surface associated to region $B$, in which case $S(A \cup B)=S(A)+S(B)$ and the mutual information vanishes. So regions that are far enough separated contain degrees of freedom that are uncorrelated.

Let us consider now two slabs of equal width and a separation between them which is of the same length or larger. If the entanglement entropy is monotonically increasing, then the mutual information vanishes for this configuration. This is the case for the total entanglement entropy and the contribution of flavor in the parallel directions shown in Figs.~\ref{fig:Sl} and \ref{fig:differences2} (left). This indicates that the theory remains ``local'', with degrees of freedom in spatially separated regions begin independent of each other, and it is the typical situation in holographic duals. 

On the other hand, we observe a strikingly different behavior in the flavor contribution to the transverse entanglement entropy, in Fig.~\ref{fig:differences2} (right). In this case the entanglement entropy is increasing for widths larger than the scale set by the temperature, but it shows a minimum and it is decreasing for smaller widths. This implies that even if the separation between the two slabs is larger than their width, the mutual information can be nonzero, with larger values reached when the width of the two slabs is very small and the separation is close to the minimum of the entanglement entropy. This indicates that the flavor degrees of freedom are correlated in the transverse direction through distances below or of the order of the temperature scale. This implies that flavor degrees of freedom separated in the transverse direction are not completely independent. It is then tempting to interpret the scale set by the minimum of the entanglement entropy as an ``effective width'' for the D5-branes that carry those degrees of freedom.

\section{Summary and discussion}\label{conclusions}

In this work we studied the holographic dual to an intersection of multiple color D3-branes and flavor D5-branes beyond the probe or equivalently the quenched approximation. We found geometries that solve the supergravity equations of motion with D3-brane fluxes and smeared D5-brane sources. In our setup the D5-branes share two spatial directions with the D3-branes and are continuously distributed along the third direction forming a multilayer structure. The solutions we found are charged black holes corresponding to a field theory configuration with non-zero temperature and chemical potential. In the dual ambient four-dimensional supersymmetric gauge field theory, the D5-branes give rise to massless flavors on a codimension one defect. 

We solved the complete set of equations of motion in a first-order perturbative expansion in two parameters.  This solution is valid when these two parameters are small, let us now comment on this further. Taking into account the parametric dependence of the densities on the number of colors $\Nc$, flavors $\Nf$, and the 't Hooft coupling $\lambda$, we can introduce the quantities
\beq
\nf=\Nf\ \bar{n}_\mt{f} \ \ , \ \ \nb=\Nf\ \bar{n}_\mt{b} \ ,
\eeq
where $\nf$ is the density of the smeared flavors in the direction transverse to the defect and $\nb$ is the baryon density. In terms of these quantities the solutions are valid when the following conditions are satisfied
\beq
 \frac{\Nc}{\lambda^{1/2}\Nf}\gg \frac{\bar{n}_\mt{f}}{T} \gg \frac{\bar{n}_\mt{b}^2}{\bar{n}_\mt{f} T^5} \ .
\label{app_conditions}
\eeq
Alternatively, the second condition can be expressed as
\beq
 \frac{\nf}{T}\gg  \frac{\nb}{T^3} \ .
\eeq
Therefore, both the defect and baryon densities (in units of the temperature) should be small, with the latter being much smaller than the former. 

In this regime of validity we have been able to obtain a consistent anisotropic thermodynamics and to compute several observables such as transport coefficients, quark-antiquark potentials, and entanglement entropies. Our results for the entropy and energy density \eqref{s_E_physical_chemical}, as well as for the pressures \eqref{pressures} reflect that there are additional degrees of freedom along the $(2+1)$-dimensional intersections between the color and flavor branes, contributing to the entropy, energy density and pressure along the intersection but not to the transverse pressure. The enhancement in the number of degrees of freedom also shows in the entanglement entropy of a slab, see Fig.~\ref{fig:Sl}, which becomes larger when the density of defects (implicit in the parameter $\epsilon$) increases. A more precise statement reflecting the same facts is captured by the formulas \eqref{entanglementS_themalS_parallel} and \eqref{entanglementS_themalS_perp}  that determine the increase of the entanglement entropy with the size of the entangling region as proportional to the thermal entropy. Curiously, in this limit the entanglement entropy shows a similar behavior independently of whether the slab is oriented parallel or transverse to the intersection. 

The results for the quark-antiquark potential, depicted in Fig.~\ref{fig:V}, indicate that the degrees of freedom at the intersections contribute significantly to the screening of color charges, in any direction, although the effect is more pronounced in the transverse direction. On the other hand, the effect of the charge density is more noticeable along the directions parallel to the defect. We did not study the effect of the charge density in hydrodynamics, but we observed that the second order transport coefficient in the shear channel along the intersection increases slightly with the number of D5-branes \eqref{transport_coeff_results}. This might be an indication of $(2+1)$-dimensional dynamics, where the value of the coefficient is larger, modifying the transport properties. Another possible indication is that the value of the speed of sound along the intersection is increased \eqref{speeds_physical}, going above the conformal value in $3+1$ dimensions but still below the conformal value in a $(2+1)$-dimensional theory (that one finds in isolated D5-branes \cite{Itsios:2016ffv}), at least in the approximation of small flavor density that we use. Since the equation of state remains conformal \eqref{eq:eos}, the transverse speed of sound goes below the conformal value in \eqref{speeds_physical}. In addition to these thermodynamic insights, we also infer from the behavior of mutual information that flavor degrees of freedom are correlated in the transverse direction, thus not being completely independent at separations close or below the temperature scale. One could take this as a signal of increasing inter-layer mobility and decreasing intra-layer mobility, which is interesting from the point of view of condensed matter physics, where multilayered materials with strongly correlated electrons constrained to move in $2+1$ dimensions show strange metallic behavior and high-$T_\mt{c}$ superconductivity.

Let us now comment on some possible extensions of our work. It would be certainly desirable to obtain a solution valid beyond the perturbative expansions performed here. Most likely this would require a numerical analysis of the equations of motion, such us the ones written in the system (\ref{eq:RedBGSystem}) for the case of vanishing chemical potential. The black hole solution studied of this system found here is not unique. Indeed, in \cite{Conde:2016hbg} an exact solution for zero chemical potential was found. This black hole geometry has a Lifshifz-like scaling symmetry,  is non-analytic in $\nf$ and, as a consequence, one cannot take the flavorless limit $\nf\to 0$. It would be interesting to construct new backgrounds interpolating between that geometry and the asymptotically AdS metric found here. More generally, in order to classify all possible solutions one should study different boundary conditions to be imposed both at the UV and at the IR. 

The solutions at nonzero baryon density display the phenomenon of having two types of charged degrees of freedom, those that are fractionalized, corresponding to those sitting behind the horizon, and coherent ones which correspond to those that are hovering above the horizon. This situation has been encountered elsewhere, in electron cloud solutions \cite{Hartnoll:2010ik,Puletti:2010de} to various probe brane configurations \cite{Bergman:2010gm,Jokela:2011eb,Bea:2014yda} with entrancing connections to quantum Hall physics. Constructing observables such as the butterfly velocity or the generalized charged entanglement entropy discussed in \cite{DiNunno:2021eyf} would yield probes for distinguishing how the degrees of freedom flow under external fields, complementarily to the obvious magnetotransport coefficients readily available. A really captivating scenario would be to establish the entanglement first law using the latter probe, addressing the long-standing open problem in the field.

The setup considered here can be generalized in several other directions. One possibility could be adding four-dimensional flavors by introducing D7-branes, which as a start could be treated as probes as in \cite{Gran:2019djz}. This analysis would shed more light to the anisotropic physics of the holographic multilayer theories. Another possible generalization could be extending the analysis to different dimensions both for the ambient theory or for the defect. For example, viable chassis for this exercise include the D2-D6 backreacted geometry of \cite{Faedo:2015ula} or the D3-D3' backreacted geometry constructed in \cite{Jokela:2021evo}.

\addcontentsline{toc}{section}{Acknowledgments}
\paragraph{Acknowledgments}
We thank Francesco Bigazzi and Aldo Cotrone for useful comments. This work has received financial support from Xunta de Galicia (Centro singular de investigaci\'on de Galicia accreditation 2019-2022), by European Union ERDF, and by the ``Mar\'\i a de Maeztu" Units of Excellence program MDM-2016-0692 and the Spanish Research State Agency. The research of A.~G. and A.~V.~R. has been funded by the Spanish grant PID2020-114157GB-100. C.~H. has been partially supported by the AEI through the Spanish grant PGC2018-096894-B-100 and by FICYT through the Asturian grant SV-PA-21-AYUD/2021/52177. N.~J. and J.~M.~P. have been supported in part by the Academy of Finland grant no. 1322307.

\appendix

\vskip 3cm
\renewcommand{\theequation}{\rm{A}.\arabic{equation}}
\setcounter{equation}{0}

\section{Equations of motion}\label{appendix_A}

In this appendix we write down in detail the equations of motion for the total action (\ref{total_action}), as well as the resulting equations for the different functions of our ansatz. Let us begin by writing the equations satisfied by the forms, which are \cite{Benini:2007kg}:
\bea
\dd(e^{2\phi}\star F_1)+e^{\phi}H_3 \wedge \star F_3-\frac{1}{3} \kappa^2_{10} T_5 \,\Xi \wedge \mathcal{F} \wedge \mathcal{F} \wedge \mathcal{F} & = & 0\rc\rc
\dd(e^\phi \star F_3)-F_5 \wedge H_3+\kappa_{10}^2\,T_5 \, \Xi \wedge \mathcal{F} \wedge \mathcal{F} & = & 0\rc\rc
\dd \star F_5-H_3\wedge F_3-2\kappa_{10}^2 T_5\,  \Xi \wedge \mathcal{F} & = & 0\rc\rc
\dd(e^{-\phi} \star H_3)+F_5 \wedge F_3+e^{\phi} \star F_3 \wedge F_1-2\kappa^2_{10}\,
\delta_{\cal F}S_{\text{branes}} & = & 0 \ .
\label{eoms_forms}
\eea
Notice that the last term in the equations  (\ref{eoms_forms}) is due to the brane sources. The equations of motion for the RR forms contain the smearing form $\Xi$ since they are coupled to the D5-brane through the WZ term (\ref{WZ_action}). 
The last term in the equation for $H_3$ in (\ref{eoms_forms}) is an eight-form which arises from the smeared DBI term of the brane action. Its expression is given below in (\ref{delta_F_S}). The RR forms have to satisfy the Bianchi identities, namely:
\bea
\dd F_1 & = & 0 \rc\rc
\dd F_3-H_3 \wedge F_1-2 \kappa^2_{10} T_5\, \Xi & = & 0 \rc\rc
\dd F_5-H_3 \wedge F_3-2\kappa^2_{10}T_5 \,\Xi \wedge \mathcal{F} & = & 0 \ ,
\label{Bianchi}
\eea
where the NSNS three-form is such that $\dd H_3=0$. The equation for the dilaton is given by:
\bea
\Box \phi=e^{2\phi}F_1^2+\frac{1}{2\cdot 3!}e^{\phi}F_3^2-\frac{1}{2\cdot 3!}e^{-\phi}H_3^2-\frac{2\kappa_{10}^2}{\sqrt{-g}}\frac{\delta S_{\text{branes}}}{\delta \phi}\,\,.
\label{dilaton_eq}
\eea
In addition, we have to impose Einstein equations, which are also written below in (\ref{Einstein_eq}), but we first need to introduce our ansatz for the metric. 

We will consider solutions for which
\beq
F_1\,=\,0\,,
\qquad\qquad
H_3\,=\,0\,\,.
\eeq
It is convenient to write the equations of motion in terms of a new radial coordinate $\sigma$, related to the $\rho$ coordinate used in the main text as:
\bea\label{eq:stor}
 e^{4\rh^4\s}=1-{\rh^4\over\r^4}\,\,.
\eea
In terms of $\sigma$, our metric ansatz  (\ref{metric_ansatz }) takes the form:
\bear
\dd s ^2_{10} & =& h^{-\frac{1}{2}}\,\big[ -b \,(\dd t)^2+(\dd x^1)^2+(\dd x^2)^2+\alpha^2(\dd x^3)^2 \,\big]\rc\rc
&&
+h^{\frac{1}{2}}\, \big[\, b\, \alpha^2 S^8F^2  \dd\sigma^2+S^2 \dd s ^2_{\rm{KE}}+F^2 (\dd \tau+{\cal A})^2\, \big]\,\,,
\eear
where the radial functions $h$, $b$, $\alpha$, $F$, and $S$ should be understood as functions of $\sigma$ through the change of variables (\ref{eq:stor}). The ansatz for the RR three-form $F_3$ has been written in (\ref{F_3_ansatz }), whereas $F_5$ can be readily obtained from (\ref{F5_ansatz })-(\ref{C_4_cp}), and is given by:
 \bear
F_5 & = & \Qc\frac{b\,\alpha^2}{h^2}\,(1+\star)\dd^4x \wedge \dd\sigma+(1+\star)\partial_{\sigma}J(\sigma)\dd\sigma \wedge \dd x^1\wedge \dd x^2 \wedge \mathrm{Re}(\hat{\Omega}_2)\rc\rc
& & -3J(\sigma)(1+\star)\dd x^1\wedge \dd x^2\wedge \mathrm{Im}(\hat{\Omega}_2)\wedge (\dd \tau +{\cal A})\ .
\eear
The Bianchi identity for $F_5$ gives the following differential equation for $J$:
\bear
\partial_{\sigma}^2J-4{\partial_{\sigma}S\over S}\,\partial_{\sigma}J-9\,\a^2b\,S^8J+3\,\Qf\, S^4 \,\partial_{\sigma}A_t=0\,\,,
\label{eom_J}
\eear
where we have used the expression (\ref{smearnig_form}) of the smearing form $\Xi$ and an ansatz  for the worldvolume gauge field strength $\mathcal{F}$, which in the $\sigma$ coordinate takes the form:
\beq
\mathcal{F}=\dd A=\partial_{\sigma}A_t(\sigma) \dd\sigma \wedge \dd t\,\,.
\eeq

Let us now write down in detail  the equation (\ref{dilaton_eq}) for the dilaton. To compute the last term in this equation we need to know the expression for the smeared DBI action. With this purpose, we define the modulus $|\Xi|$ of the smearing form as:
\bea
|\Xi|=\sqrt{\frac{1}{4!}\Xi_{\alpha_1 \alpha_2 \alpha_3 \alpha_4} \Xi_{\beta_1 \beta_2 \beta_3 \beta_4}g^{\alpha_1\beta_1}g^{\alpha_2\beta_2}g^{\alpha_3\beta_3}g^{\alpha_4\beta_4}}\,\,.
\eea
Using (\ref{smearnig_form}), we get:
\beq
 |\Xi|\,=\,{3\,\Qf\over \sqrt{2}\,\kappa_{10}^2\,T_5}\,{1\over \sqrt{h} \,\alpha\,F\,S^2}\,\,.
\label{Xi_modulus}
\eeq
Let us also introduce a shorthand
\bea
 \Lambda=\sqrt{1+e^{-\phi}\mathcal{F}_{\mu \nu}\mathcal{F}^{\mu \nu}}\,=\,
\sqrt{1-\frac{e^{-\phi} (\partial_{\sigma}A_t)^2}{\alpha^2 b^2 F^2 S^8}}\,\,.
\label{Lambda}
\eea
Then, the smeared DBI action takes the form:
\bea
S_{\rm{DBI}}^{smeared}\,=\,-T_5 \int \dd^{10}x \sqrt{-g}\,e^{\frac{\phi}{2}}\,\sqrt{2}|\Xi| \,\Lambda\,\,,
\eea
which can be expressed explicitly by plugging in the expressions in (\ref{Xi_modulus}) and (\ref{Lambda}). The resulting equation for $\phi$ is:
\bea
\partial^2_{\sigma}\phi -\alpha \,b\,F\,S^4 \left(3\frac{\Qf\, \alpha \, b\,F\,e^{\frac{\phi}{2}} \,S^6}{\sqrt{\alpha ^2 b^2 F^2 S^8-e^{-\phi}\,  (\partial_{\sigma}A_t)^2}}+\frac{F\, e^{\phi}}{2\, \alpha} \,\left(F_{123}^2 h^2 S^4+2\, \Qf^2\right)\right)=0\,\,.
\label{eom_phi}
\eea

Notice that, even though $H_3=0$, its equation of motion is non-trivial. In order to write this equation in detail we need to determine the eight-form $\delta_{\cal F}S_{\text{branes}}$, which arises when the DBI term of the brane action is varied with respect to the NSNS two-form potential. This one reads
\beq
\delta_{\cal F}S_{\text{branes}}\,=\,{1\over \kappa_{10}^2}
{3\,\Qf\,e^{-{\phi\over 2}}\,S^2\,\partial_{\sigma}A_t\over
\sqrt{\alpha^2 b^2 F^2 S^8-e^{-\phi}\, (\partial_{\sigma}A_t)^2}}\,
\dd x^1\wedge \dd x^2\wedge \dd x^3\wedge e^1\wedge e^2\wedge e^3\wedge e^4\wedge e^5
\label{delta_F_S}\,\,.
\eeq
Using this last expression, the equation for $H_3$ becomes:
\bea\label{eqAt1}
6\, \Qf\,e^{-\f/2}\,S^2 \partial_{\sigma}A_t+\,(\Qc F_{123}+6\Qf J)\sqrt{\alpha^2 b^2 F^2 S^8-e^{-\phi} (\partial_{\sigma}A_t)^2}=0\,\,.
\eea

Let us now write down the Einstein equations that follow from the action (\ref{total_action}). When $F_1=H_3=0$, these equations read
\bea
&&R_{{\mu} {\nu}}-\frac{1}{2}g_{{\mu}{\nu}}R-\frac{1}{2}(\partial_{{\mu}} \phi \partial_{{\nu}} \phi-\frac{1}{2}g_{{\mu} {\nu}} (\partial \phi)^2)\
-\frac{1}{2 \cdot 3!} e^\phi (3F_{3\,{\mu} {\rho} {\sigma}} F_{3\,{\nu}}^{\,\, \,\,{\rho} {\sigma}}-\frac{1}{2}g_{{\mu} {\nu}}F_3^2) \rc\rc
&&\qquad\qquad\qquad
 \,-\frac{1}{4 \cdot 5!}(5F_{5{\mu} {\rho} {\sigma} {\tau} {\gamma}}F_{5{\nu}}^{\ \,\,\,\,{\rho} {\sigma} {\tau} {\gamma}}-\frac{1}{2}g_{{\mu} {\nu}}F_5^2)\,=\,2\kappa^2_{10} T_{{\mu} {\nu}}\,\,,
 \label{Einstein_eq}
\eea
where $T_{{\mu} {\nu}}$ is the energy-momentum tensor for the distribution of D5-branes. Using the methods of \cite{Nunez:2010sf}, $T_{\mu\nu}$ can be easily obtained. It is useful to write it in flat components with respect to the following basis of vielbein one-forms:
\bea
&&E^{x^0}=h^{-\frac{1}{4}} b^\frac{1}{2}\dd t\,\,,\qquad\qquad
E^{x^{1,2}}=h^{-\frac{1}{4}}\dd x^{1,2}\,,\qquad\qquad
E^{x^3}=h^{-\frac{1}{4}} \alpha \dd x^3\,\,,\rc\rc
&&E^\sigma=h^{\frac{1}{4}}\alpha S^4Fb^\frac{1}{2}\dd\sigma
\,,\qquad\qquad
E^{i}=h^{\frac{1}{4}}S e^i, \ \ i=\{1,2,3,4\} \,,\rc\rc
&&E^{5}=h^{\frac{1}{4}} F e^5\,\,.
\eea
Written in flat components $T_{{\mu} {\nu}}$, it is given by:
\bear
&&T_{\hat{\mu} \hat{\nu}}=T_5 e^{\frac{\phi}{2}} \sqrt{2}\bigg[-\frac{1}{2}
|\Xi|\Lambda g_{\hat{\mu} \hat{\nu}}+\frac{|\Xi|}{2\Lambda} e^{-\phi}
(\partial_{\sigma}A_t)^2\,(E^{t}_{\ \hat{\mu}}E^{t}_{\ \hat{\nu}}g^{\s\s}+E^{\s}_{\ \hat{\mu}}E^{\s}_{\ \hat{\nu}}g^{tt} )\rc\rc
&&\qquad\qquad\qquad\qquad\qquad\qquad
+\frac{\Lambda}{12 |\Xi|} \Xi_{\hat{\mu} \hat{\alpha_1}\hat{\alpha_2}\hat{\alpha_3}}\Xi_{\hat{\nu}}^{\ \hat{\alpha_1} \hat{\alpha_2}\hat{\alpha_3}} \bigg]\,\,.
\label{T_mu_nu}
\eear
One can readily verify that the only non-vanishing components of $T_{\hat{\mu} \hat{\nu}}$ are:
\bea
T_{\hat x^0\hat x^0} & = & -T_{\hat\s\hat\s}\,=\,{3 \Qf\over 2 \kappa_{10}^2 T_5}\,
\,\frac{ e^{\frac{\phi }{2}}}{\Lambda \,\alpha \, F \sqrt{h} \,S^2}\rc\rc
T_{\hat x^1\hat x^1} & = & T_{\hat x^2\hat x^2}\,=\,-{3 \Qf\over 2 \kappa_{10}^2 T_5}\,\frac{\Lambda e^{\frac{\phi }{2}}}{\alpha \, F \,\sqrt{h} \,S^2}\rc\rc
T_{\hat 5\hat 5}=T_{\hat 6\hat 6} & = & T_{\hat 7\hat 7}=T_{\hat 8\hat 8}\,=\,-{3 \Qf\over 4 \kappa_{10}^2 T_5}\,\,\frac{\Lambda e^{\frac{\phi }{2}}} {\alpha\,  F\, \sqrt{h} \,S^2}\ .
\eea
In order to massage the equations derived from (\ref{Einstein_eq}) in more compact form, let us first define the quantities $X$, $Y$, and $Z$ as follows
\beq
X=e^{-\phi } S^2 (\partial_{\sigma}A_t)^2\,, \qquad\qquad
Y=\sqrt{\alpha ^2 b^2 F^2 e^{\phi } S^8-(\partial_{\sigma}A_t)^2}\,,\qquad 
Z=F_{123}^2 b F^2 h^2 e^{\phi } S^8\,\,.
\eeq
Then, the Einstein equations (\ref{Einstein_eq}) are equivalent to the following system of second order differential equations:
\bea	\label{eq:OriginalSystem}
\partial^2_\sigma \log F & = & \frac{1}{2} \Qf^2 b F^2 e^{\phi } S^4-\frac{3 \Qf \alpha ^2 b^2 F^2 e^{\phi } S^{10}}{2 Y}+4 \alpha ^2 b F^4 S^4-9 \alpha ^2 b J^2 S^4-\frac{Z}{4}\rc\rc
\partial^2_\sigma \log S & = & 6 \alpha ^2 b F^2 S^6-2 \alpha ^2 b F^4 S^4-\frac{9}{2} \alpha ^2 b J^2 S^4-\frac{(\partial_{\sigma}J)^2}{2 S^4}-\frac{3 \Qf X e^{\phi }}{2 Y}-\frac{Z}{4}\rc\rc
\partial^2_\sigma \log h & = & -b S^4 \left(-\frac{2 (\partial_{\sigma}J)^2}{b S^8}+\Qf^2 F^2 e^{\phi }-18 \alpha ^2 J^2\right)-\frac{3 \Qf e^{\phi } S^2 \left(\alpha ^2 b^2 F^2 S^8-\frac{2 X}{S^2}\right)}{Y}-\frac{\Qc^2 \alpha ^2\,b}{h^2}+\frac{3 Z}{2}\rc\rc
\partial^2_\sigma \log b & = & 18 \alpha ^2 b J^2 S^4+\frac{2 (\partial_{\sigma}J)^2}{S^4}+\frac{6 \Qf X e^{\phi }}{Y}+Z\rc\rc
\partial^2_\sigma \log \alpha & = & -\Qf^2 b F^2 e^{\phi } S^4+\frac{(\partial_{\sigma}J)^2}{S^4}-3 \Qf Y S^2+9 \alpha ^2 b J^2 S^4\,\,,
\eea
together with the first-order constraint:
\bea\label{eq:OriginalConstraint}
&&-\partial_\sigma \log\alpha\,\partial_\sigma \log b-\partial_\sigma \log F\,\partial_\sigma \log b-\frac{1}{2} \partial_\sigma \log h\,\partial_\sigma \log b-4 \partial_\sigma \log S\,\partial_\sigma \log b-\partial_\sigma \log \alpha\,\partial_\sigma \log h\rc
&&-2 \partial_\sigma \log F\,\partial_\sigma \log \alpha-8 \partial_\sigma \log F\,\partial_\sigma \log S-8 \partial_\sigma \log \alpha\,\partial_\sigma \log S-12 \left(\partial_\sigma \log S\right)^2+\frac{1}{2} \left(\partial_\sigma \log h\right)^2\rc
&&-Q_{f}^2 b F^2 e^{\phi } S^4-\frac{2\ 3 Q_{f} \alpha ^2 b^2 F^2 e^{\phi } S^{10}}{Y}+24 \alpha ^2 b F^2 S^6-4 \alpha ^2 b F^4 S^4-\frac{Q_{c}^2 \alpha ^2 b}{2 h^2}\rc
&&-9 \alpha ^2 b J^2 S^4+\frac{(\partial_{\sigma}J)^2}{S^4}+\frac{1}{2} (\partial_{\sigma}\phi)^2-\frac{Z}{2}=0\,\,.
\eea

The equations of motion written above contain $\partial_{\sigma}A_t$. As standard, though, it can be eliminated algebraically by using (\ref{eqAt1}) which casts $\partial_{\sigma}A_t$ in terms of  other functions. This exercise yields
\beq
\partial_{\sigma}A_t\,=\,-{(\Qc\,F_{123}\,+\,6\,\Qf\,J)\,\alpha\,b\,F\,S^4\,e^{{\phi\over 2}}\over
\Big[36\,\Qf^2\,S^4\,+\,(\Qc\,F_{123}\,+\,6 \Qf\,J)^2\Big]^{{1\over 2}}}\,\,.
\eeq

Since all the functions depend only on the coordinate $\s$ we can describe the system in terms of a one-dimensional effective action of the form:
\bea\label{eq:1dAction}
S^{\rm{eff}}=\int \dd\s\,L_{1D}\,=\,
\int \dd\s\,\,\left(L_g+L_{F_3}+L_{F_5}+L_{\rm{DBI}}+L_{\rm{WZ}}\right)\,\,,
\eea
where 
\bea
L_g & = & 24 \alpha ^2 b F^2 S^6-4 \alpha ^2 b F^4 S^4-\frac{1}{2} (\partial_{\sigma} \phi)^2-\frac{1}{2} \left(\partial_\s \log h \right)^2+12 \left(\partial_\s \log S \right)^2\rc\rc
&&\qquad+\partial_\s \log b  \left(\partial_\s \log \a +\partial_\s \log F +\frac{1}{2} \partial_\s \log h+4 \partial_\s \log S \right)\rc\rc
&&\qquad+\partial_\s \log \a \left(2 \partial_\s \log F +\partial_\s \log h+8 \partial_\s \log S \right)+8 \partial_\s \log F \partial_\s \log S \rc\rc
 L_{F_3} & = & -\frac{1}{2} b\, F^2 e^{\phi } S^4 \left(F_{123}^2 h^2 S^4+2 \Qf^2\right)\rc\rc
 L_{F_5} & = & -\frac{1}{4} \left(\frac{\alpha^2 b\, \left(\Qc^2-18 h^2 J^2 S^4\right)}{h^2}+\frac{2 (\partial_{\sigma}J)^2}{S^4}\right)\rc\rc
 L_{\rm{DBI}} & = & -6 \Qf \alpha\, b\,F\,e^{\frac{\phi }{2}} S^6\sqrt{1-\frac{e^{-\phi} (\partial_{\sigma}A_t)^2}{\alpha^2 b^2 F^2 S^8}}\rc\rc
 L_{\rm{WZ}} & = & 6 \,\Qf\, J\, \partial_{\sigma}A_t\,\,.
\eea

The one-dimensional effective action (\ref{eq:1dAction}) can be used to obtain the equation of motion of the worldvolume gauge field $A_t$. Since $A_t$ only enters $S^{\rm{eff}}$ through its derivative $\partial_{\sigma}\,A_t$, it is a cyclic variable and hence its equation of motion can be written as a conservation law, namely:
\beq
{\delta\over \delta (\partial_{\sigma}\, A_t)}\,\Big(L_{\rm{DBI}}+L_{\rm{WZ}}\Big)\,=\,C\,\,,
\label{worldvolume_eom}
\eeq
where $C$ is a constant. Remarkably, this equation of motion is consistent with the bulk equations for the backreacted ansatz. Indeed, one can directly show that (\ref{worldvolume_eom}) is equivalent to (\ref{eqAt1}) if we identify  $F_{123}$ and the constant $C$ as:
\beq
C\,=\,-\Qc\,F_{123}\,\,.
\label{C_F123}
\eeq
We regard this equivalence as a non-trivial consistency condition of our backreacted ansatz. 

The constant $F_{123}$ is related to the quark  charge density $\nq$ as we will demonstrate next. According to the holographic dictionary, the latter is given by the boundary value of the displacement field $D(\sigma)$, \ie\
\beq
\nq = D(\sigma=0) \equiv
{{\rm Vol}({\cal M}_5)\over 2\kappa^2_{10}}\,2\pi\,\alpha'\,
{\partial L_{\rm{DBI}}\over \partial(\partial_{\sigma}\,A_t)}\Bigg|_{\s\to 0}\,\,.
\eeq
Using the worldvolume equation (\ref{worldvolume_eom}) we can relate $D(\sigma)$ to the constant $C$ and the function $J(\sigma)$, since:
\beq
{\partial L_{\rm{DBI}}\over \partial(\partial_{\sigma}\,A_t)}\,=\,C\,-\,6\, \Qf\,J(\sigma)\,=\,
-\Qc\,F_{123}\,-\,6\, \Qf\,J(\sigma)\,\,.
\eeq
Therefore, taking into account that $\Qc\,{\rm Vol}({\cal M}_5)\,=\,(2\pi)^4\,g_\mt{s}\,\alpha'^2\,\Nc$ and that
$2\kappa_{10}^2\,=\,(2\pi)^7\,g_\mt{s}^2\,\alpha'^4$, we can write the displacement field as:
\beq
D(\sigma)\,=\,-{\Nc\over 4\pi^2\,g_\mt{s}\,\alpha'}\,F_{123}\,+\,{6\pi\alpha' \over \kappa_{10}^2}\,\Qf\,{\rm Vol}({\cal M}_5)
\,J(\sigma)\,\,.
\label{displacement_J}
\eeq
In our solutions the regularity condition of $A_t$ at the UV boundary requires the vanishing of the function $J(\sigma)$ as $\sigma\to 0$. Therefore, the second term in (\ref{displacement_J}) does not contribute to the quark density $\nq$, which thereby eventually leads to the relation
\beq
\nq\,=\,-{\Nc\over 4\pi^2\,g_\mt{s}\,\alpha'}\,F_{123}\,\,.
\label{n_q_F123}
\eeq

\vskip 2cm
\renewcommand{\theequation}{\rm{B}.\arabic{equation}}
\setcounter{equation}{0}

\section{Solution at vanishing chemical potential }\label{appendix_B}

In this appendix we integrate the equations of motion found in appendix \ref{appendix_A} in the case in which the chemical potential vanishes.  This particular case corresponds to taking $A_t=F_{123}=J=0$ in  eqs. (\ref{eom_J}), (\ref{eom_phi}), (\ref{eqAt1}), (\ref{eq:OriginalSystem}), and (\ref{eq:OriginalConstraint}). 
The equation for the blackening factor $b$  in (\ref{eq:OriginalSystem}) becomes  simply $\partial^2_{\sigma}\,\log b\,=\,0$, which can be integrated as:
\beq
b\,=\,e^{4\r_{\footnotesize{\textrm{h}}}^4\,\sigma}\ ,
\label{blackening_sigma}
\eeq
where we have conveniently fixed the integration constants. In terms of the radial coordinate $\rho$, this blackening factor $b$ becomes the one written in (\ref{blackening}). 

By comparing the equations satisfied by the warp factor $h$ and the dilaton $\phi$, one readily concludes that they can be related as:
\beq
h\,=\,{\Qc\over 4\rh^4}\,\Big(1\,-\,e^{4\r_{\footnotesize{\textrm{h}}}^4\,\sigma}\Big)\,e^{-\phi}\,\,,
\eeq
which, in terms of the coordinate $\rho$, is equivalent to (\ref{h_rho_phi}). Therefore, the dilaton determines the warp factor. Let us now write the reduced system of equations for $F$, $S$, and $\phi$. It is convenient to define the new functions 
$\tilde F$ and $\tilde S$ as:
\beq
\tilde F\,=\,e^{-{3\over 2}\,\phi}\,F\,\,,
\qquad\qquad
\tilde S\,=\,e^{-{3\over 2}\,\phi}\,S\,\,,
\label{tilde_F_S}
\eeq
which, together with the dilaton $\phi$,  satisfy the following system of three coupled ordinary differential equations:
\bea\label{eq:RedBGSystem}
\partial^2_\sigma \log \tilde{F} & = & \tilde{F}\tilde{S}^4 e^{4 \r_{\footnotesize{\textrm{h}}}^4 \sigma +10 \phi } \left(4 \tilde{F}^3-6 \Qf \tilde{S}^2-\Qf^2 \tilde{F}\right) \rc\rc
\partial^2_\sigma \log \tilde{S} & = & \tilde{F}\tilde{S}^4 e^{4 \r_{\footnotesize{\textrm{h}}}^4 \sigma +10 \phi } \left(6 \tilde{F}\tilde{S}^2-2 \tilde{F}^3-\frac{9}{2} \Qf \tilde{S}^2-\frac{3}{2} \Qf^2 \tilde{F}\right)\rc\rc
\partial^2_\sigma\phi & = & \Qf \tilde{F}\tilde{S}^4 e^{4 \r_{\footnotesize{\textrm{h}}}^4 \sigma +10 \phi } \left(\Qf \tilde{F}+3 \tilde{S}^2\right)\,\,.
\eea
We will solve the system (\ref{eq:RedBGSystem}) at first-order in the parameter $\epsilon$ defined in 
(\ref{epsilon_def}). Let us represent $\phi$, $\tilde{F}$, and $\tilde{S}$ in terms of the reduced radial coordinate $\tilde\rho$ introduced in (\ref{tilde_rho_def}):
\bea
\f & = & \eps\,\f_{1}(\rt)\rc\rc
\tilde{F} & = & \rh\, \rt\left(1+\eps\,\tilde{F}_1(\rt)\right)\rc\rc
\tilde{S} & = & \rh\, \rt\left(1+\eps\,\tilde{S}_1(\rt)\right)\ .
\label{phi_F_S_expan_def}
\eea
In (\ref{phi_F_S_expan_def})  the zeroth order term corresponds to the $AdS_5\times {\cal M}_5$ geometry with a constant dilaton (set to zero). The first-order contribution $\f_{1}(\rt)$ to the dilaton satisfies a second-order differential equation which can be obtained by plugging the expansions (\ref{phi_F_S_expan_def}) into the system (\ref{eq:RedBGSystem}) and then keeping first-order terms. We find
\beq\label{eq:OrdEpsSys}
\f_1''+\frac{\left(5 \rt^4-1\right) \f_1'}{\rt \left(\rt^4-1\right)}-\frac{15 \rt}{\rt^4-1}=0\ ,
\eeq
where the primes denote derivatives with respect to $\tilde\rho$. The general solution of (\ref{eq:OrdEpsSys}) is
\be
\f_1=c_2+\frac{1}{4} \left(c_1\log \left(\frac{\rt^4}{\rt^4-1}\right)+5 \log \left(\frac{\rt-1}{\rt+1}\right)+10 \arctan(\rt)\right)\ , \label{phi_1_sol}
\ee
where $c_1$ and $c_2$ are integration constants. The equations for $\tilde{F}_1$ and $\tilde{F}_2$ derived from the system (\ref{eq:RedBGSystem}) are coupled. They can be decoupled by introducing new functions $f_1$ and $f_2$ as follows
\be
\tilde{F}_1(\tilde\rho)={1\over\sqrt{2}}f_1(\tilde\rho)-{4\over\sqrt{17}}f_2(\tilde\rho)\,\,,
\qquad\qquad \tilde{S}_1(\tilde\rho)={1\over\sqrt{2}}f_1(\tilde\rho)+{1\over\sqrt{17}}f_2(\tilde\rho)\ .
\label{decoupling_F1_S_1}
\ee
The decoupled equations for $f_1$ and $f_2$ are thus
\bea
f_1''+\frac{\left(5 \rt^4-1\right) f_1'}{\rt \left(\rt^4-1\right)}-\frac{32 \rt^2\, f_1}{\rt^4-1}+\frac{24 \sqrt{2} \rt}{\rt^4-1}-\frac{40 \sqrt{2} \rt^2\f_1}{\rt^4-1} & = & 0\rc
f_2''+\frac{\left(5 \rt^4-1\right) f_2'}{\rt \left(\rt^4-1\right)}-\frac{12 \rt^2f_2}{\rt^4-1}-\frac{3 \sqrt{17}\rt}{2 \left(\rt^4-1\right)} & = & 0
\,\,.
\label{eqs_f1_f2}
\eea
The equation for $f_1$  contains $\phi_1$ as a source and can be integrated as:
\bea
 f_1 & = & \frac{1}{\sqrt{2}}\left(\rt^3+\frac{1}{4} \left(\rt^4-13\right) \left(\log \left(\frac{\rt-1}{\rt+1}\right)+2 \arctan(\rt)\right)\right)\rc\rc
& &-\frac{c_3}{2}-2 c_4+\rt^4 \left(\frac{5 c_1}{2 \sqrt{2}}-\frac{5 c_2}{\sqrt{2}}+c_3\right)+\log \left(\frac{\rt^4-1}{\rt^4}\right) \left(\frac{5 c_1 \rt^4}{4 \sqrt{2}}-2 c_4 \rt^4+c_4\right)\ ,
\eea
where $c_3$ and $c_4$ are new constants of integration. In order to integrate the equation satisfied by $f_2$, 
let us define functions $P(z)$ and $Q(z)$ as follows
\bear
P(z) & = & F\Big(-{1\over 2}\,,\,{3\over 2}\,; 1\,;\,1-z^4\Big)\rc\rc
Q(z) & = & (2z^4-1)^{-{3\over 2}}\,
F\Bigg({5\over 4}\,,\,{3\over 4}\,; 2\,;\,{1\over (2z^4-1)^2}\Bigg)\ .
\label{P_Q_def}
\eear
The functions $F(\ldots)={}_2F_1(a,b;c;z)$ here are standard hypergeometric functions, but we have dropped the subscripts to avoid extra clutter.
The functions $P(\tilde \rho)$ and $Q(\tilde \rho)$ are two independent solutions of the homogeneous equation  for $f_2(\tilde \rho)$ in (\ref{eqs_f1_f2}). To find a solution of the inhomogeneous equation we apply the Wronskian method. The general solution for $f_2(\tilde \rho)$ is
\beq
f_2(\tilde \rho)\,=\,c_5\,P(\tilde \rho)\,+\,c_6\,Q(\tilde \rho)\,+\,I_1(\tilde \rho)\,P(\tilde \rho)\,+\,I_2(\tilde \rho)\,Q(\tilde \rho)\,\,.
\label{f_2_general_sol}
\eeq
In (\ref{f_2_general_sol}) $c_5$ and $c_6$ are constants of integration and $I_1(\tilde\rho)$ and $I_2(\tilde\rho)$ are the following integrals
\beq
I_1(\tilde\rho)\,=\,\int_{\tilde\rho}^{\infty}\,
{Q(z) \,g( z)\over W (P,Q)}\,d z\,\,,
\qquad\qquad
I_2(\tilde\rho)\,=\,\int_{1}^{\tilde\rho}\,
{P(z) \,g(z)\over W (P,Q)}\,\dd z\,\,,
\eeq
where $g(z)$ is the inhomogeneous term in (\ref{eqs_f1_f2}):
\beq
g(z)\,=\,{3\sqrt{17}\,z\over 2(z^4-1)}\,\,,
\eeq
and $ W (P,Q)$ is the Wronskian of the functions $P(z)$ and $Q(z)$, given by
\beq
W (P,Q)\,=\,P(z)\,Q'(z)\,-\,Q(z)\,P'(z)\,=\,{8\,\sqrt{2}\over \pi\,z\,(1-z^4)}\,\,.
\eeq
We have found a solution for $\phi_1$, $F_1$, and $S_1$ which depends on six integration constants $c_1\,,\,\cdots\,,c_6$. These constants can be determined by requiring the fulfilment of the constraint (\ref{eq:OriginalConstraint}), as well as  regularity both at the horizon $\tilde\rho=1$ and at the boundary $\tilde\rho\to\infty$. Doing this we get
\beq\label{eq:FixedConst}
c_1=5,\qquad c_2=-{5\p\over 4},\qquad c_3=-\frac{25+13 \pi }{2 \sqrt{2}},\qquad c_4=\frac{13}{4 \sqrt{2}},
\qquad c_5= c_6=0\,\,.
\eeq
Let us now write the final form of our ${\cal O}(\epsilon)$ solution. The dilaton $\phi(\tilde\rho)$, the warp factor $h (\tilde\rho)$, and the functions $\alpha(\tilde\rho)$ and $G(\tilde\rho)$ are written in (\ref{eq:OrderEpsMetricFuncSubsConst}) in terms of the function $\Omega(\tilde\rho)$ defined in (\ref{Omega_def}). The internal deformation functions $F(\tilde\rho)$ and $S(\tilde\rho)$ are written in (\ref{F_S_BHsolution}) in terms of the first-order functions $F_1(\tilde \rho)$ and $S_1(\tilde \rho)$. 
In order to present compact expressions for these functions, let us define the integrals 
${\cal I}_{P}(x)$ and ${\cal I}_{Q(x)}$ as
\beq
{\cal I}_{P}^{ (1)}(x)\,=\,\int_{1}^{x} z^2 P(z)\,\dd z\,\,,
\qquad\qquad
{\cal I}_{Q}^{(1)}(x)\,=\,\int_{x}^{\infty} z^2 Q(z)\,\dd z\,\,.
\eeq
This procedure finally yields
\bear
F_1(\tilde \rho) & = & {3\sqrt{2}\,\pi\over 8}\,\Big(P(\tilde\rho)\,{\cal I}_{Q}^{(1)}(\tilde\rho)+
Q(\tilde\rho)\,{\cal I}_{P}^{(1)}(\tilde\rho)\Big)+{1\over 2}\,\tilde\rho^3-{1\over 8}\,+\,{1\over 10}\,
(\tilde\rho^4+2)\,\Omega(\tilde \rho)\rc\rc
S_1(\tilde \rho) & = & -{3\sqrt{2}\,\pi\over 32}\,\Big(P(\tilde\rho)\,{\cal I}_{Q}^{(1)}(\tilde\rho)+
Q(\tilde\rho)\,{\cal I}_{P}^{ (1)}(\tilde\rho)\Big)+{1\over 2}\,\tilde\rho^3-{1\over 8}\,+\,{1\over 10}\,
(\tilde\rho^4+2)\,\Omega(\tilde \rho)\ .
\label{F_1_S_1_explicit}
\eear

\vskip 2cm
\renewcommand{\theequation}{\rm{C}.\arabic{equation}}
\setcounter{equation}{0}

\section{Solution with non-zero chemical potential}\label{appendix_C}

Let us now add non-zero chemical potential to the black hole solution of appendix \ref{appendix_B}. We will first expand the dilaton in powers of $\epsilon$ and $\delta$ as:
\beq
\f=\eps\,\f_{1}(\rt)\,+\,\epsilon\,\delta^2\,\f_{2}(\rt)\,\,,
\eeq
where $\phi_1(\rt)$ is the solution found in appendix \ref{appendix_B},in (\ref{phi_1_sol}). It turns out that the equation for $\phi_2$ is completely decoupled and given by
\beq
\f_{2}''+\frac{\left(5 \rt^4-1\right) }{\rt \left(\rt^4-1\right)}\f_{2}'-\frac{\Qc^2}{120\, \rh^6 \,\rt^3 \left(\rt^4-1\right)}=0\,\,.
\label{phi_2_eq}
\eeq
The solution for this equation is:
\beq
\f_{2}=d_2+d_1 \log \left(1-\frac{1}{\rt^4}\right)-\frac{\Qc^2}{480 \rh^6 \rt} \left(\rt \left(\log \left(\frac{\rt-1}{\rt+1}\right)+2 \arctan(\rt)\right)+4\right)\ ,
\eeq
where $d_1$ and $d_2$ are integration constants. Similarly, we expand $\alpha(\rt)$ and $b(\rt)$ as:
\beq
\alpha(\tilde\rho)\,=\,\alpha_0\,+\,\epsilon\,\alpha_1(\tilde\rho)\,+\,\epsilon\,\delta^2\,\alpha_2(\tilde\rho)\,\,,
\qquad\qquad
b(\tilde\rho)\,=\,b_0(\tilde\rho)\,+\,\epsilon\,\delta^2\,b_2(\tilde\rho)\,\,,
\eeq
where $\alpha_0$, $\alpha_1(\tilde\rho)$, and $b_0(\tilde\rho)$ are the zeroth and first order terms in (\ref{eq:OrderEpsMetricFuncSubsConst}) per subscript labeling. Notice that $b(\tilde\rho)$ does not contain ${\mathcal O }(\epsilon))$ terms, in agreement with our result (\ref{eq:OrderEpsMetricFuncSubsConst})  for the black hole with vanishing chemical potential. The equations for $\alpha_2(\rt)$ and $b_2(\rt)$ are:
\bea
\a_{2}''+\frac{5 \rt^4-1 }{\rt \left(\rt^4-1\right)}\a_{2}'-\frac{\Qc^2 }{120 \rh^6\, \rt^3 \left(\rt^4-1\right)} & = & 0  \nonumber \\
 b_{2}''-\frac{5 \rt^4-9 }{\rt \left(\rt^4-1\right)}b_{2}'+
\frac{16}{\rt^2 \left(\rt^4-1\right)^2}b_{2}-\frac{\Qc^2}{30\, \rh^6\, \rt^7} & = & 0\ ,
\eea
and, remarkably, can be solved in terms of $\phi_2(\rt)$ as 
\beq
\alpha_2(\tilde\rho)\,=\,\phi_2(\tilde\rho)\,\,,
\qquad\qquad
b_2(\tilde\rho)\,=\,4\Big(1-{1\over \tilde\rho^4}\Big)\,\phi_2(\tilde\rho)\,\,.
\eeq
Similarly, we can expand the warp factor $h(\tilde\rho)$ as:
\beq
h(\tilde\rho)\,=\,h_0(\tilde\rho)\,+\,\epsilon\,h_1(\tilde\rho)\,+\,\epsilon\,\delta^2\,h_2(\tilde\rho)\,\,,
\eeq
where $h_0(\tilde\rho)$  and $h_1(\tilde\rho)$ can be extracted from (\ref{eq:OrderEpsMetricFuncSubsConst}). The second order contribution $h_2(\tilde\rho)$ satisfies the equation
\bear
 &&h_{2}''+\frac{13 \rt^4-9 }{\rt \left(\rt^4-1\right)}h_{2}'-\frac{16 }{\rt^2 \left(\rt^4-1\right)}h_{2} \rc\rc
 &&\quad\qquad
 +\frac{4 \Qc\, \rt^2 }{\rh^4 \left(\rt^4-1\right)^2}b_{2}+\frac{8 \Qc }{\rh^4\, \rt^2 \left(\rt^4-1\right)}\a_{2}-\frac{\Qc^3}{160 \rh^{10} \,\rt^7 \left(\rt^4-1\right)}=0\,\,,
\eear
which can also be solved in terms of the dilaton function  $\f_{2}(\tilde\rho)$ as:
\beq
h_{2}(\tilde\rho)=\frac{3 \Qc}{4 \rh^4 \rt^4}\,\f_{2}(\tilde\rho)\,\,.
\eeq

Let us next represent $F(\tilde \rho)$ and $S(\tilde \rho)$ as in (\ref{F_S_chemical}). The differential equations for $F_2(\tilde\rho)$ and $S_2(\tilde\rho)$ are coupled and contain $\alpha_2(\tilde\rho)$ and $b_2(\rt)$ as sources. These equations can be decoupled by performing the same linear combinations as in (\ref{decoupling_F1_S_1}), \ie\ by introducing new functions $g_1(\tilde\rho)$ and $g_2(\tilde\rho)$ as
\bea
F_2(\tilde\rho)={1\over\sqrt{2}}g_1(\tilde\rho)-{4\over\sqrt{17}}g_2(\tilde\rho)
\qquad\qquad S_2(\tilde\rho)={1\over\sqrt{2}}g_1(\tilde\rho)+{1\over\sqrt{17}}g_2(\tilde\rho)\,\,.
\eea
The equations satisfied by $g_1$ and $g_2$ are:
\bea
&& g_1''+\frac{1-5 \rt^4}{\rt-\rt^5}\,g_1'-\frac{32 \rt^2 }{\rt^4-1}\,g_1-
{24\,\sqrt{2}\,\tilde\rho^2\over \tilde\rho^4-1}\Big(d_1\log \left(1-\frac{1}{\rt^4}\right)+d_2\Big)\rc\rc
&&\qquad
+ {\Qc^2\,\tilde\rho^3\over 10\sqrt{2}\,\rh^6\,(\tilde\rho^4-1)}\Big(
 \log \left(\frac{\rt-1}{\rt+1}\right)+2\arctan (\tilde\rho)\Big)+
  {\Qc^2\,(80\tilde\rho^4+3)\over 
  200\sqrt{2}\,\rh^6\,\rt^3 (\rt^4-1)}=0
\rc\rc
&& g_2''+\frac{1-5 \rt^4 }{\rt-\rt^5}\,g_2'-\frac{12 \rt^2 }{\rt^4-1}\,g_2+\frac{\sqrt{17}\, \Qc^2 }{1200\,\rh^6 \,\rt^3 \left(\rt^4-1\right)}=0\,\,.
\label{g_1_g_2_eq}
\eea
The equation for $g_1$ can be explicitly integrated as:
\bea
 g_1(\tilde \rho) & = & -\frac{1}{2400 \sqrt{2}\rh^6 \rt} \Bigg(7200\rh^6 \rt^5  \left(d_1 \log \left(1-\frac{1}{\rt^4}\right)+2 d_1+d_2\right)+4 \Qc^2 \left(5 \rt^4-9\right) \rc\rc
&&+5 \Qc^2 \rt \left(\rt^4-2\right)  \left(\log \left(\frac{\rt-1}{\rt+1}\Bigg)+2 \arctan(\rt)\right)\right)+d_3 \left(1-2 \rt^4\right)\rc\rc
&&+\frac{1}{2} d_4 \left(2-\left(1-2 \rt^4\right) \log \left(1-\frac{1}{\rt^4}\right)\right)\,\,,
\label{g_1_sol}
\eea
where $d_3$ and $d_4$ are new integration constants. To integrate the equation for $g_2$ in (\ref{g_1_g_2_eq}) we again make use of the Wronskian method. The homogeneous part of the equation is the same as the one in the equation for $f_2$ in (\ref{eqs_f1_f2}) and, therefore, is solved by the functions $P$ and $Q$ defined in (\ref{P_Q_def}).  To write a special solution of the second-order inhomogeneous equation in (\ref{g_1_g_2_eq}), we define new functions 
${\cal I}_{P}^{ (2)}(x)$ and ${\cal I}_{Q}^{(2)}(x)$ as
\beq
{\cal I}_{P}^{ (2)}(x)\,=\,\int_{1}^{x} {P(z)\over z^2 }\,\dd z\,\,,
\qquad\qquad
{\cal I}_{Q}^{(2)}(x)\,=\,\int_{x}^{\infty} { Q(z)\over z^2 }\,\dd z\,\,.
\eeq
Then, the general solution for $g_2(\rt)$ can be written as:
\beq
g_2(\tilde \rho)=d_{5} \, P(\tilde \rho)+d_{6} \, Q(\tilde \rho)+{\sqrt{17}\,\pi\,\Qc^2
\over 9600\,\sqrt{2}\,\rh^6}
\Big(P(\tilde \rho)\,{\cal I}_{Q}^{(2)}(\tilde \rho)\,+\,Q(\tilde \rho)\,{\cal I}_{P}^{ (2)}(\tilde \rho)\Big)\,\,,
\eeq
where $d_5$ and $d_6$ are two new integration constants. By imposing regularity of $\phi_2$, $g_1$, and $g_2$, both at the horizon $\tilde\rho=1$ and at the UV $\tilde\rho\to\infty$, we can determine the constants of integration, which can be shown to be given by:
\bear
&&d_1\,=\,{d_2\over \pi}\,=\,{\Qc^2\over 480\,\rh^6}\ \ ,
\qquad\qquad
d_3\,=\,-{3+2\pi\over 480\,\sqrt{2}}\,{\Qc^2\over \rh^5}\rc\rc
&&d_4\,=\,{\Qc^2\over 120\,\sqrt{2}\,\rh^5}\ \ ,
\qquad\qquad
d_5\,=\,d_6\,=\,0\,\,.
\eear
Using these values for the $d_i$'s we get the expressions of $\phi(\rt)$, $\alpha(\rt)$, $b(\rt)$, and $h(\rt)$ displayed in eqs. 
(\ref{b_chemical}) and (\ref{phi_alpha_h_G_chemical}) of the main text. The expression for $G(\rt)$ in \ref{phi_alpha_h_G_chemical}) can be obtained from the other functions by using (\ref{G_gauge}).  The functions $F(\rt)$ and $G(\rt)$ can be written as in (\ref{F_S_chemical}), with $F_2(\tilde \rho)$ and $S_2(\tilde \rho)$ given by
\bear
F_2(\tilde \rho) & = & -{\pi\over 4\sqrt{2}}\,
\Big(P(\tilde\rho)\,{\cal I}_{Q}^{(2)}(\tilde\rho)+
Q(\tilde\rho)\,{\cal I}_{P}^{(2)}(\tilde\rho)\Big)\,-\,
{5\over 2}\,\tilde\rho^3+{5\over 8}\,+\,{9\over 2\,\tilde\rho}+
{1\over 2}
(2-\tilde\rho^4)\,\Omega(\tilde \rho)\rc\rc
S_2(\tilde \rho) & = & {\pi\over 16\sqrt{2}}\,
\Big(P(\tilde\rho)\,{\cal I}_{Q}^{(2)}(\tilde\rho)+
Q(\tilde\rho)\,{\cal I}_{P}^{ (2)}(\tilde\rho)\Big)\,-\,
{5\over 2}\,\tilde\rho^3+{5\over 8}\,+\,{9\over 2\, \tilde\rho}+
{1\over 2}
(2-\tilde\rho^4)\,\Omega(\tilde \rho)\ .
\label{F_2_S_2_explicit}
\eear
To complete the solution it remains to find $j(\rt)$. Its equation at leading order can be easily found from (\ref{eom_J}) and is given by
\be\label{eq:jExpanded}
j''+\frac{\rt^4+3}{\rt \left(\rt^4-1\right)}j'-\frac{9\, \rt^2 }{\rt^4-1}j-\frac{\Qc\, \rt}{2 \rh \left(\rt^4-1\right)}=0\ .
\ee 
The homogeneous version of this equation has the following two independent solutions:
\bea
&&J_1(\rt)=F\left(-\frac{3}{4},\frac{3}{4};1;1-\rt^4\right)\,\,,\qquad 
J_2(\rt)=\left(\rt^4-1\right)^{-3/4}\, \, F\left(\frac{3}{4},\frac{3}{4};\frac{5}{2};\frac{1}{1-\rt^4}\right)\,\,.\qquad\qquad
\eea
The general solution for $j(\rt)$ can be found by adding a linear combination of $J_1(\rt)$ and $J_2(\rt)$ and a special solution of the complete equation obtained again by the Wronskian method. Notice that $J_1(\rt)$ is regular at $\rt=1$ and blows up at $\rt\to\infty$, whereas $J_2(\rt)$ has the opposite behavior and diverges at the horizon and is regular at the UV. By taking these facts into account, it is not difficult to find the regular solution for $j(\rt)$. If we define ${\cal I}_{1}(\rt)$ and ${\cal I}_{2}(\rt)$ as
\beq
{\cal I}_{1}(\rt)\,=\,\int_1^{\rt}{J_1(z)\over z^2}\dd z\,\,,
\qquad\qquad
{\cal I}_{2}(\rt)\,=\,\int_{\rt}^\infty{J_2(z)\over z^2}\dd z
\eeq
then we have
\bea\label{eq:jSol}
j(\rt)\,=\,-\frac{\Qc\big[\Gamma 
\left(\frac{3}{4}\right)\big]^2}{8 \sqrt{\pi }\, \rh}\left(J_1(\rt)\, {\cal I}_{2}(\rt)+J_2(\rt)\, {\cal I}_{1}(\rt)\right) \ .
\eea
On the other hand,  we can use these results to find the worldvolume gauge field $A_t'$. Let us write:
 \bea
{1\over 2\pi\alpha'}A_t'(\rt)={1\over 2\pi\alpha'}\Big(A_t'^{\,(0)}(\rt)+\eps\,A_t'^{\,(1)}(\rt)\Big)=\tilde{\delta}\,\left(a_0(\rt)+\eps\,a_1(\rt)\right)\ ,
\label{At_prime_sol}
\eea
where
\bea\label{eq:at1}
a_0(\rt) & = & -\frac{\rh }{\sqrt{6} \pi \alpha'}{1\over \rt^2}\rc\rc
a_1(\rt) & = & -\frac{ \sqrt{3}\,\rh}{16\,\alpha' \rt^2}  \Big(P(\rt)\, I_Q(\rt)+Q(\rt) \,I_P(\rt)-\frac{\sqrt{2} \left(3 \left(1-4 \rt^3\right)-\left(3 \rt^4+1\right) {4\over 5}\Omega(\rt)\right)}{3 \pi }\rc\rc
&& +\frac{80\sqrt{2}\, \rh}{ \pi  \,\Qc}  j(\rt)\Big)\ .
\eea
The UV value of $A_t/(2\pi\alpha')$ is related to the chemical potential $\mu$. This value can be obtained by integrating (\ref{At_prime_sol}) with respect to $\rt$.  Doing this integration with the condition that $A_t(\rt=1)=0$, we get
\beq
{A_{t,UV}\over 2\pi\alpha'}\,=\,{A_t(\rt\to\infty)\over 2\pi\alpha'}\,=\,{1\over 2\pi\alpha'}
\int_{1}^{\infty}\,A_t'(\rt)\,\dd\rt\,=\,
\tilde{\delta}\,\left(a_{0, UV}\,+\,
\eps\,a_{1, UV}\right)\,\,,
\eeq
where $a_{0, \text{UV}}$ and $a_{1, \text{UV}}$ are given by
\bear
a_{0,\text{UV}} & = & \int_{1}^{\infty}\,a_0(\tilde\rho)\,\dd\tilde\rho\,=\,-
{\rh\over \sqrt{6}\,\pi\alpha'}\rc\rc
a_{1,\text{UV}} & = & \int_{1}^{\infty}\,a_1(\tilde\rho)\,\dd\tilde\rho\,\approx\,-1.06\,
a_{0, \text{UV}}\ ,
\label{A_t_UV}
\eear
The integration of $a_{0, \text{UV}}$ in (\ref{A_t_UV}) can be done analytically, whereas the one needed to get $a_{1, \text{UV}}$ has to be done numerically. By comparing these results with the chemical potential calculated in (\ref{mu_eps_delta}), we notice that the ${\cal O}(\epsilon\tilde\delta)$ terms are different (see (\ref{A_t_mu})). This difference can be attributed to the backreaction of the flavors.

\vskip 2cm
\renewcommand{\theequation}{\rm{D}.\arabic{equation}}
\setcounter{equation}{0}

\section{Dimensional reduction}\label{appendix_D}

The ten-dimensional supergravity action  can be dimensionally reduced both to four and five dimensions. 
The details of these constructions were given in \cite{Penin:2017lqt} and will be briefly reviewed in this appendix. Since this reduced theory corresponds to vanishing chemical potential, in this appendix we restrict ourselves to this $\mu=0$ case. In the reduction to five dimensions we adopt the following ansatz for the metric
\beq
\dd s ^2_{10}=e^{{10\over 3}\g}g_{pq}\dd z^p\dd z^q+e^{-2(\g+\l)}\dd s ^2_{\rm{KE}}+e^{2(4\l-\g)}\left(\dd \t+{\cal A}\right)^2\,\,,
\eeq
where $g_{pq}$ is a  metric for the coordinates $z^{p}=(t,x^1, x^2,x^3, \rho)$ and $(\gamma, \lambda)$ are scalar fields depending on the 5d coordinates $z^p$. In our ansatz $\gamma$ and $\lambda$ depend only on the radial coordinate $\rho$ and are related to the functions of the ten-dimensional metric as
\beq
 \l={1\over5}\left(\log F-\log S\right)\,\,,
\qquad\qquad \g={1\over 10}\left(-2\log F-8\log S-{5\over 2}\log h\right)\,\,.
\eeq
Moreover, the reduced 5d metric takes the form
\beq
\dd s ^2_{5}=-c_1^2\,\dd t^2+c_2^2\left((\dd x^1)^2+(\dd x^2)^2\right)+c_3^2\,(\dd x^3)^2+c_{\rho}^2\,\dd \r^2\,\,,
\eeq
where the different coefficients are:
\bear
&&c_1=F^{1/3} S^{4/3}b^{1/2}h^{1/6}\,\,,\qquad \qquad c_2=F^{1/3} S^{4/3}h^{1/6}\rc\rc
&& c_3=\a\,F^{1/3} S^{4/3}h^{1/6}\,\,,\qquad \qquad c_{\rho}=\frac{\a\, F^{4/3} h^{2/3} S^{16/3}}{\r^{5}b^{1/2}}\,\,.
\label{cs_5d}
\eear
We will use the reduced theory to compute the VEV of the stress-energy tensor. This calculation makes use of the holographic renormalization formalism to regulate the on-shell action. The VEV of the stress-energy tensor of the dual theory is obtained by taking the functional derivative of the renormalized theory with respect to the boundary metric. The result is \cite{Penin:2017lqt}:
\bea
\langle T^\m{}_\n \rangle={V_5\over 2\k_{10}^2}\sqrt{\g}\left(-2K^\m{}_\n+\d^\m{}_\n\left(2K+W_{5d}\right)\right)\Big|_{\r\to\infty}+\Qf{V_5\over \k_{10}^2}\sqrt{\hat{\g}}e^{2\g+2\l+{\f\over 2}}\d^\m{}_\n\Big|_{\r\to\infty}\,\,,
\label{VEV_T_5d}
\eea
where $\gamma$ is the determinant of the metric induced on constant $\rho$ slices, $K^\m{}_\n$ is the extrinsic curvature of these constant $\rho$ surfaces ($K=K^\m{}_\m$), and $W_{5d}$ is the superpotential, given by
\bea
 W_{5d}=-6e^{{8\over 3}\g-4\l}-4e^{{8\over 3}\g+6\l}+\Qc e^{{20\over 3}\g}\,\,.
\eea
In the last term in (\ref{VEV_T_5d}), $\hat\gamma$ is the determinant of the metric induced on  constant $\rho$ and $x^3$ slices. It is not difficult to show that, at first order in $\epsilon$, that
\beq
\langle T^\m{}_\n \rangle=\text{diag}(-\mathcal{E}, p_{xy}, p_{xy}, p_z)=
\mathcal{E}\,\text{diag}\Bigg(-1\,,\,{1\over 3}\,+\,{10\over 9}\eps\,,\,{1\over 3}\,+\,{10\over 9}\eps\,,\,{1\over 3}\,-\,{20\over 9}\eps\Bigg)\,\,,
\label{VEV_T_result}
\eeq
where $\mathcal{E}$ is the energy density:
\bea
\mathcal{E}={3\,{\rm Vol} \big({\cal M}_5\big)\over (2\pi)^7\,\alpha'^4\,g_\mt{s}^2}\,\rh^4
\left(1\,-\,\frac{5  }{6}\epsilon\right)\ .
\eea
This result matches with the one found by thermodynamic methods in section \ref{Thermo}.

\subsection{Hydrodynamics in the shear channel}\label{hydro_appendix}

The five-dimensional supergravity theory can be further reduced along the $x^3$-direction. The $5d\to 4d$ ansatz for the reduction takes the form\cite{Penin:2017lqt}:
\beq
 \dd s ^2_5\,=\,e^{-\beta}\,\dd s ^2_4\,+\,e^{2\beta}\,(\dd x^3)^2\ ,
\eeq
where $\beta$ is a new scalar field of the $4d$ supergravity. For our ansatz  $e^{\beta}\,=\,c_3$, where $c_3$ is the function displayed in (\ref{cs_5d}). This $4d$ theory can be used to compute the VEV of the stress-energy tensor of the dual theory along the $(t,x^1, x^2)$-directions. The result of this calculation coincides with the one displayed in (\ref{VEV_T_result}) and will not be repeated here. Moreover, we can use the $4d$ theory to obtain the transport coefficients for perturbations propagating along the ($x^1,x^2$)-plane. In this appendix we will analyze the so-called shear channel. With this purpose we fluctuate the $4d$ metric about the background solution via the substitution
\beq
g_{mn}\,\to\,g_{mn}\,+\,h_{mn}\,\,,
\eeq
where $g_{mn}$ is the reduced metric of the first-order solution at zero chemical potential. We will work in the radial gauge for the metric,  in which
\beq
h_{m\rho}=0\,\,,
\qquad\qquad
(m=t, x^1, x^2, \rho)\,\,.
\label{radial_gauge}
\eeq
Without loss of generality we consider a perturbation propagating along the $x^2$-direction. In the shear channel only the metric fluctuations $h_{t x^1}$ and $h_{x^1\,x^2}$ are excited. Let us assume that these fluctuations (in the Fourier space) have frequency $\omega$ and momentum $q$ and, accordingly, let us parametrize them as
\beq
h_{t x^1}=e^{-i\left(\omega t-qx^2\right)}c_2^2(\rho)\,H_{t x^1}(\r)\,\,,
\qquad\qquad
h_{x^1\,x^2}=e^{-i\left(\omega t-qx^2\right)}c_2^2(\rho)\,H_{x^1\,x^2}(\r)\,\,.
\eeq
It turns out that the equations for the fluctuations can be reduced to a single second-order differential equation for a gauge invariant combination $X=X_0+\eps\,X_1$, defined as
\be
X(\r)=q\,H_{tx^1}(\r)+\omega H_{x^1\,x^2}(\r)\,\,.
\ee
In order to write the fluctuation equation in a simple form, let us perform the following change of the radial variable
\be
x\,=\,\sqrt{b(\r})\,=\,\sqrt{1-{\rh^4\over \r^4}}\,\,.
\ee
In this new variable, the horizon is at $x=0$, while the boundary resides at $x=1$. Moreover, let us work with the dimensionless frequency and momentum $\hat{q}$ and $\hat{\w}$ defined in (\ref{hat_q_hat_omega_def}). Then, if now the prime denotes derivative with respect to $x$, the fluctuation equation becomes:
\bea\label{eq:Xeqx}
X''\,-\,\frac{\hat{q}^2 x^2+\hat{\w}^2}{x \left(\hat{q}^2 x^2-\hat{\w}^2\right)}\,X' \,+\,
\Gamma\,\frac{ \hat{q}^2 x^2-\hat{\w}^2}{x^2 \sqrt{1-x^2}\,k(x)}\,X=0\,\,,
\eea
where $\Gamma$ is the constant:
\bea
\Gamma=\frac{ \pi^2  T^2\Qc}{ \rh^2}=4\,-\,15 \,\epsilon\,\,,
\eea
and $k(x)$ is the following function 
\bea
k(x)=4\left( x^2-1\right)+5 \epsilon  \left(x^2-1+4 \left(1-x^2\right)^{1/4}+{4\over 5}\,x^2\, 
\Omega\left({1\over 
\left(1-x^2\right)^{1/4}}
\right)
\right)\,\,.
\eea
Notice that $\Gamma$ and $k(x)$ contain terms which are zeroth-order in the flavor parameter, together with the ones that are first-order in $\epsilon$. We want to find solutions of (\ref{eq:Xeqx}) which satisfy infalling boundary conditions at the horizon $x=0$, as well as Dirichlet boundary conditions at the UV boundary $x=1$. These solutions only exist if certain dispersion relation $\hat \omega=\hat\omega(\hat q)$ is satisfied. Our objective in this section is to find such dispersion relation in a power series of $\hat q$. Actually, we will only keep quadratic and quartic terms of $\hat q$, as in (\ref{hatted_disp_rel_shear}), which contains two transport coefficients $\hat D_{\eta}$ and $\hat\tau_s$.  

In order to impose infalling boundary conditions at the horizon $x=0$, we adopt an ansatz for $X(x)$
\beq
X(x)\,=\,x^{-i\hat \omega}\,S(x)\,\,,
\label{infalling_ansatz }
\eeq
where $S(x)$ must be regular at $x=0$. Let us expand $S(x)$ in powers of $\hat q$ as:
\beq
S(x)\,=\,S_0(x)\,+\,\hat q^2\,S_2(x)\,+\,\ldots\,\,.
\label{S_q_expansion}
\eeq
Plugging (\ref{infalling_ansatz }) and the expansion (\ref{S_q_expansion}) into (\ref{eq:Xeqx}) we get the following two differential equations for $S_0(x)$ and $S_2(x)$:
\bea
S_0''-{1\over x}\,S_0'  & = & 0\rc\rc
S_2''-\frac{1}{x}S_2'   & = & -\Bigg[{2\hat{D}_\eta\over x^2}+{\Gamma\over \sqrt{1-x^2}\,k(x)}\Bigg]S_0-
\frac{2\hat{D}_\eta\,\left(\hat{D}_\eta-x^2\right)}{x^3}\,S_0'\,\,.
\label{S_0_2_eq}
\eea
We can immediately  integrate the equation for $S_0$ in (\ref{S_0_2_eq}), yielding
\bea
S_0(x)\,=\,c_1{x^2\over 2}\,+\,c_2\ ,
\label{S_0_integrated}
\eea
where $c_1$ and $c_2$ are integration constants.  The differential equation for $S_2$ in (\ref{S_0_2_eq}) contains terms which depend on the flavor expansion parameter $\epsilon$. Since we are working at first order in $\epsilon$, it is convenient to expand the transport coefficients in powers of $\epsilon$ as:
\beq
\hat{D}_\eta\,=\,\hat D_0\,+\,\epsilon\,\hat D_1\,\,,
\qquad\qquad
\hat\tau_s\,=\,\hat\tau_0\,+\,\epsilon\,\hat\tau_1\,\,,
\eeq
as well as for the function $S_2$:
\be
S_2(x)\,=\,s(x)+\eps\,\tilde{s}(x)\,\,.
\ee
Proceeding in this way we generate two second-order differential equations for $s(x)$ and $\tilde{s}(x)$, which can be analytically integrated (these functions have rather long expressions and will not be written here).  This integration generates two new integration constants $c_3$ and $c_4$ for $s(x)$ and another two $c_5$ and $c_6$ for $\tilde{s}(x)$. Expanding our analytic result for $s(x)$ near $x=0$, up to second order in $x$, we get
\bear
s(x) & = & \frac{1}{2} \Big(2 \hat D_0  (c_1 \hat D_0+c_2)\log x-c_1-c_2+2 c_4\Big)\rc\rc
&&+ \frac{1}{4} x^2 \left(c_1 \hat D_0 (2 \log x-1)+c_1+2 c_2 \log \left(\frac{x}{2}\right)+c_2+2 c_3\right)+\ldots\,\,.
\label{s_x}
\eear
Similarly, $\tilde{s}(x)$ near $x=0$ takes the form
\bear
\tilde{s}(x) & = & -
\Big[(6 \hat D_1+5i\pi-5-10\log2)c_1+\big((15+25i)\pi-70+40\log 2\big)c_2-12c_5\Big]{x^2\over 24}\rc\rc
&&
+\Big[{c_1\over 2} x^2+2c_1\hat D_0+c_2\Big]\,\hat D_1\,\log x\,+\,
{1\over 12}\big(10 i\pi+35-20\log 2)c_1\rc\rc
&&\qquad\qquad\qquad\qquad
+{1\over 12}\big(5i\pi-5-10\log 2\big)c_2\,+\,c_6 +  \ldots \ .
\label{tilde_s_x}
\eear
Combining all these equations, we get $S(x)=S_0(x)+\hat q^2 (s(x)+\epsilon \tilde s(x))$ for small $\hat q$ and small $x$. Let us now obtain an expression by expanding $S(x)$ in opposite order and demand that it should be consistent with the above. We first expand near $x=0$ and write
\bea\label{eq:Ssigmas}
S(x) = 1+\s_2 \,x^2+\s_4\,x^4 + \ldots \ .
\eea
Plugging the expansion (\ref{eq:Ssigmas}) into (\ref{eq:Xeqx}) we get $\s_2$ and $s_4$ in terms of $\hat\omega$ and $\hat q$. Expanding this result for small $\hat{q}$ and $\eps$, $\s_2$, we obtain
\bea
\s_2 & = & -\frac{1}{2 \hat D_0}+\frac{1}{4} \hat{q}^2 (2 \hat \t_0-1)+\epsilon  \left(\frac{\hat D_1}{2 \hat D_0^2}+\frac{\hat{q}^2 \hat \t_1}{2}\right)+\ldots\rc\rc
\s_4 & = & -\frac{(4-12\hat D_0) \hat{q}^2}{64 \hat D_0}-\epsilon \frac{\hat{q}^2  \left(5 \pi  \hat D_0^2-30\hat D_0^2+ 30\hat D_0^2\log 2 -4 \hat D_1\right)}{64 \hat D_0^2}+\ldots \ .
\eea
Thus, an expansion for small $x$ and small $\hat{q}$ \eqref{eq:Ssigmas} leads to the following expressions for $S_0(x)$ and $S_2(x)$:
\bear
S_0(x) & = & 1-\frac{x^2}{2 \hat D_0}+\eps\,\frac{\hat D_1 x^2}{2 \hat D_0^2}\rc\rc
S_2(x) & = & {12 x^4\hat D_0^2\,+\,16(2\hat \tau_0-1)\hat D_0^2\,x^2\,-\,4\hat D_0\,x^4\over 
64\hat D_0^2}\rc\rc
&&
+{\epsilon\over 64\,\hat D_0^2}\,\Big[
(30-5\pi-30\log 2)\hat D_0^2\,x^4\,+\,4\hat D_1\,x^4\,+\,32\hat\tau_1\,\hat D_0^2\,x^2\Big]\ .
\label{S_0_S_second_expansion}
\eear
Let us now compare (\ref{S_0_S_second_expansion}) and (\ref{S_0_integrated}). Since the right-hand side of (\ref{S_0_integrated}) does not depend on $\epsilon$, we conclude that
\beq
\hat D_1=0\,\,.
\eeq
Moreover, by comparing the expression for $S_2$ in (\ref{S_0_S_second_expansion})  and the one that follows from (\ref{s_x}) and (\ref{tilde_s_x}), we can relate the different integration constants and the transport coefficients as
\bea\label{eq:ctesHydro}
&&c_1=-{1\over \hat D_0}\,\,,\qquad  c_2=1 \qquad c_3=\frac{\hat D_0(2 \hat \t_0-3+\log 4)+1}{2 \hat D_0}\,\,,
\qquad c_4={1\over 2}-{1\over2\hat D_0}\rc\rc
&&c_5=\frac{\hat D_0 (12\hat  \t_1+\pi  (15+25 i)-70+20 \log 4)+5 (1-i \pi +\log 4)}{12 \hat D_0}\rc\rc
&&c_6=\frac{5 (\hat D_0(2-2 i \pi +\log (16))+14+4 i \pi -\log (256))}{24 \hat D_0}\,\,.
\eea
With these constants, the function $S_0$ becomes
\be
S_0(x)=1-{1\over 2\,\hat D_0}x^2\,\,.
\ee
Imposing the Dirichlet boundary condition   $S_0(x=1)=0$  at $x=1$ leads to the result for $\hat D_0$,
\beq
\hat D_0={1\over 2}\,\,.
\label{hat_D0_result}
\eeq
Using this new condition and \eqref{eq:ctesHydro} we arrive at the following expressions for $s(x)$ and $\tilde{s}(x)$
\bea
 s(x) & = & \frac{1}{2} \Bigg(x^2 \left( \t_0-\log \big(x/ 2\big)\right)-1+\sqrt{1-x^2}+{1\over 2}\log \left(\frac{1-\sqrt{1-x^2}}{\sqrt{1-x^2}+1}\right)\rc\rc
&&
+\left(1-x^2\right) \atanh\left(\sqrt{1-x^2}\right)\Bigg)\rc\rc
\tilde{s}(x) & = & -\frac{5 \left(1-x^2\right)^{3/4} }{x^2}\left(\frac{1}{3} \, F\left(1,1;\frac{7}{4};1-\frac{1}{x^2}\right)-\frac{1}{21} 2 \left(1-x^2\right) \, F\left(1,2;\frac{11}{4};1-\frac{1}{x^2}\right)+\frac{2 x^2}{3}\right)\rc\rc
&&
+\frac{1}{4} x^2 (2 \t_1+5 (\log (4)-2))-\sqrt{1-x^2} \left(\Omega \left(\frac{1}{
(1-x^2)^{{1\over 4}}}\right)+\frac{15}{4}\right)\rc\rc
&&
+\frac{5}{4} \left(x^2+2\right) \atanh \left(\sqrt{1-x^2}\right)-\frac{5}{4} (\pi -5+\log (4))\ .
\eea
Requiring that $s(x=1)=0$ determines the value of $\hat\tau_0$, which turns out to be
\beq
\hat \t_0=1-\log 2\ .
\eeq
Moreover, the Dirichlet boundary condition $\tilde s(x=1)=0$ is satisfied if $\hat \t_1$ is given by
\beq
\hat \t_1=\frac{5}{2} (\pi -3)\ .
\eeq

\bibliographystyle{JHEP}
\bibliography{biblio}

\end{document}